\documentclass[twocolumn,showkeys,showpacs,prd,amsmath,amssymb,aps,floats,floatfix]{revtex4}

\usepackage[caption=false]{subfig}
\usepackage{epsfig}
\usepackage{graphicx}
\usepackage{dcolumn}
\usepackage{bm}
\usepackage[usenames]{color}
\usepackage{tabularx}
\def\be{\begin{equation}}
\def\ee{\end{equation}}
\def\bea{\begin{eqnarray}}
\def\eea{\end{eqnarray}}
\def\barr{\begin{array}}
\def\earr{\end{array}}
\def\ben{\begin{enumerate}}
\def\een{\end{enumerate}}
\def\nn{\nonumber}

\def\gev{\; {\rm GeV} }

\def\bea{\begin{eqnarray}}
\def\eea{\end{eqnarray}}
\def\bsub{\begin{subequations}}
\def\esub{\end{subequations}}

\def\mtt{m_{t\bar{t}} }
\newcommand{\mttb}{\ensuremath{m_{t\bar{t}}\, }}
\def\ttbar{t\bar{t}}
\def\tt{$t\bar{t}$ }
\newcommand{\du}{\ensuremath{d_{\cal U}\, }}

\newcommand{\dut}{\ensuremath{d_{\cal U}^T\, }}
\def\ou{{{\mathcal O}_{\cal U}}}
\def\scaleu{\Lambda }
\def\la{\Lambda}
\newcommand{\lau}{\ensuremath{\Lambda_{\cal U}} }
\newcommand{\afbt}{\ensuremath{A_{_{FB} }^{t \bar t}\, } }
\newcommand{\cfbt}{\ensuremath{C_{_{FB} }^{t \bar t}\, }}
\newcommand{\spincorr}{\ensuremath{C^{t \bar t}\, }}
\newcommand{\crtt}{\ensuremath{\sigma^{t \bar{t}}\, }}
\newcommand{\lrr}{\ensuremath{\lambda_{RR}^V }}
\newcommand{\lall}{\ensuremath{\lambda_{LL}^V }}
\newcommand{\lrl}{\ensuremath{\lambda_{RL}^V }}
\newcommand{\llr}{\ensuremath{\lambda_{LR}^V }}

\newcommand{\laa}{\ensuremath{\lambda_{AV}^V }}
\newcommand{\lrrs}{\ensuremath{(\lambda_{RR}^V)^2 }}
\newcommand{\lalls}{\ensuremath{(\lambda_{LL}^V)^2 }}
\newcommand{\lrls}{\ensuremath{(\lambda_{RL}^V)^2 }}
\newcommand{\llrs}{\ensuremath{(\lambda_{LR}^V)^2 }}

\newcommand{\lrrts}{\ensuremath{(\lambda_{RR}^T)^2 }}
\newcommand{\lallts}{\ensuremath{(\lambda_{LL}^T)^2 }}
\newcommand{\lrlts}{\ensuremath{(\lambda_{RL}^T)^2 }}
\newcommand{\llrts}{\ensuremath{(\lambda_{LR}^T)^2 }}

\newcommand{\lrrt}{\ensuremath{\lambda_{RR}^T }}
\newcommand{\lallt}{\ensuremath{\lambda_{LL}^T }}
\newcommand{\lrlt}{\ensuremath{\lambda_{RL}^T }}
\newcommand{\llrt}{\ensuremath{\lambda_{LR}^T }}

\newcommand{\lprr}{\ensuremath{(g^{ut}_{R})^2 }}
\newcommand{\lpll}{\ensuremath{(g^{ut}_{L})^2 }}

\newcommand{\lplr}{\ensuremath{g^{ut}_L g^{ut}_R }}

\newcommand{\lprrsq}{\ensuremath{(g^{ut }_{R})^4 }}
\newcommand{\lpllsq}{\ensuremath{(g^{ut }_{L})^4 }}
\newcommand{\lprlsq}{\ensuremath{(g^{ut }_{L}g^{ut }_{R})^2 }}

\newcommand{\ct}{\ensuremath{c_\theta}}
\newcommand{\st}{\ensuremath{s_\theta}}
\newcommand{\bt}{\ensuremath{\beta_t}}
\newcommand{\cfb}{\ensuremath{C_{FB }^{t} }}
\begin{document}

\title{Constraining Unparticles from Top Physics at TeVatron}
\author{Mamta Dahiya$^a$}
\author{Sukanta Dutta$^a$}
\author{Rashidul Islam$^{a,b}$}\email{rislam@physics.du.ac.in}
\affiliation{$^a$SGTB Khalsa College, University of
Delhi. Delhi-110007. India.}
\affiliation{$^b$Department of Physics and Astrophysics, University of
Delhi. Delhi-110007. India.}
\begin{abstract}
We study and analyze the recent observations of the top pair production $\sigma \left(p\bar p\to t \bar t\right)$  at
TeVatron through flavor conserving and flavor violating channels ${\it
  via} $ vector and tensor unparticles. The unparticle sector is
considered with the possibility of being a color singlet or octet. The modified unparticle propagator is used to investigate the contribution of these
unparticles to the observed $A_{FB}^{t\bar t } $  (forward backward asymmetry in top pair production)  and the spin
correlation at TeVatron. We have also studied the impact of the flavor violating  couplings of unparticles to the third generation quarks on (a)
pair production of same sign tops/antitops $\sigma \left(p\bar p\to tt+\bar t \bar t\right)$ at TeVatron and (b) the partial top decay width for $\Gamma_{\cal U}\left(t\to u\,{\cal U}^V\right)$. 
\par We find that  a large region of parameter space  is
consistent with the measurements of \tt production cross-section, \afbt and spin correlation coefficient at TeVatron and observe  that   the top decay width measurement constrains the flavor violating coupling of vector unparticles more severely  than the same sign top/antitop production at TeVatron.  We also predict the best  point-set  in the model parameter space for specific choices of \du corresponding to $\chi^2_{\rm min}$ evaluated  using  the  \mttb ~ spectrum of \afbt  from the data set of Run II of TeVatron at the integrated luminosity  $8.7$~fb$^{-1}$. Our results and analysis are consistent  even with unparticle theories  having broken scale invariance as long as the infrared cut-off scale is much less than the top pair production threshold.  \end{abstract} 
\maketitle
\section{Introduction} 
\label{sec:intro} 
The study of the top quark production and related discrepancies at TeVatron and LHC might hold key to new physics beyond standard model (SM). Both CDF and D\O\, collaborations have consistently measured values of $t\bar t$  production cross-sections through various decay channels~\cite{cdf-top-cross,cdf-cross,d0-cross}
and they are all consistent with the theoretical predictions at NNLO
level \cite{kidonakis-tcross,Baernreuther:2012ws}. On the other hand the top quark
forward-backward asymmetry is observed to be significantly larger than
what the SM predicts~\cite{Abazov:2011rq,afbt-cdf-mar2012,AFBt-cdf-comb,AFBt-cdf1,Aaltonen:2011kc}.
The recent measurements of  $\afbt $ at
CDF obtains parton level asymmetry to be $0.296 \pm 0.067 $
for $\mtt >450$ GeV with $8.7$~fb$^{-1}$ of data  in contrast to the
NLO QCD prediction of $0.100$~\cite{afbt-cdf-mar2012}.  It has also been
observed that in the $t \bar t$ rest frame, the asymmetry
increases with the $t \bar t$ rapidity difference and with the
invariant mass. If this asymmetry is true, it should
indicate the presence of new physics. Hence the study of this subject
has drawn a lot of attention and various explanations have been given
for the observed deviations in the context of different new physics
scenarios~\cite{afb-models,Jung:2009pi, Shu:2011au,Gresham:2011pa, AguilarSaavedra, Chen:2010hm}. 
Any model trying to account for the high values of \afbt is constrained by
 the SM consistency in the measured cross-section and $m_{t \bar t}$
 spectrum of \afbt.

In the present article we examine the forward-backward (FB) asymmetry
and spin correlations in top-pair
production at TeVatron with the possibility of existence of
a conformally invariant hidden sector containing unparticles coupling weakly with SM fields~\cite{Georgi:2007ek}. The effective couplings of unparticle
with SM fields are likely to interfere with the SM processes and 
hence affect the $\afbt$ and the  top spin correlations.
Effects of unparticles on top-antitop quark pair production process at
hadron colliders and ILC have been studied in references~\cite{Choudhury:2007cq,unp-top,Arai:2009cp}.

We organize the paper as follows: Calculation of forward-backward
asymmetry \afbt and that of spin correlation \spincorr are discussed in sections~\ref{sec:afb} and 
\ref{sec:spin-corr} respectively. We  review the unparticle 
scenario in section~\ref{sec:unp}.  Numerical results 
for the flavor conserving and violating interactions  are given in section~\ref{sec:num-results}. Section \ref{SST} analyses the constraints to the flavor violating channels emerging from two important experimental signatures (a) same sign top production and the (b) top decay width. Section \ref{analysis} presents the $\chi^2$ analysis of the model and exhibits  the $m_{t\bar t}$ distribution of \afbt. This section also addresses the implication of the broken scale invariance by introduction of a mass gap and finally summarizes the observations.
\section{The observables}
\subsection{Forward-Backward Asymmetry} 
\label{sec:afb} 
The $t \bar t$ differential charge asymmetry at the partonic level is defined as $  A_C (\cos{\theta}) =
\frac{N_t(\cos{\theta}) -N_{\bar t}(\cos{\theta})
 }{N_t(\cos{\theta}) +N_{\bar t}(\cos{\theta}) }$, where $N_t(\cos{\theta}) = d\sigma^{t\bar t}/d(\cos{\theta})$, $\theta $ being  the polar angle of the top quark momentum with
respect to the incoming parton in the $t \bar t$ rest frame (which is
 same as $q \bar q$ rest frame for $q \bar q \to t \bar t$, the dominant 
production process at the TeVatron) while in the lab frame it will
 correspond to the polar angle between the top quark and the proton beam.
 The
charge conjugation invariance of the strong interaction would imply 
$N_t(\cos{\theta}) = N_{\bar t}(-\cos{\theta}) $ and the 
difference in the production of top quarks in the forward and backward
hemispheres is equivalent to the difference in the production of top
and antitop quarks in the forward hemisphere. Thus the integrated FB asymmetry in the lab frame is
equivalent to the charge asymmetry  and can be written as 
\be 
\afbt= \frac{N_t(\cos{\theta}\ge 0) -N_t(\cos{\theta}\le 0)
 }{N_t(\cos{\theta}\ge 0) +N_t(\cos{\theta}\le 0) } 
\label{eq:int-afb-def-cos} 
\ee
 Collinear initial-state radiation (ISR) makes the fundamental $q \bar q$ frame inaccessible in both experiment
and simulation, leaving a choice between the $t \bar t$ rest frame or the lab ($p \bar p$) frame. The direction of top quark in the  lab frame can be determined using cosine of the polar angle between
 the hadronically decaying top quark and the proton beam. However the information on fundamental production asymmetry in the lab frame is diluted because of  an uncontrolled longitudinal boost from the
 rest frame of primary $q \bar q$ interaction to the
 laboratory frame.
In the $t \bar t$ rest frame, a measurement of the variable $\cos{\theta}$ in  equation ~\eqref{eq:int-afb-def-cos}
requires reconstruction of the initial parton ($q\bar q$) rest frame 
which is not accessible experimentally. 
Hence instead of  $\cos{\theta}
$, the rapidity difference $\Delta y$ between top quark $y_t$ and the anti-top quark 
$y_{\bar t}$ is considered as the sensitive variable for the measurement 
of \afbt experimentally in the $t \bar t$ rest frame. This variable being Lorentz invariant can be
measured in the lab frame and the shape of rapidity
distributions $d\sigma/dy$ in the two frames would remain the same if the
 boost is by a constant velocity.  Experimentally the rapidity difference $\Delta y$
is calculated from the rapidity difference $y_{{l^+_t (l^-_{\bar t})}} - y_{h_{ \bar t} (h_{t})}$ of
the semileptonically decaying top and the hadronically decaying
top~\cite{Aaltonen:2011kc} and the top rapidity in the $t \bar t$ rest frame
 comes out to be $\frac{1}{2} \Delta y$ in the limit 
of small $p_T$. The variables $\Delta y$ and $\cos{\theta}$ are directly 
related  by the relation 
\bea
 y_{_t} -  y_{_{\bar t}} = \Delta y &=& 2\, {\rm tanh}^{-1}\left(\frac{\cos{\theta}}{\sqrt{1 - \frac{4 m_t^2}{\hat s} }} \right)
\label{eqn:rapidity-diff}
\eea
$m_t$ being the top mass and $\hat s$ the square of the
center of masss energy of the $t \bar t $ pair. Thus $\Delta y$
 is a close estimate of the production angle in the
$t \bar t $ frame.  Also the sign of rapidity difference is same as
$\cos{\theta}$ and thus the asymmetry in $\Delta y$ is identical to
asymmetry in top production angle $\cos{\theta}$ in the $t \bar t $
rest frame, allowing an
effective measurement in the  $t \bar t $ frame. Hence D\O\, and CDF have considered the following observable
which also reflects the integrated charge asymmetry in \tt production at
TeVatron:
\bea
\afbt &=& \frac{N\left[\Delta y\, \ge 0\right] -
N\left[\Delta y\, \le 0\right] 
 }{N\left[\Delta y \, \ge 0\right] +
 N\left[\Delta y \, \le 0\right]  }= \frac{N_+ -N_-}{N_+ +N_-}
\label{eq:diff-afb-rapidity}
\eea
$N_+(N_-)$ being the number of events with a positive (negative)
rapidity difference.  
\par
\begin{figure*}[!ht] 
\begin{center}
\includegraphics[width=16cm]{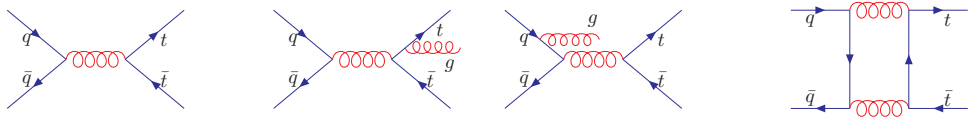}
\vskip -18 cm
\caption{{\em QCD diagrams contributing to charge asymmetry in top pair
  production}.\label{fig:feyn-diag} }
\end{center}
\end{figure*}

  Within SM the 
differential distributions of top and anti-top are identical at
leading order in $\alpha_s$. At next-to-leading order (NLO) 
 (${\cal O}(\alpha_s^3)$) the dominant positive FB asymmetry is generated from  the interference  between the Born amplitude and two-gluon exchange
(box) along  with a negative asymmetry from the interference of
initial state and final state gluon bremsstrahlung
(figure ~\ref{fig:feyn-diag}).  The inclusive charge asymmetry receives
contribution from the radiative corrections to quark antiquark
annihilation (mentioned above) and also from interference between
various amplitudes contributing to quark-gluon scattering ($q g \to t
\bar t q$ and $\bar q g \to t \bar t \bar q$), the latter contribution
being much smaller than the former. Gluon-gluon fusion
remains charge symmetric and $gg$ initial state does not contribute to
this asymmetry at any order in perturbation theory, due to the fact
that the gluon distribution is the same for protons and
antiprotons. It  however lowers the average value.  QCD predicts
the size of this asymmetry to be 5 to 8\%~\cite{afb-sm,Kidonakis:2011zn}. 
The measurement of the total FB asymmetry has been carried out by CDF
and D\O\, collaborations at TeVatron in both laboratory frame ($p\bar
p$ frame) as well as the center-of-mass frame of the top pair
($\ttbar$
frame)~\cite{Abazov:2011rq,AFBt-cdf1,Aaltonen:2011kc} (see
table~\ref{tab:afb-val}).  While the other TeVatron measurements of top
quark properties (e.g. production cross-section) are all consistent
with the SM, all the past measurements of CDF and D\O\, have
consistently yielded a higher $\afbt$ than SM prediction by
more than $2\sigma$ deviation (see table~\ref{tab:afb-val}).
\begin{table}[!ht]
\begin{center}
\begin{tabular}{|c|c|}
\hline
\hline
 {\bf Description }& {\bf Value of \afbt in specific frame } 
\\
& {\bf ($p \bar p$ rest frame or $t \bar t$ rest frame )} \\
\hline
CDF ($L = 8.7 $\, fb$^{-1}$) & $0.162 \pm .047^{[{\rm  stat}+{\rm syst}]} (t \bar t$)\cite{afbt-cdf-mar2012}\\
(semileptonic)&\\
\hline
CDF ($L \sim 5$\, fb$^{-1}$) & $0.201 \pm .067^{[{\rm  stat}+{\rm syst}]}$\cite{AFBt-cdf-comb}\\
(combined)& $0.158 \pm .072^{(\rm
  stat)}\pm .017^{(\rm syst)} $ ($t \bar t$)  \\
\hline
CDF($L = 5.1$\, fb$^{-1}$)   & $0.42 \pm .15^{\rm  stat}
\pm .05^{(\rm syst)} $ ($t \bar t$)  \cite{AFBt-cdf1} 
 \\(dileptonic) &     
 \\
 \hline
D\O\,($L = 5.4$\, fb$^{-1}$) & ($15.2 \pm 4.0$)\% ($t \bar t$) \cite{Abazov:2011rq} 
 \\(lepton+jets) &   
 \\
\hline\hline
&\\
SM (NNLO) &$0.052^{+0.000}_{-0.006}$ ($p \bar p$) \cite{Kidonakis:2011zn}  \\
 \hline
\hline
\end{tabular}
\caption{ \label{tab:afb-val} {\it Values of Forward-Backward Asymmetry as measured at CDF and D\O\, along with SM theoretical Value.}}
\end{center}
\end{table}
CDF has also released the
functional dependence  of the $t \bar t$ rest frame asymmetry on
$\Delta  y$  and on $m_{t \bar t}$ ~\cite{afbt-cdf-mar2012}. 

 There are many
efforts~\cite{afb-models} in the literature  to explain the
large excess of top asymmetry observed at TeVatron.   A challenge is to realize a model which can 
  generate a large $\afbt$ without making an appreciable change in    the observed $t \bar{t}$ production cross section or
invariant mass spectrum.  The all channel measurement from CDF with
4.6~fb$^{-1}$ of data~\cite{cdf-top-cross} is $\sigma(t \bar{t}) = 7.5
\pm 0.31$~(stat) $\pm 0.34$~(syst) $\pm 0.15\, ({\rm Z\, theory})$ pb for $m_t =
172.5$~GeV, in good agreement with the SM prediction of $\sigma(t
\bar{t})^{NNLO}_{SM} = 7.08^{+0.00+0.36}_{-0.24-0.27}$~pb for $m_t =
173$~GeV~\cite{kidonakis-tcross} and is consistent with measurements
from D\O\,~\cite{d0-cross}. 

References \cite{Shu:2011au} and \cite{Gresham:2011pa} have compared
various models in light of CDF measurement of $\afbt$ and invariant
mass distributions. They show that axigluon and heavy $Z'$ models are
highly constrained while reference~\cite{AguilarSaavedra} further demonstrate that $Z'$ and  $W'$ models are disfavored by the LHC measurements.

Reference~\cite{Chen:2010hm}
studies the $\afbt$ in the framework of  unparticles. They have only
considered a colored octet vector unparticle in the s-channel having
flavor conserving couplings to quarks. We shall compare our results
for this particular case with them.

\subsection{Spin Correlation} 
\label{sec:spin-corr} 
Due to its large mass the top quark decays before hadronization
leading to preservation of its spin information. In the leading order
in SM, top quarks remain unpolarized because they are mainly produced
by parity conserving QCD interaction. The decay of top quarks {\it via}
electroweak interaction which is negligible with respect to the strong
interaction leaves very small effect on top polarization~\cite{ew-tt}.
The angular distribution of the partial top quark decay width $\Gamma $ in $t
\rightarrow W^+ +b$ followed by $W^+ \rightarrow l^+ +\nu$ or $\bar{d}
+u$ are correlated with the top spin axis~\cite{spin-corr-expt,spin-corr-th} as follows:
\begin{eqnarray}
\frac{1}{\Gamma_t} \frac{d \Gamma}{d \cos \chi_i} 
= \frac{1}{2}(1+\alpha_i \cos \chi_i); \alpha_i=\left\{ \begin{array}{l}
+1\,{\rm for}\,l^+, \bar{d}\\
-.31 \,{\rm for}\,\bar\nu , u \\
-.41\,{\rm for}\,  b
  \end{array} \right.
\end{eqnarray}
where $\Gamma_t$ is the total top decay width and $\chi_i$ is the angle between the $i$-th decay product and the 
top quark spin axis in the top quark rest frame.
Since net polarization of the top quarks is negligible in leading order, the correlation between 
the $i$-th decay product of the top and $\bar{\imath}$-th decay 
product of the antitop can be expressed by
\begin{eqnarray}
\frac{1}{\sigma_{t\bar t}} 
\frac{d^2 \sigma_{i\bar i}}{d\cos\chi_i d\cos\bar{\chi}_{\bar{i}}} 
= \frac{1}{4} (1+  
C^{t\bar{t}} ~\alpha_i ~\bar{\alpha}_{\bar{i}}~\cos \chi_i 
\cos \bar{\chi}_{\bar{i}} ).
\end{eqnarray}
with
\begin{eqnarray}
C^{t\bar{t}} = 
\frac{
\sigma_{\uparrow \uparrow} 
+ \sigma_{\downarrow \downarrow} 
-\sigma_{\uparrow \downarrow} 
-\sigma_{\downarrow \uparrow} 
} {
\sigma_{\uparrow \uparrow} 
+ \sigma_{\downarrow \downarrow} 
+\sigma_{\uparrow \downarrow} 
+\sigma_{\downarrow \uparrow} 
} .
\end{eqnarray}
$\sigma_{\uparrow/\downarrow~ \uparrow/\downarrow}$ is the 
production cross section for top quark pairs where the
top quark has spin up or down with respect to the top spin axis 
and the antitop has spin up or down with respect to the antitop 
spin axis.

Mahlon et al discussed  the right choice for the spin axes of
the top quark pair since a poor choice of spin axes can lead 
to a small value of $C^{t\bar{t}}$ ~\cite{mahlon}. 
They proposed three choices for the spin axis,  namely helicity basis, beamline basis  and Off-diagonal basis.

We choose the helicity basis (top quark momentum is chosen as spin quantization axis) for our calculation. In $t \bar t$ rest frame the quarks move back-to-back and the same
spin (S = 1) states are those with opposite helicity so that in helicity basis
\be
\spincorr =  \frac{
\sigma_{RL} 
+\sigma_{LR} 
-\sigma_{R R} 
- \sigma_{LL} 
} {
\sigma_{R R} 
+ \sigma_{LL} 
+\sigma_{RL} 
+\sigma_{LR} 
} 
\label{eq:def_spin_corr}
\ee
The dominant production mechanism for the $t\bar t$ pairs at the TeVatron is $q q
\to t \bar t$ with a $J=1$ gluon exchanged in $s$-channel. Near threshold, the $t \bar t$ pair is produced in $S=1$ state with the eigenstates $|++ >, \,\, \frac{1}{\sqrt 2}(|+ -> +
|-+>) \,\,, |- ->$.  Two of the three states have the opposite helicity (same spin), hence the
correlation near threshold is $C_{t\bar{t}}= 33\%$ while 
helicity conservation at high energy ensures that the $t$ and $\bar t$ are
produced with the opposite helicity and $C_{t\bar{t}} =100\%$ at very high
 energies \cite{Stelzer:1995gc}.
CDF has reported the value of top spin correlation in the helicity basis to be
$0.60 \pm 0.50\ (stat) \pm 0.16\ (syst) $ ~\cite{Aaltonen:2010nz} and in the beam basis to be $0.042^{+ 0.563}_{-.562}$~\cite{CDFNote10719}. The D\O measurement for spin correlation coefficient is $-.66<-\spincorr<0.81$ in the beam basis~\cite{Bloom:2011br}.  The
SM values for the same in beam and helicity basis are -0.614 and 0.299 respectively~\cite{Bernreuther:2010ny}. 
\par The terms linear in $\cos{\theta}$ do not contribute to $C^{t\bar{t}}$ after integration over $\cos{\theta}$. Thus $C^{t\bar{t}}$ and \afbt are sensitive to different terms. To relate the dependence of the two observables on the chiral structure of the quark couplings to any new physics sector, recently Ko {\it et. al.} proposed a new spin-spin FB asymmetry~\cite{Jung:2009pi} which is defined as
\bea
C^{t\bar{t}}_{FB} &\equiv& C^{t\bar{t}}(\cos{\theta} \ge 0 ) - C^{t\bar{t}} (\cos{\theta} \le 0 )
\label{eq:def_FBspin_corr}
\\
 &=&   \afbt ({\rm opp \,hel}) -\afbt ({\rm same\, hel}) \label{cfb_def}
\eea
while  the charge asymmetry may be written as
\bea
\afbt = \afbt ({\rm opp \,hel}) +\afbt ({\rm same\, hel})
\eea
 Since there is no contribution to \afbt in SM from same helicity states,
 for SM $C^{t\bar{t}}_{FB} =  \afbt ({\rm opp\, hel}) = \afbt $.
The correlation between \cfbt and \afbt can help to distinguish between
 various new physics scenarios.
\par New physics can show up both at the production of top quarks as well as
decay. If the new physics contains chiral asymmetry then it will affect
the  top spin correlation appreciably  and also the forward-backward spin correlation asymmetry~\cite{Arai:2009cp,np-polarization,spin-corr}.
\section{The Model} 
\label{sec:unp} 
 It is known that the visible particle sector is based on theories
 that are free in infrared and/or have a mass gap.  In contrast, an
 exact conformal invariance would require the mass spectrum to be
 either continuous or all zero masses.  Although there exist
 interacting conformal theories that have an infra red fixed point but
 such theories do not have asymptotically free in and out states and
 the traditional $S$-matrix description does not work.  The theory of
 unparticles as an conformally invariant sector that is weakly coupled
 to the SM particles was proposed by Georgi \cite{Georgi:2007ek} which
 was motivated by Banks-Zaks theory~\cite{Banks:1981nn}. This assumes
 the existence of a hidden sector with non-trivial infrared fixed point
 (e.g. Banks-Zaks type) that interacts with the SM through the
 exchange of messenger field with a large mass $M$.  Below the scale
 $M$, one can integrate out the heavy field giving rise to the
 effective non-renormalizable couplings of the form
\be
\frac{C_i}{M^{d_{\rm UV}+d^i_{\rm SM}-4}}{\mathcal O}^i_{\rm SM}{\mathcal O}_{\rm UV},
\label{op-uv}
\ee
where $C_i$ are the dimensionless coupling constants, ${\mathcal
O}^i_{\rm SM}$ and ${\mathcal O}_{\rm UV}$ are respectively the local
  operators built out of SM fields and hidden sector fields having 
  scaling dimensions $d^i_{\rm SM}$ and $d_{\rm UV}$ respectively.
  The hidden sector has an infrared fixed point and becomes conformal
  at some scale say $\sim \scaleu < M$.  Below the energy scale
  $\scaleu$, the renormalizable couplings of the hidden sector fields
  cause dimensional transmutation. In the effective theory the high
  energy operators ${\mathcal O}_{\rm UV}$ above this scale match onto
  the unparticle operators $\ou$ (the operator ${\mathcal O}_{\rm UV}$
  becomes $\scaleu^{d_{\rm UV}-\du} \ou$) below this scale and the
  interactions of equation~\eqref{op-uv} now take the form
\be
\frac{C_i \scaleu^{d_{\rm UV}-\du}}{M^{d_{\rm UV}+d^i_{\rm SM}-4}}{\mathcal O}^i_{\rm SM}\ou 
= \frac{C_i}{ \Lambda_{\cal U}^{d_{\rm SM}^i+\du -4 }} {\mathcal O}^i_{\rm SM} \ou \,,
\label{op-unp}
\ee
where $\du$ is the scaling dimension of the operator $\ou$.  
$M$, $\scaleu$, $\du$ and $d_{\rm UV}$ are the hidden sector
parameters while the exponent of $\Lambda_{\cal U}$ depends upon the dimension 
of the SM operator and is given by
\be \Lambda_{\cal U}^{d_{\rm SM}^i+\du -4 } = \frac{M^{d_{\rm UV}+d^i_{\rm
      SM}-4}}{ \scaleu^{d_{\rm UV}-\du}}
\ee
The operators with different mass dimension are likely to couple with
different strengths. The couplings $C_i$ and the scale $\scaleu$ only
appear in a combination given by~(\ref{op-unp}) and there is no
guiding theoretical principle to fix $C_i$.  In addition to the
interaction of SM fields with unparticle which can be probed at high
energy colliders (below the scale $\scaleu$), the elimination of heavy
fields also induces the contact interaction among SM fields which are
suppressed by powers of $M$ and have the effect of drowning down any
unparticle effects for $d_{\cal U}\gg 1$. 
\par Unparticle effects are detected in the colliders either through missing energy distributions or by the   interference effects with SM amplitudes. But the constraints from astrophysics and cosmology \cite{astro-unp-constrain} would render them practically undetectable in the collider experiments unless otherwise they break the scale invariance at $ \gtrsim 1$ GeV. The conformal
invariance may be broken at a scale $\mu $ by the higgs-unparticle coupling $H^\dagger H
\ou$ which introduces a scale in the theory once higgs acquires vacuum
expectation value.  For consistency $\mu \le \scaleu$
and the two scales should be well separated to give a window where the
sector is conformal. Thus for scales $\scaleu$ and $M$ to be experimentally accessible, the higgs-scalar unparticle coupling is assumed to
break scale invariance at electroweak scale. However, if only vector unparticles
are present, scale invariance is broken by higher dimensional
operators leading to the breaking of the scale invariance  below the electroweak scale~\cite{Barger:2008jt}. 
\par Initially  we have not introduced the effect of the breaking   scale invariance with respect to the study of the top pair production and the same sign top production at TeVatron. Nonetheless we have shown its effect in the evaluation of the top decay width. Introduction of such  modifications in the theory do not change the cross sections appreciably if $\mu \ll 2\, m_t$. The phenomenological lower bound on the scale invariance breaking scale
comes from the  BBN and SN 1987A where  $\mu$ is required to be sufficiently large compared to the relevant energy scales  $\simeq $ 1 MeV and $\simeq$  30 MeV respectively \cite{Barger:2008jt}. We dwell on  these issues in the appropriate sections and also in the subsection \ref{mass_gap}.  
  \begin{figure*}[!ht]
  \centering
  \subfloat[$\lambda^V_{RR} \ne 0$ and $\lambda^V_{LR}=\lambda^V_{RL}=\lambda^V_{LL} =0 $]{\label{fig:FC_vec_sing_sigma-1-2-a}\includegraphics[width=0.5\textwidth]{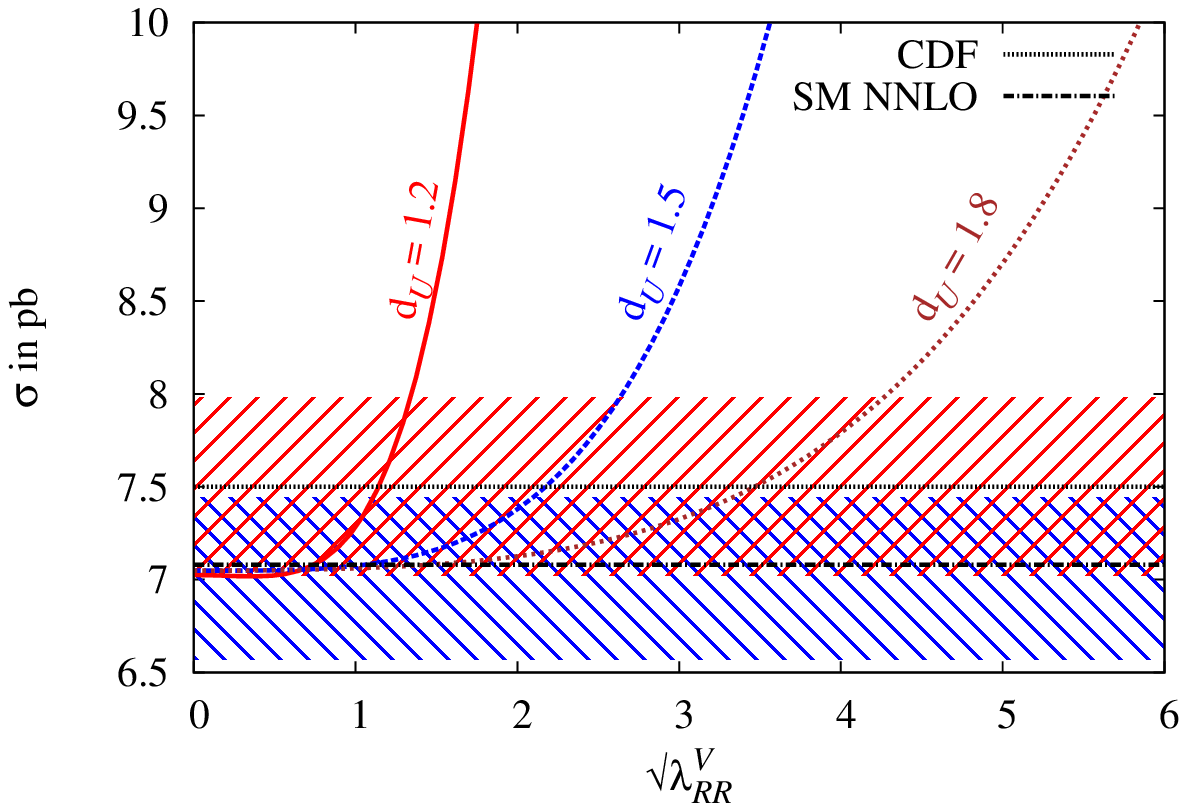}}
  \subfloat[$\lambda^V_{LL} = \lambda^V_{RR} = \lambda^V_{RL}  = \lambda^V_{LR} =\lambda^V_{VV}$]{\label{fig:FC_vec_sing_sigma-1-2-c}\includegraphics[width=0.5\textwidth]{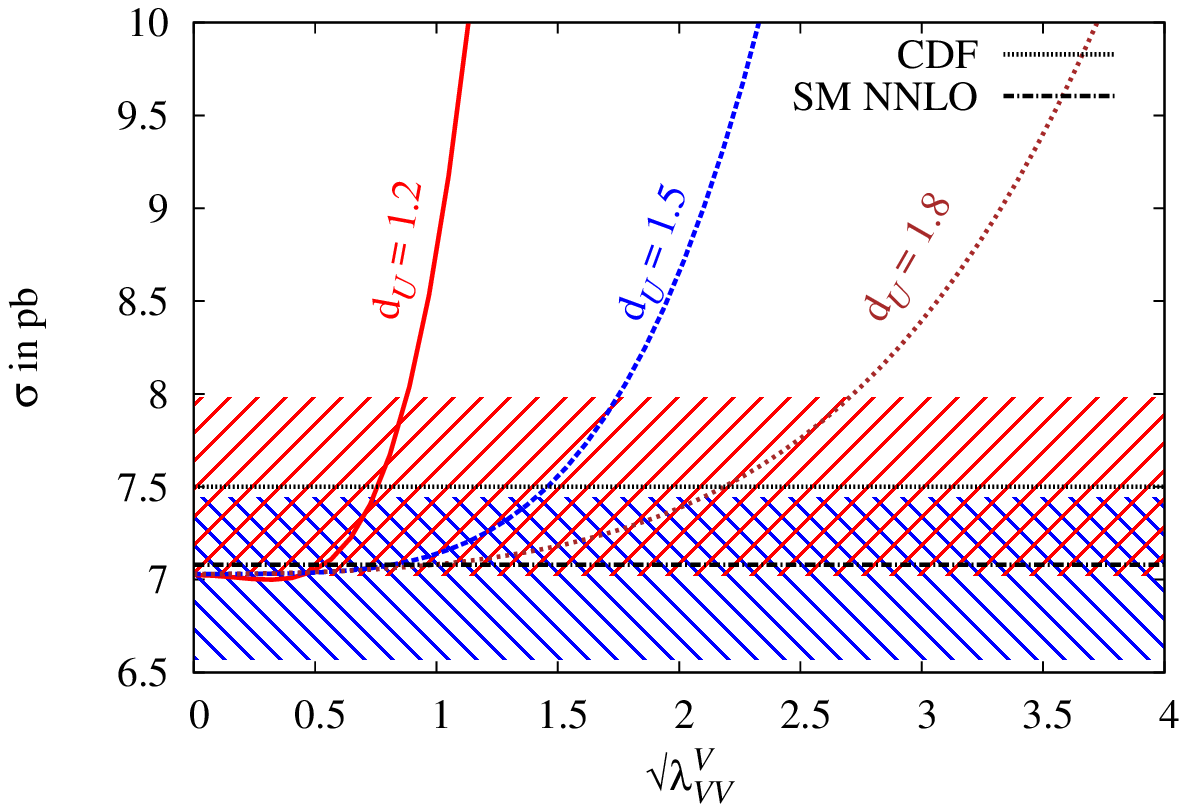}} \\
  \subfloat[$\lambda^V_{LL} = \lambda^V_{RR} = -\lambda^V_{RL}  = -\lambda^V_{LR} =\lambda^V_{AV}$]{\label{fig:FC_vec_sing_sigma-1-2-d}\includegraphics[width=0.5\textwidth]{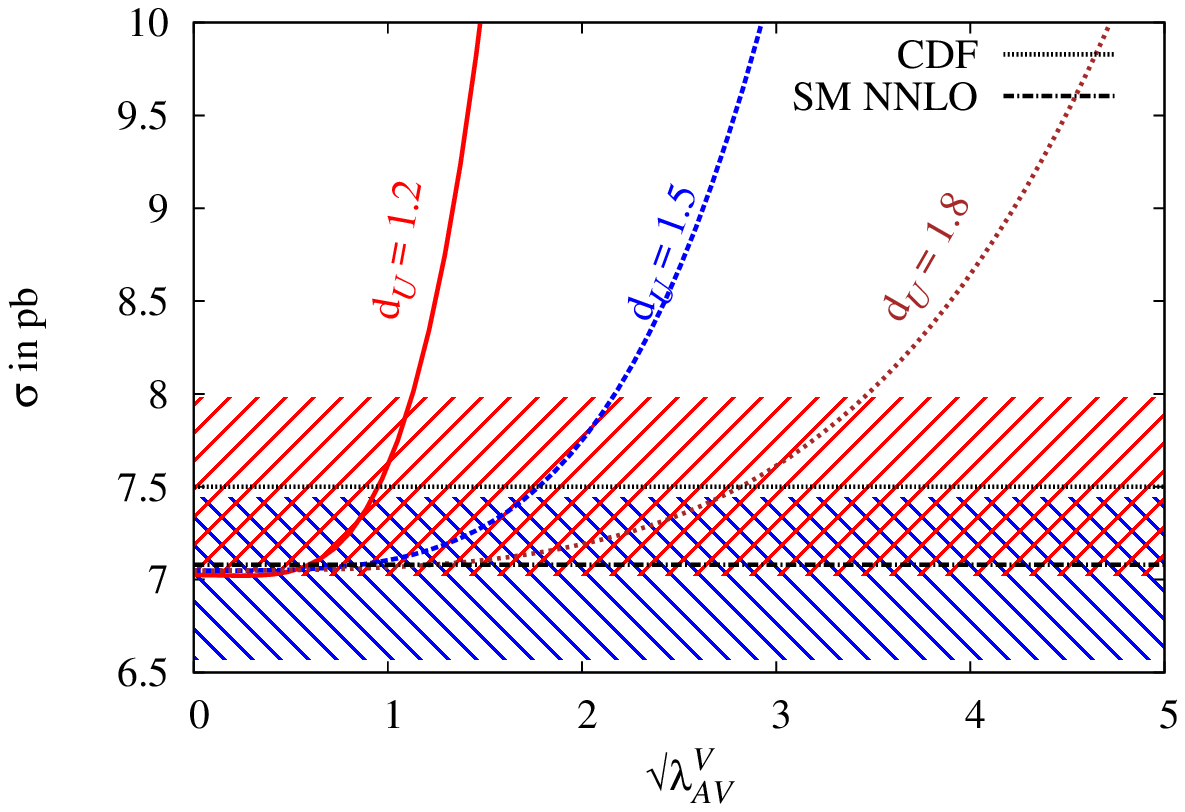}}
  \subfloat[$\lambda^V_{RR} \ne 0$ and $\lambda^V_{LR}=\lambda^V_{RL}=\lambda^V_{LL} =0 $]{\label{fig:FC_vec_sing_sigma-3-4-a}\includegraphics[width=0.5\textwidth]{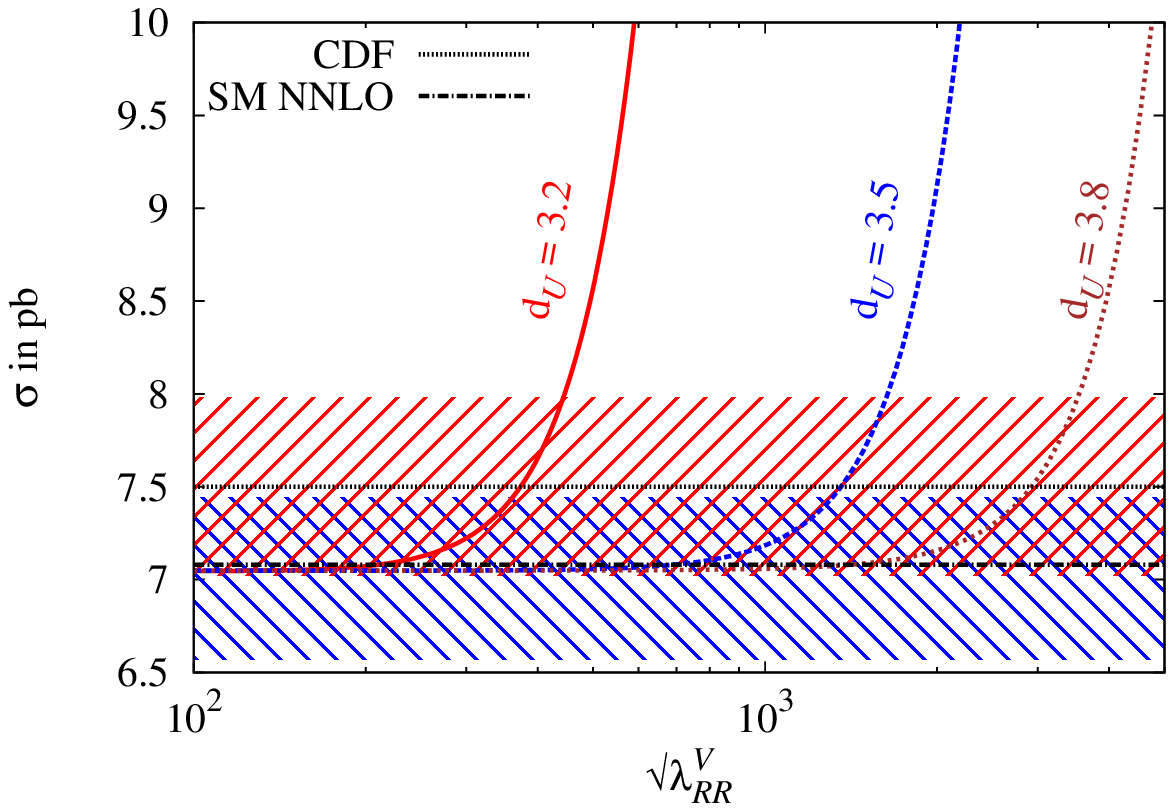}}
\caption{\small \em {Variation of the cross-section $\sigma \left( p\bar p \to t\bar t\right)$ with  couplings
$\surd\lambda^V_{i j}$ for
color singlet flavor conserving vector unparticles corresponding to  different values of  $d_{\cal U}$ at fixed $\Lambda_{\cal U}=1$ TeV . The  upper dotted line with a red band  depicts the  cross-section  $7.50 \pm 0.48$~pb from CDF (all channels) ~\cite{cdf-top-cross}, while the lower dot-dashed  line with a blue band show theoretical estimate $7.08\pm 0.36$~pb at NNLO  ~\cite{kidonakis-tcross}. The panels 'a', 'b' and 'c' show the variation of $\sigma$ for different combinations of couplings (the cases (a) (or b), (c) and (d) of the text respectively) and for \du in the range $1<\du<2$ while the panel 'd' corresponds to the variation in the \du range $3<\du<4$ for the coupling combination (a) mentioned in the text.}}
\label{fig:FC_vec_sing_sigma}
\end{figure*}
\par The two-point function of unparticle operators~\cite{Georgi:2007si} is
written as
\begin{eqnarray}
 \langle 0|{\cal O}_{\cal U}(x){\cal O}_{\cal U}^\dagger(0)| 0 \rangle
 =\int \frac{d^4 p }{ (2\pi)^2}e^{-ip\cdot x}\rho(p^2), \label{op1}
\end{eqnarray}
where $\rho(p^2)=(2\pi)^2\int d\lambda \delta^4(p-p_\lambda)\left\vert\langle
0|{\cal O}_{\cal U}|\lambda \rangle\right\vert^2$.  The spectral function
$\rho(p^2)$ is determined by scale invariance to be
$\rho(p^2)=A_{d_{\cal U}}\theta(p^0)\theta(p^2)(p^2)^{d_{\cal U}-2}$,
where $A_{d_{\cal U}}$ is the normalization factor.  This factor is
fixed by identifying $\rho(p^2)$ with $d_{\cal U}$-body phase space of
massless particles to be
\begin{eqnarray}
 A_{d_{\cal U}}=\frac{16\pi^2 \sqrt{\pi}}{(2\pi)^{2d_{\cal U}}}
 \frac{\Gamma(d_{\cal U}+1/2)}{\Gamma(d_{\cal U}-1)\Gamma(2d_{\cal U})}. 
 \label{adu-def}
\end{eqnarray} 
The scaling dimension $\du$ can have non-integral values as well.
With the use of the spectral function $\rho(p^2)$ and requiring scale
 invariance, the Feynman propagators for vector and tensor  unparticle are defined to be~\cite{Georgi:2007si}
\begin{eqnarray}
 \Delta_{\mu\nu}(p^2)&=\frac{iA_{d_{\cal U}}}
  {2\sin(d_{\cal U}\pi)}(-p^2)^{d_{\cal U}-2} \Pi_{\mu\nu}(p), \nn\\
 \Delta_{\mu\nu,\, \rho\sigma}(p^2)&=\frac{iA_{d_{\cal U}}}
  {2\sin(d_{\cal U}\pi)}(-p^2)^{d_{\cal U}-2} {\cal T}_{\mu\nu,\, \rho\sigma}(p) 
  \label{prop-vec-tensor}
  \eea
where
\begin{widetext}
\bea 
\Pi^{\mu\nu}(p)= -g^{\mu\nu}+ \frac{p^\mu p^\nu}{ p^2}, \,\,\,\, {\cal T}^{\mu\nu,\, \rho\sigma}(p) = \frac{1}{2} \left\{
  \Pi^{\mu\nu}(p)\Pi^{\nu\sigma}(p) +
  \Pi^{\mu\sigma}(p)\Pi^{\nu\rho}(p) -\frac{2}{3} \Pi^{\mu\nu}(p)
  \Pi^{\rho\sigma}(p) \right\}
  \label{vec-tensor}
\eea
\end{widetext}
and $(-p^2)^{d_{\cal U}-2}$ is  interpreted as
\bea
(-p^2)^{d_{\cal U}-2} = 
\begin{cases}
& \left\vert p^2\right\vert^{\du-2} \qquad\quad {\rm for }\,\, p^2<0   \\
& \left\vert p^2\right\vert^{\du-2} e^{i\du\pi} \,\,\, {\rm for }\,\, p^2>0 . \label{def-Psq}
\end{cases}
\eea
\noindent Propagators are chosen such that they satisfy the relations $p_\mu \Pi^{\mu\nu}(p)=0$, $p_\mu
    {\cal T}^{\mu\nu,\, \rho\sigma}(p)=0 $, and ${\cal
      T}_{\mu}^{\mu,\, \rho\sigma}(p)=0 $. Also the unparticle
    operators are all taken to be hermitian, $\ou^\mu$ and
    $\ou^{\mu\nu}$ are assumed to be transverse and the spin-2
    unparticle operator is taken to be traceless $\ou^\mu_\mu =0$
\par The negative sign in front of $p^2$ of the second term gives
rise to a unique phase factor for time like virtual unparticles but
not for the space like. This leads to interesting interference
effects with  SM process which will be discussed later.
\par Based on this template of the unparticle formalism, possible signatures of scalar, vector and tensor un-particles at
colliders and their effects on low energy phenomenology have been 
studied~\cite{Choudhury:2007cq,unp-top,unp_pheno}. Astrophysics  and cosmology
 also put strong constraints on unparticles~\cite{astro-unp-constrain,unp-astro}.  Various
 theoretical aspects of unparticles have been also 
studied~\cite{unp-theory,Grinstein:2008qk}.

\par Grinstein et al~\cite{Grinstein:2008qk} have revisited the computation of two point function for unparticles  demanding rigid conformal invariance in the hidden sector and shown that the unitarity of  conformal algebra imposes lower bounds on scaling dimensions of the vector and tensor  operators~\cite{Grinstein:2008qk}. Thus $\du^V \ge 3$ and $\du^T \ge 4$ for vector unparticle and symmetric traceless tensor unparticle respectively. The primary vector operator $\ou^\mu$ with $\partial_\mu \ou^\mu= 0$ corresponds to particle with $du^V = 3$, while the vector  unparticle corresponding to  $\partial_\mu \ou^\mu\ne 0$ and $d_u >3$ get a modified propagator
\begin{eqnarray}
  \Delta^{\mu\nu}(p)=\frac{-iA_{d_{\cal U}}}
  {2\sin(d_{\cal U}\pi)}(-p^2)^{d_{\cal U}-3}
  \left(p^2 g^{\mu\nu}- a \,p^\mu p^\nu \right),\
\label{gprop-vec}
\end{eqnarray}
with $a= 2(\du -2)/(\du -1)$. It is to be noted that this differs from the propagator given in equation (\ref{prop-vec-tensor}) not only in the relative size of the terms but also by  an overall extra phase factor $e^{-i\pi}$. This extra phase would affect the sign of interference term with SM processes. In fact this contradicts the many observations made in the literature with respect to the unparticles.
Similarly the tensor unparticle propagators is modified to~\cite{Lee:2009dda}
\begin{eqnarray}
 \Delta_{\mu\nu,\alpha\beta} (p^2)& =
         \frac{A_{d_{\cal U}}}{2 \sin (d_{\cal U} \pi)} \, (-p^2)^{d_{\cal U} -2} \, 
         {\cal T}_{\mu\nu,\alpha\beta} \;,
\label{gprop-tensor}\\
{\rm where}\quad\quad &\nn
\end{eqnarray}
\vskip -1.1cm
\begin{widetext}
\vskip -0.7cm
\begin{eqnarray}
{\cal T}_{\mu\nu,\alpha\beta}(p)&=&
d_{\cal U}(d_{\cal U}-1)(g_{\mu\alpha}g_{\nu\beta}+\mu\leftrightarrow\nu)
+\left[2-\frac{d_{\cal U}}{2}(d_{\cal U}+1)\right]g_{\mu\nu}g_{\alpha\beta} \nn \\
&& -2(d_{\cal U}-1)(d_{\cal U}-2)\left(g_{\mu\alpha}\frac{p_\nu p_\beta}{p^2}
   +g_{\mu\beta}\frac{p_\nu p_\alpha}{p^2}+\mu\leftrightarrow\nu\right)\nn\\
&&+4(d_{\cal U}-2)\left(g_{\mu\nu}\frac{p_\alpha p_\beta}{p^2}+g_{\alpha
\beta}\frac{p_\mu p_\nu}{p^2}\right)
+8(d_{\cal U}-2)(d_{\cal U}-3)\frac{p_\mu p_\nu p_\alpha p_\beta}{(p^2)^2}. 
\end{eqnarray}
\end{widetext}
We have considered the symmetric structure for the tensor-propagator.
\par In this article we explore the phenomenologically viable and
 interesting  scale invariant (not strictly conformally invariant) hidden
 sector where the bounds on $\du ^V,\,\, \du^T$ are relaxed. We do
 investigate the  sensitivity of the observables with the full conformal
 invariance also though it should be kept in mind that for such large
 values of $\du$, the SM contact interactions (induced at scale $M$ by the
 exchange of ultra heavy particle) dominate over the unparticle-SM
 interference effects.
\par The relevant unparticle-SM flavor conserving (FC) and flavor violating (FV) interaction Lagrangian is given in  equations
 (\ref{lag-V1}-\ref{lag-T2}) at the Appendix.
We consider the vector and tensor unparticles  with the possibility of
each being a singlet or octet under $SU(3)_C$. Since gluons interact with  vector unparticles  through a derivative term $G{_\mu \alpha} G^{\alpha}_\nu \partial^\mu {\cal O}_{\cal U}^\nu$ and not with the primary field ${\cal O}_{\cal U}^\nu$, it is 
suppressed by a factor of $\Lambda_{\cal U}$ compared  to the interaction of vector
unparticle with quarks. Moreover, the gluon flux at TeVatron is low and so it is further
subdued. Hence we do not consider such interaction terms.

\par We compute the  helicity amplitudes with appropriate color factors 
corresponding to Lagrangian and present them  in the appendix \ref{app:helamp}. In calculating the amplitudes for unparticle we have used the improved
propagator of equation~\eqref{gprop-vec} for vector and of
equation~\eqref{gprop-tensor} for the tensor unparticle. 
\par The  phenomenological consequences of   the color  octet unparticles are addressed and validated in  section \ref{mass_gap}. There
 have been various attempts to provide a complete gauge theory of
 unparticles~\cite{Cacciapaglia:2007jq,gauge-unp}, however in the absence of any well established approach to
 a full theory of unparticle interactions we study the topic from a
 phenomenological point of view.
\section{Sensitivity of the model parameters}                   
\label{sec:num-results}
We perform a parton level and leading order calculation, where  we have used
 CTEQ6L LHApdf parton distribution function, the top mass $m_t = 173 \gev/c^2$ and
 $\alpha_s =\alpha_s(m_Z)= 0.13$. We evolve $\alpha_s$ to get $ \alpha_s(2 m_t)$ using CTEQ6L LHApdf. 
The tree-level SM
 cross-section was obtained to be $\sigma_{\rm SM}^{\rm tree} = 4.26$~pb, which is consistent with the results available from Herwig++ \cite{herwig}, CalcHEP~\cite{Belyaev:2012qa} and MadGraph/Madevent \cite{madgraph} for the given choice of parameters.  
We have also matched our results with that existing literature~\cite{Chen:2010hm,Arai:2009cp}.
\par We normalize our tree level SM cross-section to $7.08$~pb to include the NNLO corrected matrix element squared \cite{kidonakis-tcross} with $m_t = 173 \gev/c^2$. Accordingly we normalise the SM amplitude by a factor $k = \sqrt{7.08/4.26}
 = \sqrt{1.663}$. The $k$ factor is sensitive to the choice of renormalisation and factorization scale but  eventually it stabilses with the inclusion of higher order corrections of the matrix element at the NNLO level \cite{Baernreuther:2012ws}. 
As the NLO corrections in the new physics sector is not feasible within the scope of our article, we work with the LO contribution only in the unparticle physics sector.
Thus after
 scaling we may write the squared matrix element as
\bea
|{\cal M}^{t \bar t}|^2 = |k{\cal M}^{\rm SM}_{\rm tree}|^2 &&+ 2 \left({\cal R}e
(k {\cal M}_{\rm tree}^{\rm SM})^\dagger ({\cal M}^{\rm unp} ) \right) \nn\\
&& + | {\cal M}^{\rm unp}|^2
\label{eq:matsq-scaled}
\eea
\par As mentioned earlier  SM inclusive processes can generate \afbt  which is   at most $5-8\%$ through the radiative diagrams at NLO level, it is worth to examine the additional contribution to the integrated and differential \afbt  generated by the  LO top pair production processes mediated through unparticles. The contribution of unparticles to cross-section, charge asymmetry and spin
 correlation coefficient come from the matrix element square term
 containing only unparticles and the interference of unparticles with SM QCD and electroweak matrix elements.
 To see the correlation between \cfbt and \afbt and dependence on the
 chiral structure of the theory, we divide these two into same helicity
 and opposite helicity contributions as
\bea
\left\vert {\cal M}^{\rm unp}\right\vert^2 = \left\vert {\cal M}^{\rm unp}\right\vert^2_{\rm same\,\, hel} + \left\vert {\cal M}^{\rm unp}\right\vert^2_{\rm opp\,\, hel}\nn\\
2 {\cal R}e \Bigl[\bigl({\cal   M}^{\rm SM}_{\rm tree}\bigr)^\dagger
{\cal M}^{\rm unp}\Bigr] = 2 {\cal R}e \Bigl[\bigl({\cal   M}^{\rm SM}_{\rm tree}\bigr)^\dagger
{\cal M}^{\rm unp}\Bigr]\Big\lvert_{\rm same\, hel} \nn\\
+ 2 {\cal R}e \Bigl[\bigl({\cal   M}^{\rm SM}_{\rm tree}\bigr)^\dagger
{\cal M}^{\rm unp}\Bigr]\Bigr\rvert_{\rm opp\, hel} \nn\\
\label{eq:mat_div}
\eea
\begin{figure*}[!ht]
  \centering
  \subfloat[$\lambda^V_{RR} \ne 0$ and $\lambda^V_{LR}=\lambda_{RL}=\lambda^V_{LL} =0 $]{\label{fig:FC_lambda_vec_sing_afb-1-2-a}\includegraphics[width=0.49\textwidth]{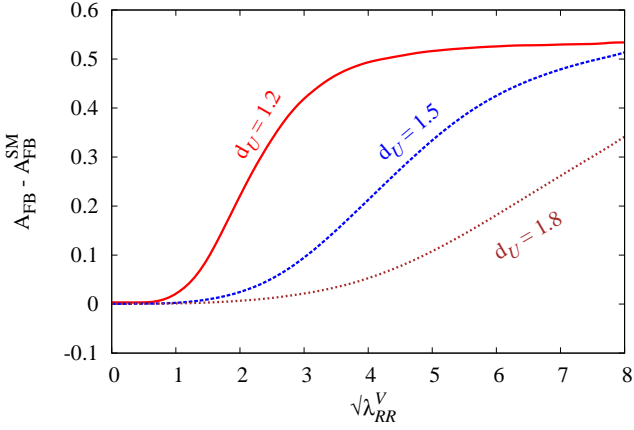}}
  \subfloat[$\lambda^V_{RR} \ne 0$ and $\lambda^V_{LR}=\lambda^V_{RL}=\lambda^V_{LL} =0 $]{\label{fig:FC_lambda_vec_sing_afb-3-4-a}\includegraphics[width=0.49\textwidth]{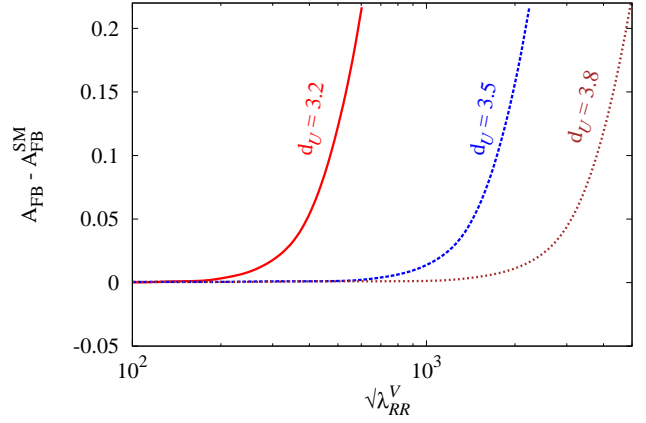} }
  \caption{\small\em{ Variation of the $A_{FB} - A_{FB}^{SM}$ (the unparticle contribution to \afbt) with  couplings $\surd \lambda^V_{RR}$ (corresponding to the case (a) of the text) for color singlet flavor conserving vector unparticles for various value of $\du$ at fixed $\Lambda_{\cal U}=1$ TeV. This is evaluated using the 1d differential distribution of rapidity in the $t\bar t$ rest frame.}}
  \label{fig:FC_vec_sing_afb}
\end{figure*}

 The
helicity amplitudes (as given in the appendix) involve the product of
the couplings $g^{{\cal U}_{(V/T)}^{\bf n}\bar q q}_i\, g^{{\cal
    U}_{(V/T)}^{\bf n}\bar t t}_j=\lambda^{(V/T)}_{ij}\, \, \left({\rm
  with}\, \,\, i\,\,\, {\rm and}\,\,\, j\,\equiv L\, {\rm or}\, R\right)$ for flavor conserving (FC) vector/tensor unparticles and $g^{ut}_{i}$ for flavor violating (FV) vector unparticles.  Therefore
the observables  are analyzed with
the independent free parameters : the scaling dimension $\du$ and  $\lambda^{(V/T)}_{ij} $ ($ g^{ut}_{i}$) for FC vector/tensor  (FV vector) unparticles. Throughout our analysis we have fixed $\Lambda_{\cal U}$ = 1 TeV $\gg $\mttb .
 In all the figures showing the cross-section, we  plot the central values with the respective error band  for
 CDF \crtt $=7.5\pm0.48$~pb ~\cite{cdf-top-cross} and   the SM NNLO  $7.0
8+0.36-0.51$~pb~\cite{kidonakis-tcross}. We bound  the total contribution from SM and unparticles to the \crtt for a given \du at fixed $\Lambda_{\cal U}=1$ TeV within the error bars of CDF data, which in turn gives the upper cut-off for the couplings.
\subsection{Flavor Conserving  Unparticle}
\label{subsec:FC}
The FC couplings for the light quarks are tightly constrained from the
measurement of dijet events at TeVatron \cite{Aaltonen:2008dn}.  Therefore
the FC couplings of the light quarks $g^{{\cal U}_{(V/T)}^{\bf n}\bar
  q q}_{(L/R)}$ with the vector and tensor unparticles are taken to be
order of magnitude smaller than those of third generation quarks or
antiquarks $g^{{\cal U}_{(V/T)}^{\bf n}\bar t t}_{(L/R)}$. A detailed discussion on this can be found in the reference \cite{Agarwal:2009vh} for $d_{\cal U}\ge 3$. We have also  verified the  dijet spectrum corresponding to $1\le d_{\cal U}\le 3$ for most of the parameter region probed and it is consistent with the observed data. 
For FC processes, the  products of couplings $\lambda^{(V/T)}_{ij} $ can be
broadly classified into four combinations for a given $\du$ and
$\Lambda$. 
 \ben
\item[(a)] $g^{{\cal U}_{(V/T)}^{\bf n}\bar q q}_L = g^{{\cal
    U}_{(V/T)}^{\bf n}\bar t t}_L = 0$ i.e
  $\lambda^{(V/T)}_{LR}=\lambda^{(V/T)}_{RL}=\lambda^{(V/T)}_{LL} =0 $
  and $\lambda^{(V/T)}_{RR} \ne 0$
\item[(b)] $g^{{\cal U}_{(V/T)}^{\bf n}\bar q q}_R = g^{{\cal
    U}_{(V/T)}^{\bf n}\bar t t}_R = 0$ i.e
  $\lambda^{(V/T)}_{LR}=\lambda^{(V/T)}_{RL}=\lambda^{(V/T)}_{RR} =0 $
  and $\lambda^{(V/T)}_{LL} \ne 0$
\item[(c)] $g^{{\cal U}_{(V/T)}^{\bf n}\bar q q}_L = g^{{\cal
    U}_{(V/T)}^{\bf n}\bar q q}_R$, $g^{{\cal U}_{(V/T)}^{\bf n}\bar t
  t}_L = g^{{\cal U}_{(V/T)}^{\bf n}\bar t t}_R$
  i.e. $\lambda^{(V/T)}_{LL} = \lambda^{(V/T)}_{RR} =
  \lambda^{(V/T)}_{RL} = \lambda^{(V/T)}_{LR}= \lambda_{VV/TT}$
\item[(d)] $g^{{\cal U}_{(V/T)}^{\bf n}\bar q q}_L =- g^{{\cal
    U}_{(V/T)}^{\bf n}\bar q q}_R$, $g^{{\cal U}_{(V/T)}^{\bf n}\bar t
  t}_L =- g^{{\cal U}_{(V/T)}^{\bf n}\bar t t}_R$
  i.e. $\lambda^{(V/T)}_{LL} = \lambda^{(V/T)}_{RR} =
  -\lambda^{(V/T)}_{RL} = -\lambda^{(V/T)}_{LR} =\lambda_{AV/AT}$
\een
These four combinations correspond to the pure right handed (RH), left handed (LH), vector/tensor and axial vector/axial tensor couplings respectively.
The symmetry in the helicity amplitudes  given in equations
\eqref{app-vec-fc1}$-$\eqref{app-vec-fc4} and \eqref{app-ten-fc1}$-$\eqref{app-ten-fc4} renders the same new physics
 contributions for the combination (a) and (b). 


\subsubsection{Color Singlet Vector Unparticles}
We study the effect of presence of color singlet unparticles on
 the three observables, namely, \crtt, $\afbt$ and \spincorr. The variation of these observables with the
product of couplings is shown in figures \ref{fig:FC_vec_sing_sigma},
\ref{fig:FC_vec_sing_afb} and \ref{fig:FC_vec_sing_spcr} respectively corresponding to
 the specific cases.  
\begin{figure*}[!ht]
  \centering
  \subfloat[$\lambda^V_{RR} \ne 0$ and $\lambda^V_{LR}=\lambda^V_{RL}=\lambda^V_{LL} =0 $]{\label{fig:FC_lambda_vec_sing_spcr-1-2-a}\includegraphics[width=0.5\textwidth]{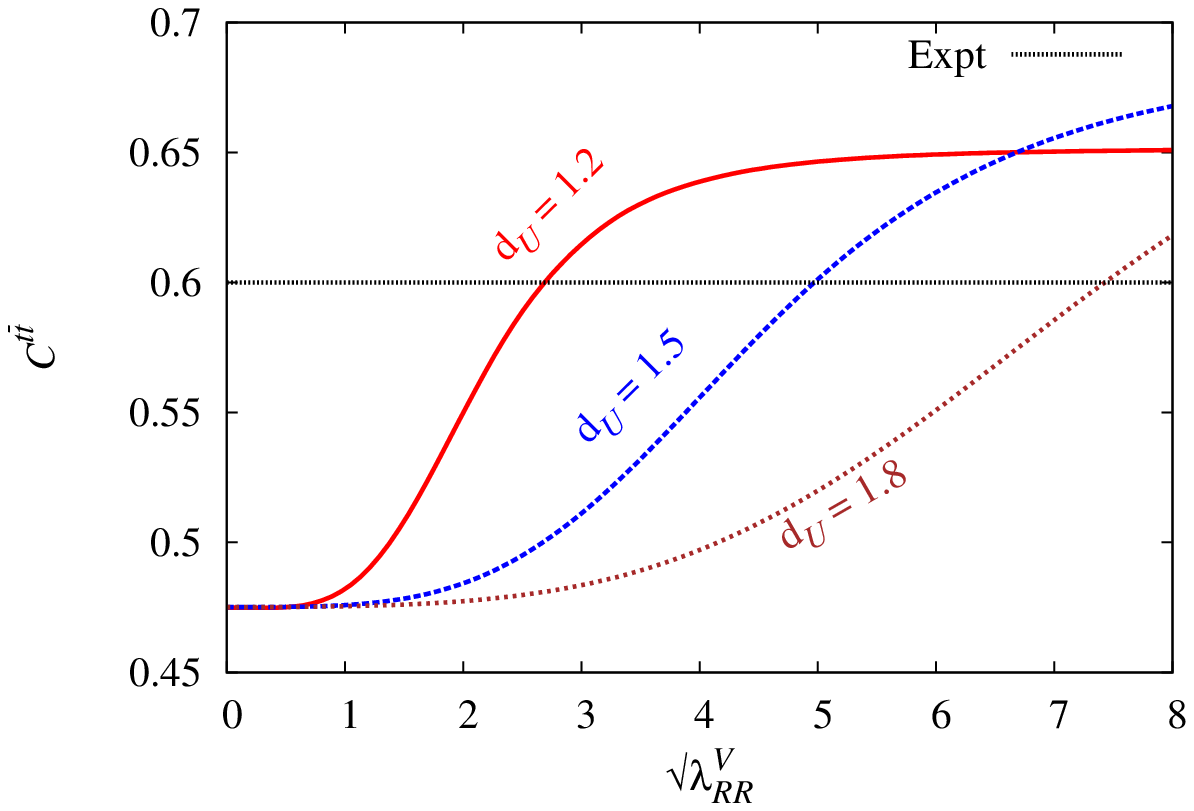}}
  \subfloat[$\lambda^V_{LL} = \lambda^V_{RR} = \lambda^V_{RL}  = \lambda^V_{LR} =\lambda_{VV}$]{\label{fig:FC_lambda_vec_sing_spcr-1-2-c}\includegraphics[width=0.5\textwidth]{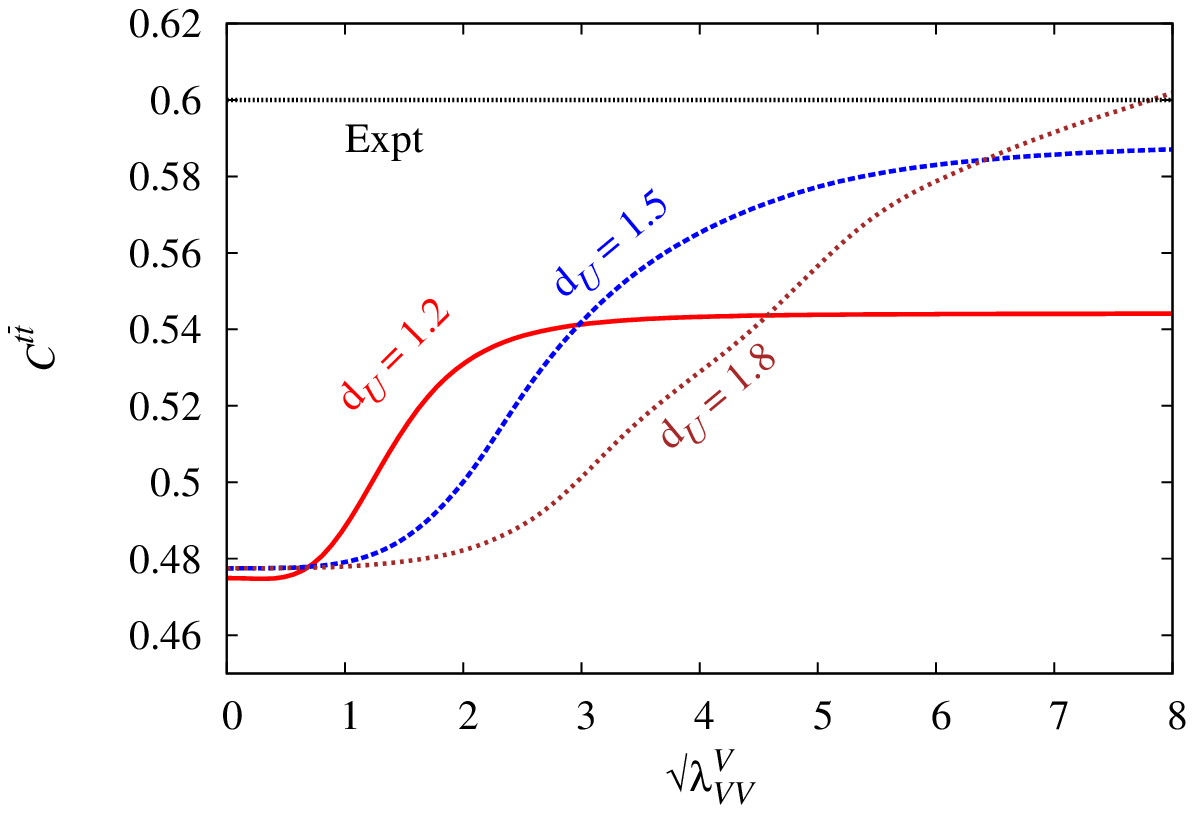}} \\
  \subfloat[$\lambda^V_{LL} = \lambda^V_{RR} = -\lambda^V_{RL}  = -\lambda^V_{LR} =\lambda_{AV}$]{\label{fig:FC_lambda_vec_sing_spcr-1-2-d}\includegraphics[width=0.5\textwidth]{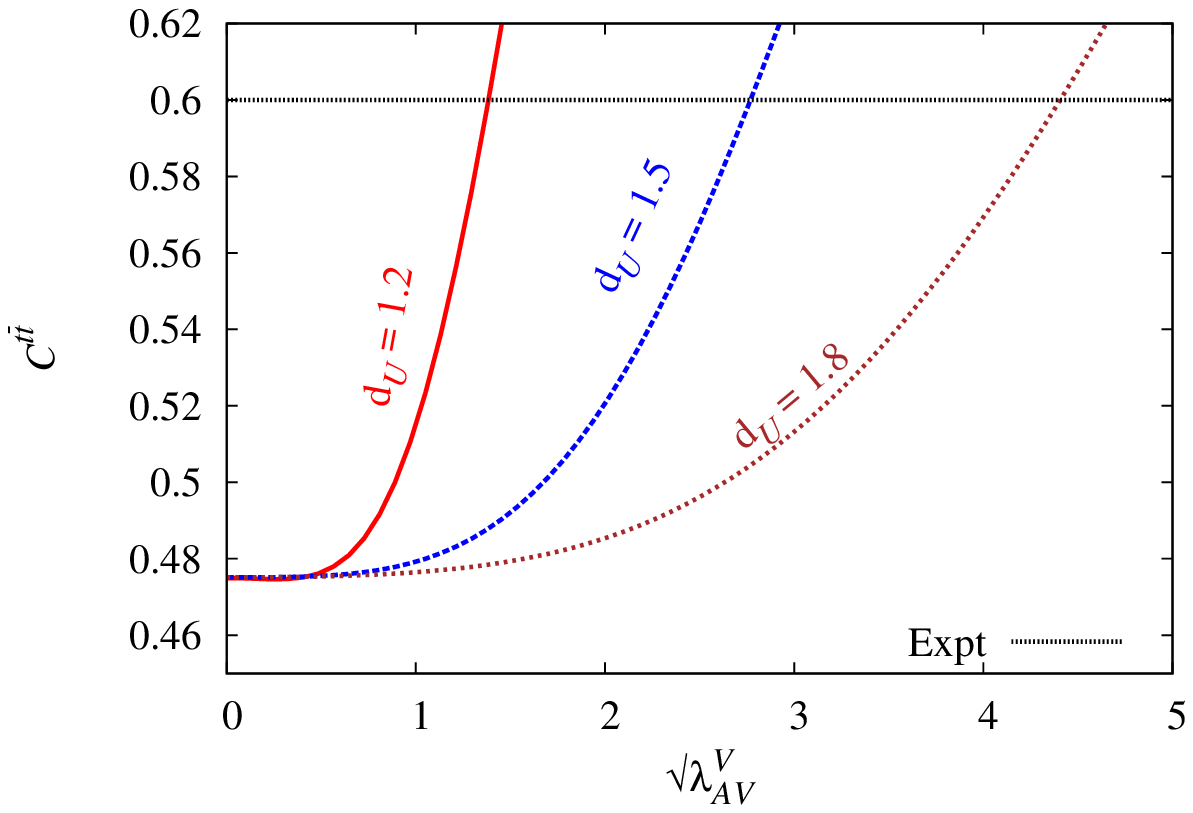}}
  \subfloat[$\lambda^V_{RR} \ne 0$ and $\lambda^V_{LR}=\lambda^V_{RL}=\lambda^V_{LL} =0 $]{\label{fig:FC_lambda_vec_sing_spcr-3-4-a}\includegraphics[width=0.5\textwidth]{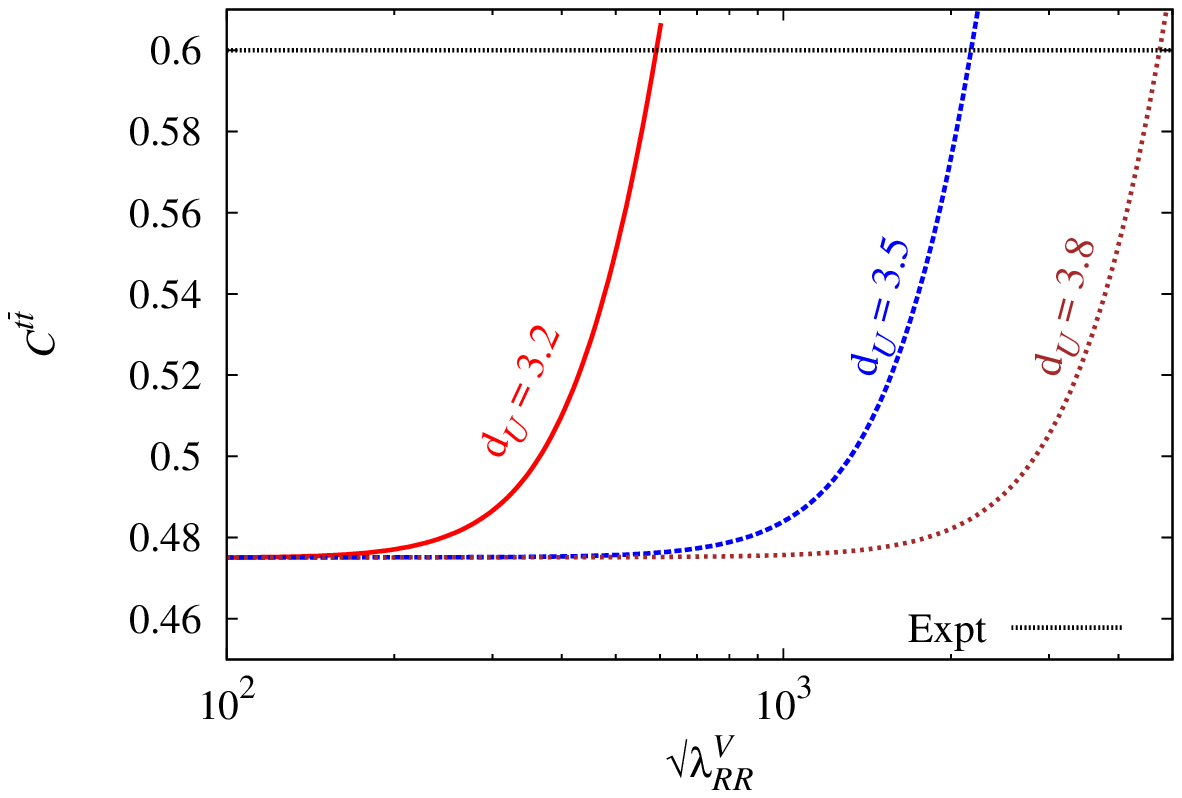}}
  \caption{\small \em{Variation of the spin-correlation coefficient \spincorr with  couplings $\surd \lambda^V_{i j}$ for color singlet flavor conserving vector unparticles for various value of $\du$ at fixed $\Lambda_{\cal U}=1$ TeV. The experimental value is depicted with a dot-dashed line  at $0.60 \pm 0.50\ (stat) \pm 0.16\ (syst) $ ~\cite{Aaltonen:2010nz}. The panels 'a', 'b' and 'c' respectively correspond to the cases (a), (c) and (d) for the values of \du in the range $1<\du<2$ while the panel 'd' gives the variation for case (a) with \du in the range $3<\du<4$.}}
  \label{fig:FC_vec_sing_spcr}
\end{figure*}
\par  Since there is no interference of the color singlet FC vector
 unparticle with QCD and the squared term of the unparticle  dominates
 over  the interference with the electroweak sector, the behavior of
 observables for singlet unparticles are determined by the contribution of unparticle alone.
For the FC vector singlet unparticle, the contribution of unparticle is given by
\begin{widetext}
\bea
 \left\vert {\cal M}_{_{\rm FCV}}^{\rm unp}\right\vert^2_{\rm same\,\, hel} &=& B^{\rm unp}_{_{\rm FCV}} \biggl[ \frac{1}{2} (1- \beta_t^2)
\left\{(\lrr +\lrl)^2 +(\lall + \llr)^2 \right\} s_\theta^2\biggr] \\
 \left\vert {\cal M}_{_{\rm FCV}}^{\rm unp}\right\vert^2_{\rm opp\,\, hel}  &=& B^{\rm unp}_{_{\rm FCV}} \biggl[ \frac{1}{2}\biggl\{(1+ \beta_t^2)
\bigl[(\lrr)^2 +(\lall)^2 +(\lrl)^2 + (\llr)^2\bigr] 
+ (1 - \bt)^2 (\lrr \lrl + \lall \llr) \biggr\}  (1+c_\theta^2) 
\nn\\
&& \hskip 3cm
+\,\, 2 \,\beta_t \left(\lrrs +\lalls -\lrls - \llrs \right) c_\theta\biggr]\nn\\
{\rm with} \quad B^{\rm unp}_{_{\rm FCV}} &=& g_s^2\Bigl(
\frac{\hat s}{\Lambda_{\cal U}^2}\Bigr)^{2(\du-1)} \Bigl[
\frac{A_{\du}}{2\sin{(\du \pi)}} \Bigr]^2 
\label{eq-matsq-FCV}
\eea
\end{widetext}

We highlight  following observations with  regard  to the spin correlation \spincorr and charge asymmetry $\afbt$:
\begin{enumerate}
\item $\afbt$ gets contribution only from the opposite helicity amplitudes which is proportional to $2\beta_t\left[\lrrs +\lalls -\lrls - \llrs \right]$. Hence $\afbt $ vanishes for  $\beta_t =0$ (i.e. when top is produced at threshold)  and also for
 $\lall =\lrr =\pm \llr =\pm \lrl$ where $+(-)$ corresponds to  case
(c) and (d) respectively. However, the small contribution will come from the interference with
electroweak  part. 
We show the variation of \crtt and \spincorr for this case in figures \ref{fig:FC_vec_sing_sigma-1-2-d} and \ref{fig:FC_lambda_vec_sing_spcr-1-2-d}.
\item As  $\beta_t$ varies from 0 to 1 we observe that the spin-correlation varies from a large negative value to a positive number. This is attributed  to the fact  that the opposite (same) helicity contribution increases (decreases) with increasing $\beta_t$. Also,   the contribution to \crtt ,\afbt  and \spincorr decreases with increasing $\du$ for a given  coupling because of the term $\Lambda^{(\du-1)}$ in the denominator.
\item 
\afbt  and \spincorr  initially increase quadratically
 with the product of  couplings and then becomes constant as for the large couplings the   unparticle squared term takes over the SM, resulting into the cancellation of the coupling dependence in the numerator and denominator of these observables.
\item The $\cfbt \approx \afbt=\afbt $ (opposite helicity)  for FC vector unparticles as the interference term of the
 unparticles with EW part is negligibly small. 
\end{enumerate}
\begin{figure*}[!ht]
  \centering
  \subfloat[$\lambda^V_{RR} \ne 0$ and $\lambda^V_{LR}=\lambda^V_{RL}=\lambda^V_{LL} =0 $]{\label{fig:FC_vec_oct_sigma-1-2-a}\includegraphics[width=0.5\textwidth]{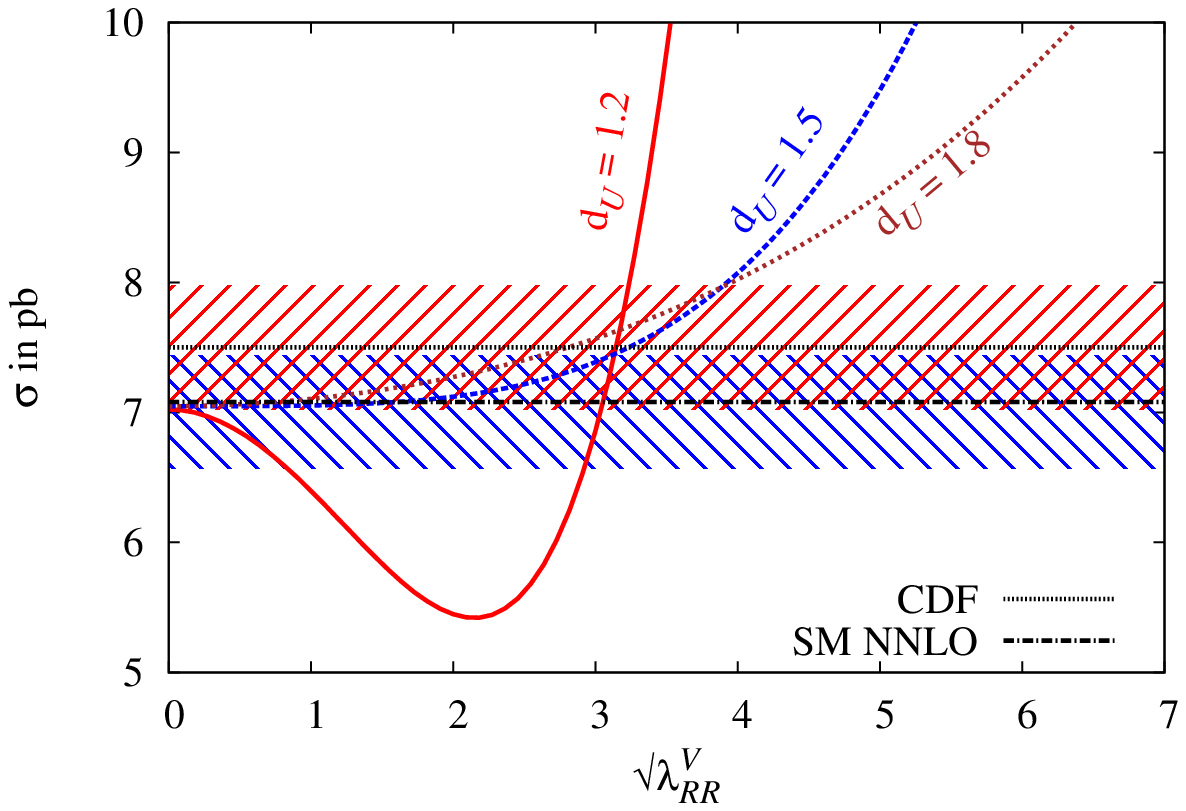}}
  \subfloat[$\lambda^V_{LL} = \lambda^V_{RR} = -\lambda^V_{RL}  = -\lambda^V_{LR} =\lambda^V_{AV}$]{\label{fig:FC_vec_oct_sigma-1-2-d}\includegraphics[width=0.5\textwidth]{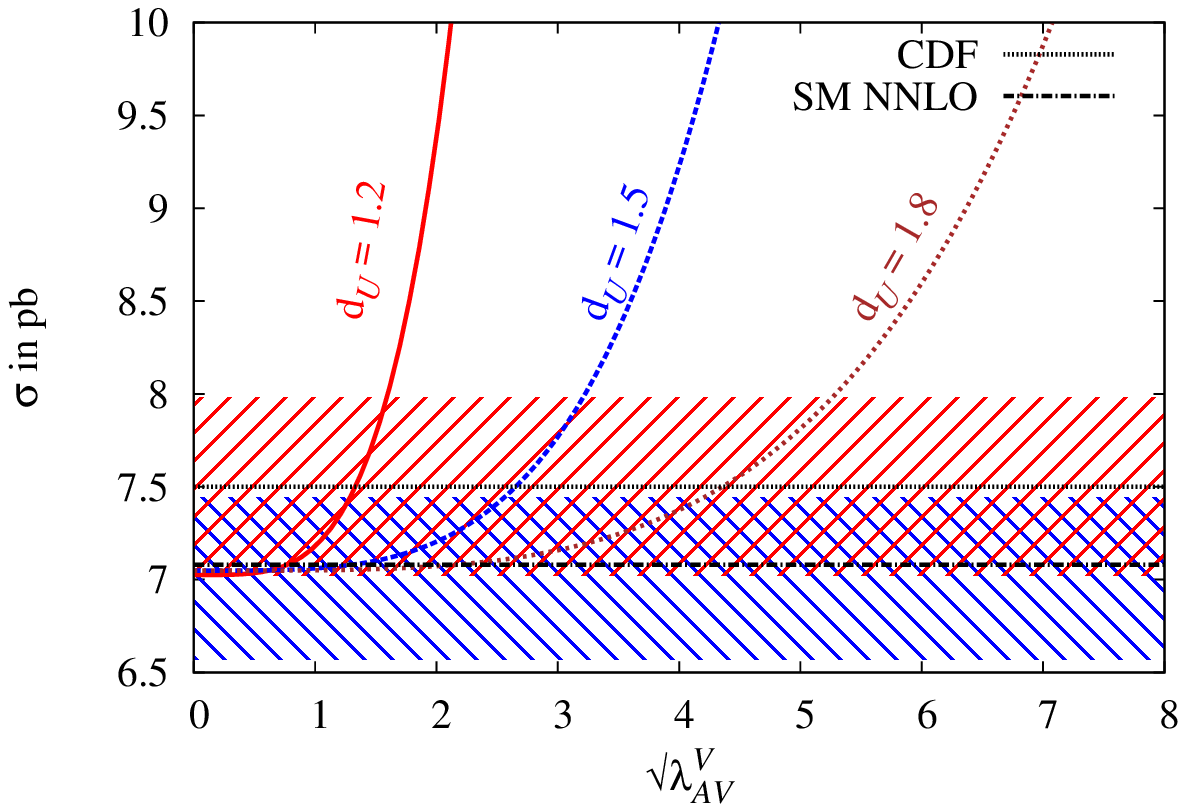}} \\
  \subfloat[$\lambda^V_{RR} \ne 0$ and $\lambda^V_{LR}=\lambda^V_{RL}=\lambda^V_{LL} =0 $]{\label{fig:FC_vec_oct_sigma-3-4-a}\includegraphics[width=0.5\textwidth]{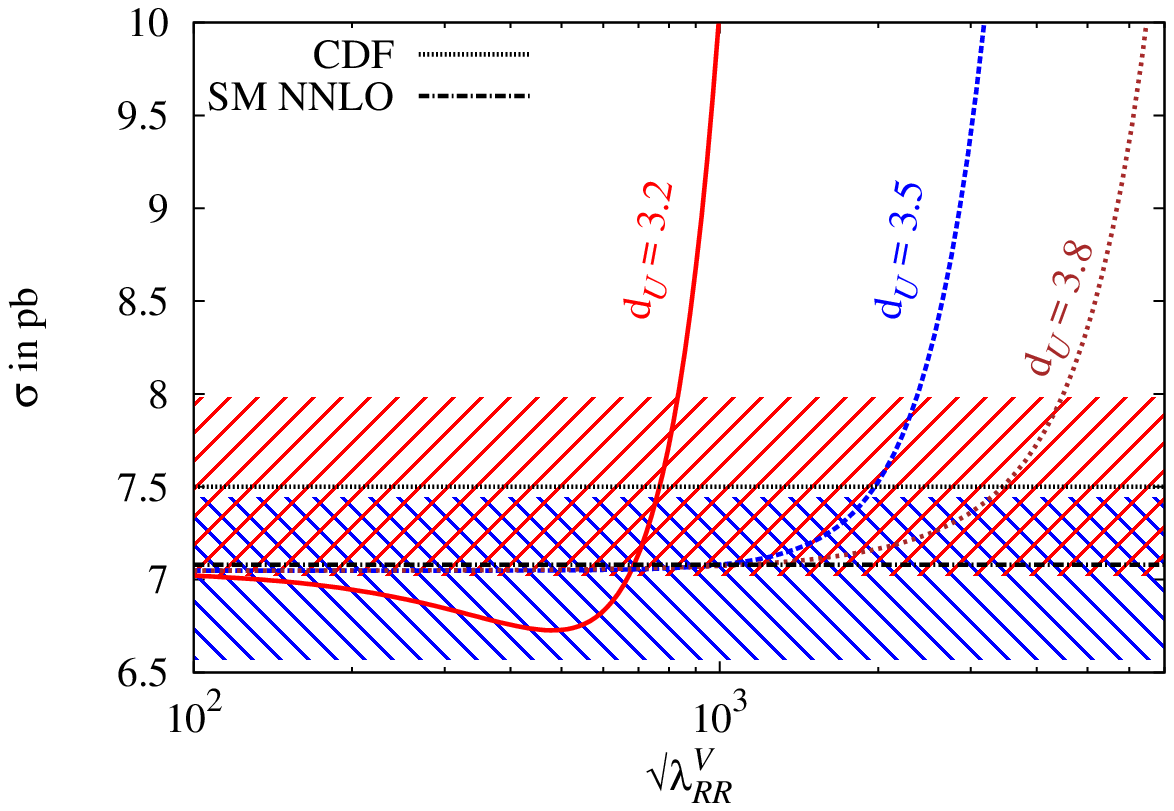}}
  \subfloat[$\lambda^V_{LL} = \lambda^V_{RR} = -\lambda^V_{RL}  = -\lambda^V_{LR} =\lambda^V_{AV}$]{\label{fig:FC_vec_oct_sigma-3-4-d}\includegraphics[width=0.5\textwidth]{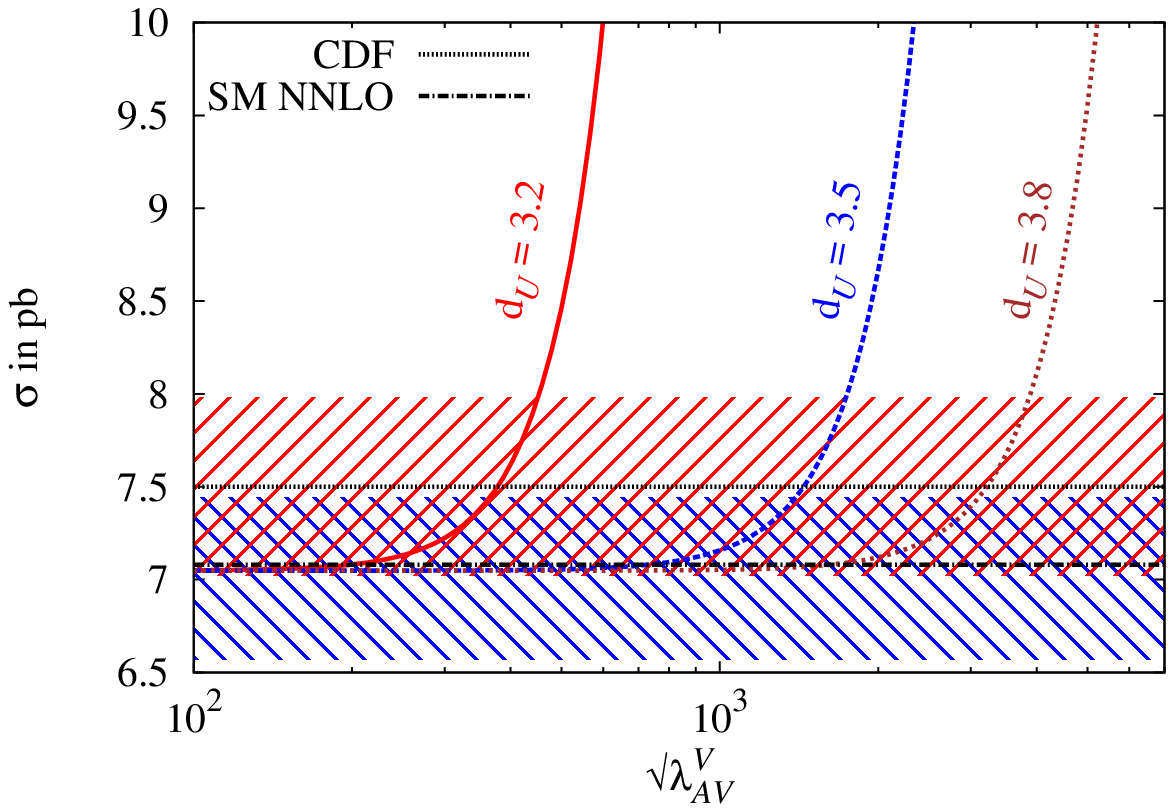}}
  \caption{\small \em{Variation of the cross-section $\sigma \left( p\bar p \to t\bar t\right)$ with  couplings
$\surd\lambda^V_{i j}$ for
color octet flavor conserving vector unparticles corresponding to  different values of  $d_{\cal U}$ at fixed $\Lambda_{\cal U}=1$ TeV . The  upper dotted line with a red band  depicts the  cross-section  $7.50 \pm 0.48$ pb from CDF (all channels) ~\cite{cdf-top-cross}, while the lower dot-dashed  line with a blue band show theoretical estimate $7.08\pm 0.36$ pb at NNLO  ~\cite{kidonakis-tcross}. The panels 'a' and 'b' show the variation of $\sigma$ for the cases (a) and (d) of the text for the various values of \du in the range $1<\du<2$ while the panels  'c' and 'd' give the same variation for the \du values in the range $3<\du<4$.}}
  \label{fig:FC_vec_oct_sigma}
\end{figure*}
\begin{figure*}[!ht]
  \centering
  \subfloat[$\lambda^V_{RR} \ne 0$ and $\lambda^V_{LR}=\lambda^V_{RL}=\lambda^V_{LL} =0 $]{\label{fig:FC_vec_oct_afb-1-2-a}\includegraphics[width=0.5\textwidth]{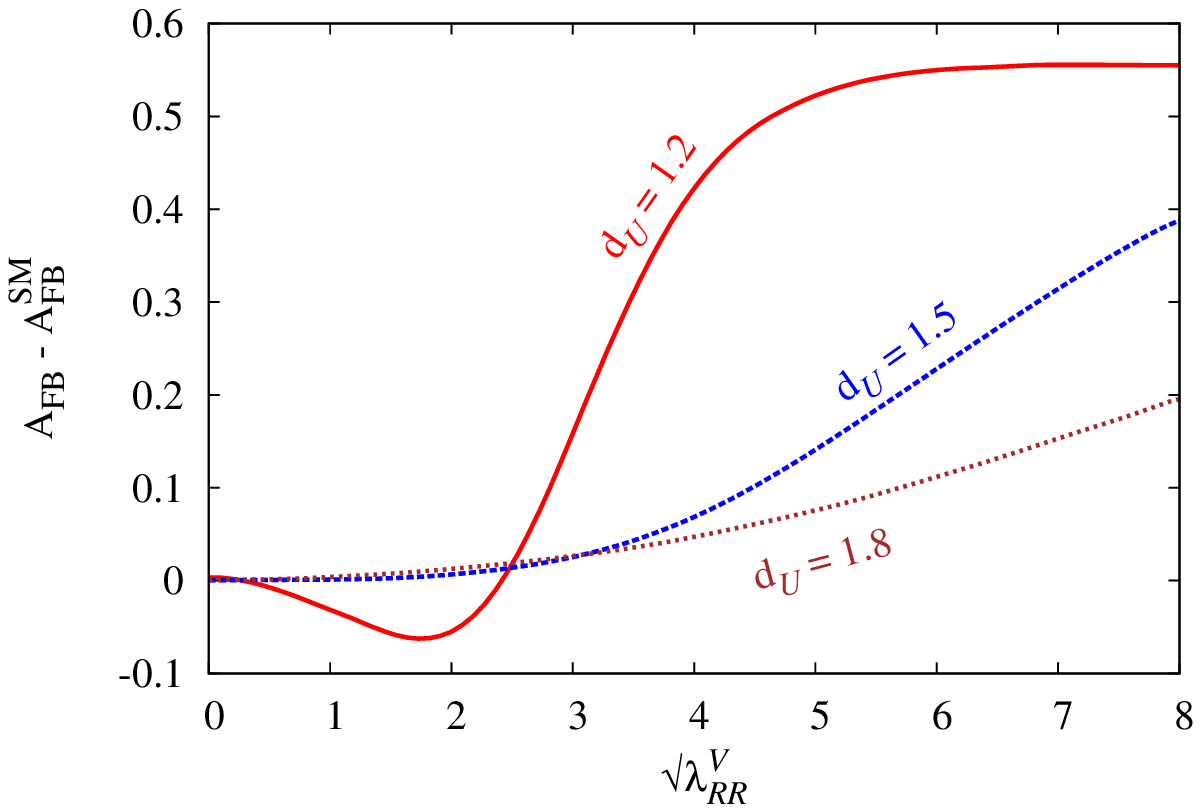}}
  \subfloat[$\lambda^V_{LL} = \lambda^V_{RR} = -\lambda^V_{RL}  = -\lambda^V_{LR} =\lambda^V_{AV}$]{\label{fig:FC_vec_oct_afb-1-2-d}\includegraphics[width=0.5\textwidth]{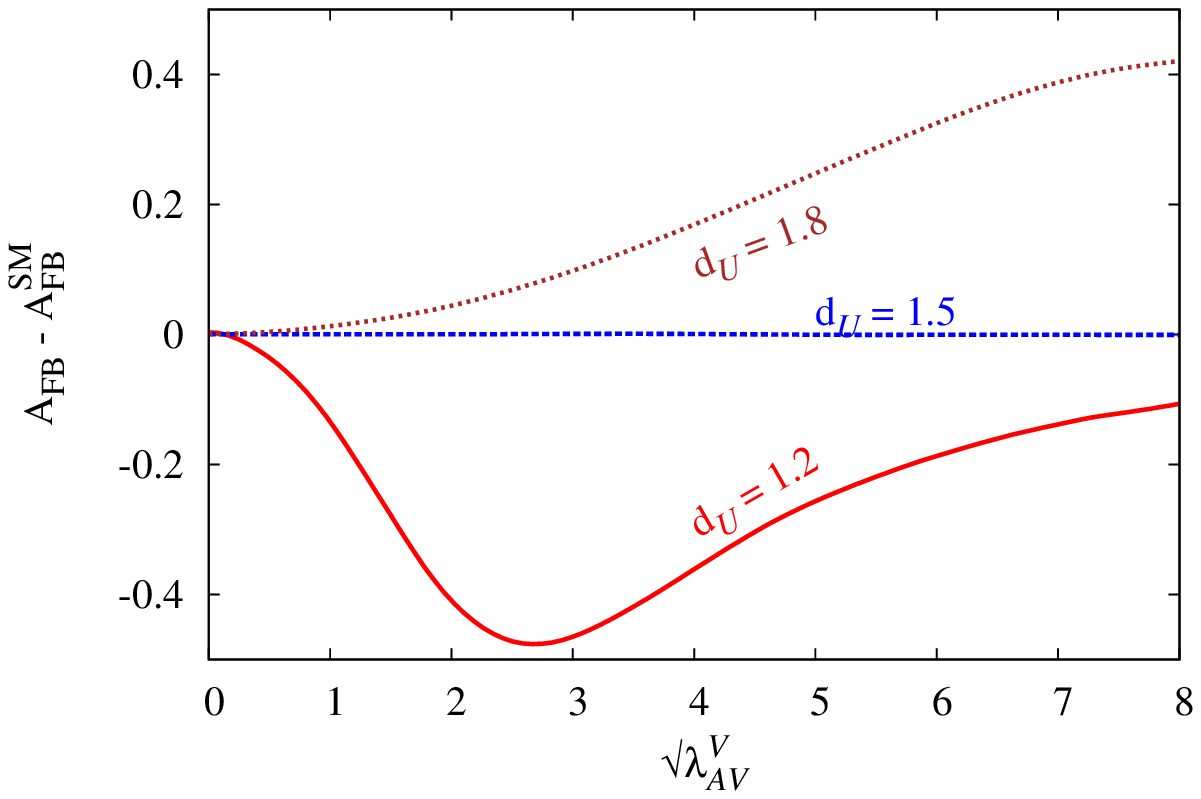}} \\
  \subfloat[$\lambda^V_{RR} \ne 0$ and $\lambda^V_{LR}=\lambda^V_{RL}=\lambda^V_{LL} =0 $]{\label{fig:FC_vec_oct_afb-3-4-a}\includegraphics[width=0.5\textwidth]{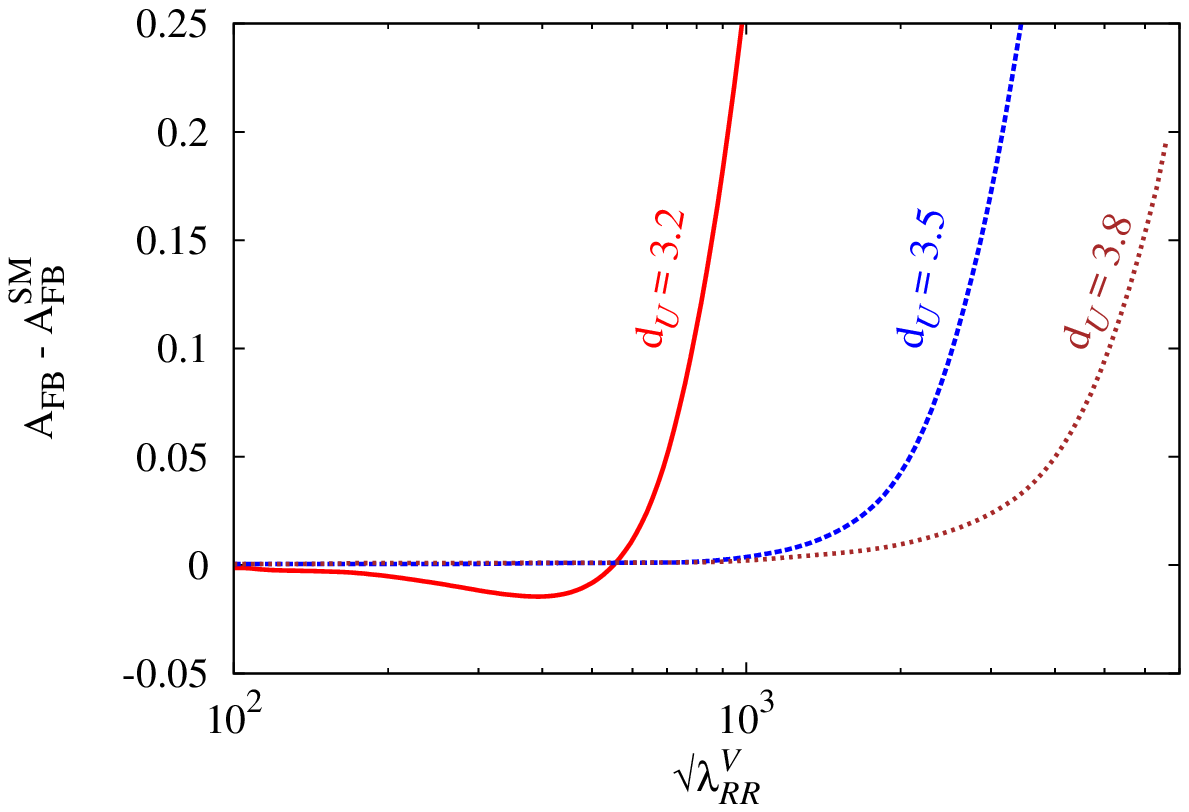}}
  \subfloat[$\lambda^V_{LL} = \lambda^V_{RR} = -\lambda^V_{RL}  = -\lambda^V_{LR} =\lambda^V_{AV}$]{\label{fig:FC_vec_oct_afb-3-4-d}\includegraphics[width=0.5\textwidth]{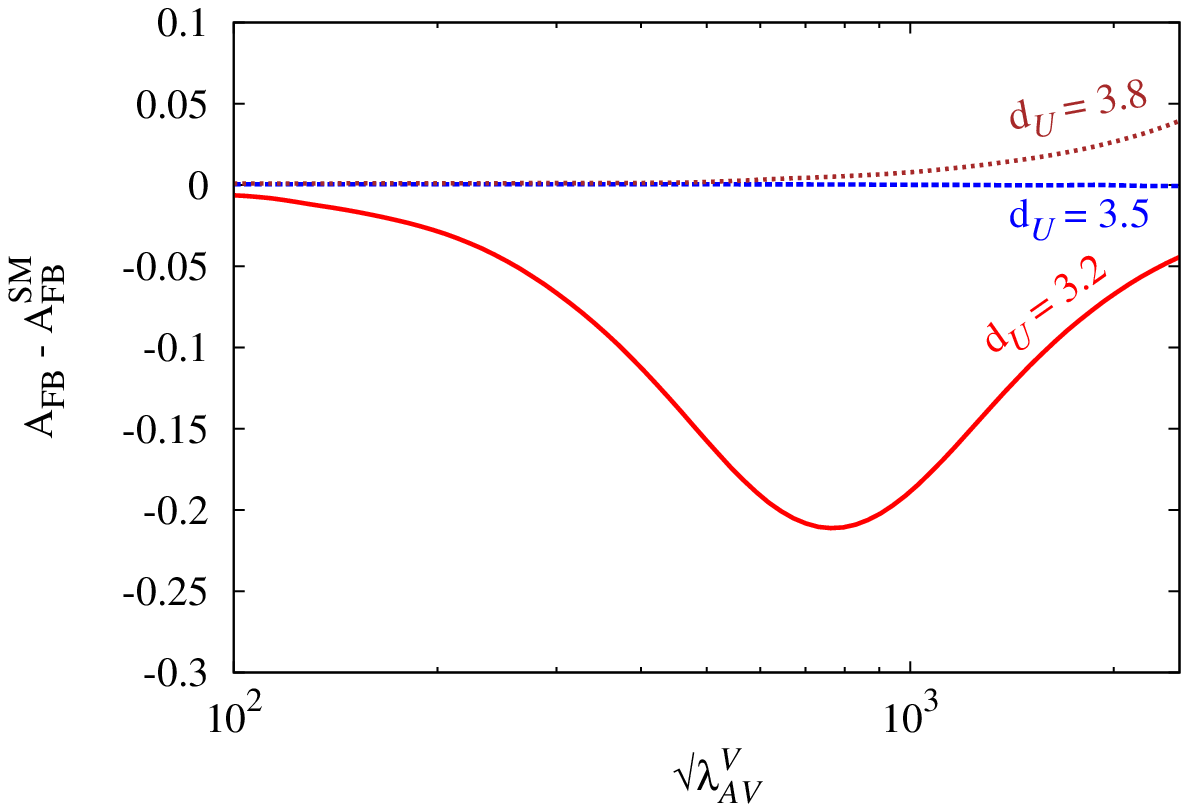}}\\
  \subfloat[$g_{L/R}^{{\cal U}_V \bar q q} \, g_{L/R}^{{\cal U}_V \bar t t} < 0 $, the choice of coupling combination favored by the non-universal axigluon models]{\label{fig:FC_vec_oct_afb-1-2-e}\includegraphics[width=0.5\textwidth]{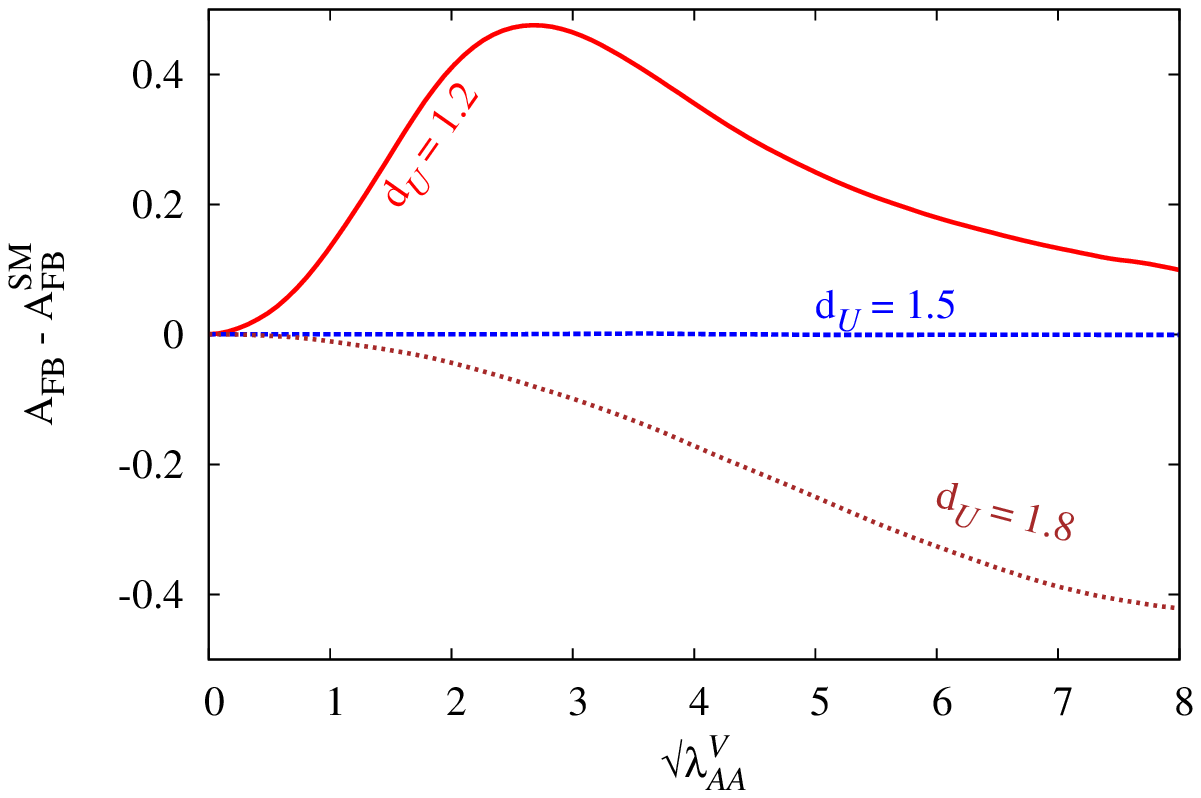}}
  \caption{\small \em{Variation of the $A_{FB} - A_{FB}^{SM}$(the unparticle contribution to \afbt) with  couplings $\surd \lambda^V_{i j}$  for color octet flavor conserving vector unparticles  for various values of $\du$ at fixed $\Lambda_{\cal U}=1$ TeV and for different coupling combinations mentioned in the text.  This is evaluated using the 1d differential distribution of rapidity in $t\bar t$ rest frame. }}
  \label{fig:FC_vec_oct_afb}
\end{figure*}
These graphs  exhibit  that it
is possible to get appreciable $\afbt$ (keeping the cross section and
spin correlation in agreement with the experimental values) for the
cases (a) and (b). This higher value of $\afbt$ for the first two
cases can be attributed to the complete asymmetry in the left and
right couplings. We find that for the other two cases (pure vector in case
(c) and pure axial vector in case (d)) there is no parameter region
that agrees with experimental value of \crtt and at the same time giving
appreciable $\afbt$. 
\par In figures \ref{fig:FC_vec_sing_sigma-3-4-a}, \ref
{fig:FC_lambda_vec_sing_afb-3-4-a} and
 \ref{fig:FC_lambda_vec_sing_spcr-3-4-a}, we show the variation of \crtt,
 \afbt and \spincorr respectively with the coupling for $\du > 3$ which is allowed by the
 unitarity of completely conformally invariant hidden sector. For such large values of \du the unparticle effects are pronounced for very high values of couplings and hence the interference with electroweak sector plays a crucial role. It is to be noted that
 the SM contact terms induced at large scale
 can also become important in this region of parameter space. 
\subsubsection{Color Octet Vector Unparticle}
\begin{figure*}[!ht]
  \centering
  \subfloat[$\lambda^V_{RR} \ne 0$ and $\lambda^V_{LR}=\lambda^V_{RL}=\lambda^V_{LL} =0 $]{\label{fig:FC_lambda_vec_oct_spcr-1-2-a}\includegraphics[width=0.5\textwidth]{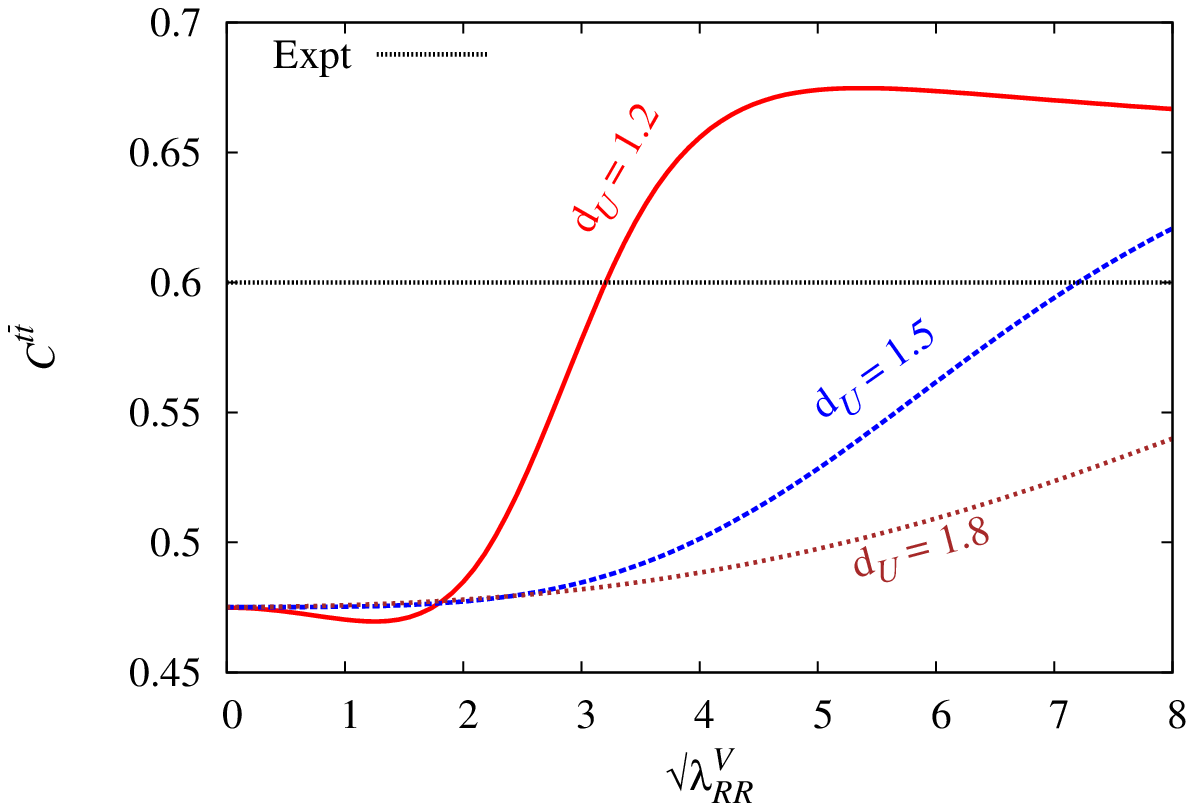}}
  \subfloat[$\lambda^V_{LL} = \lambda^V_{RR} = -\lambda^V_{RL}  = -\lambda^V_{LR} =\lambda^V_{AV}$]{\label{fig:FC_lambda_vec_oct_spcr-1-2-d}\includegraphics[width=0.5\textwidth]{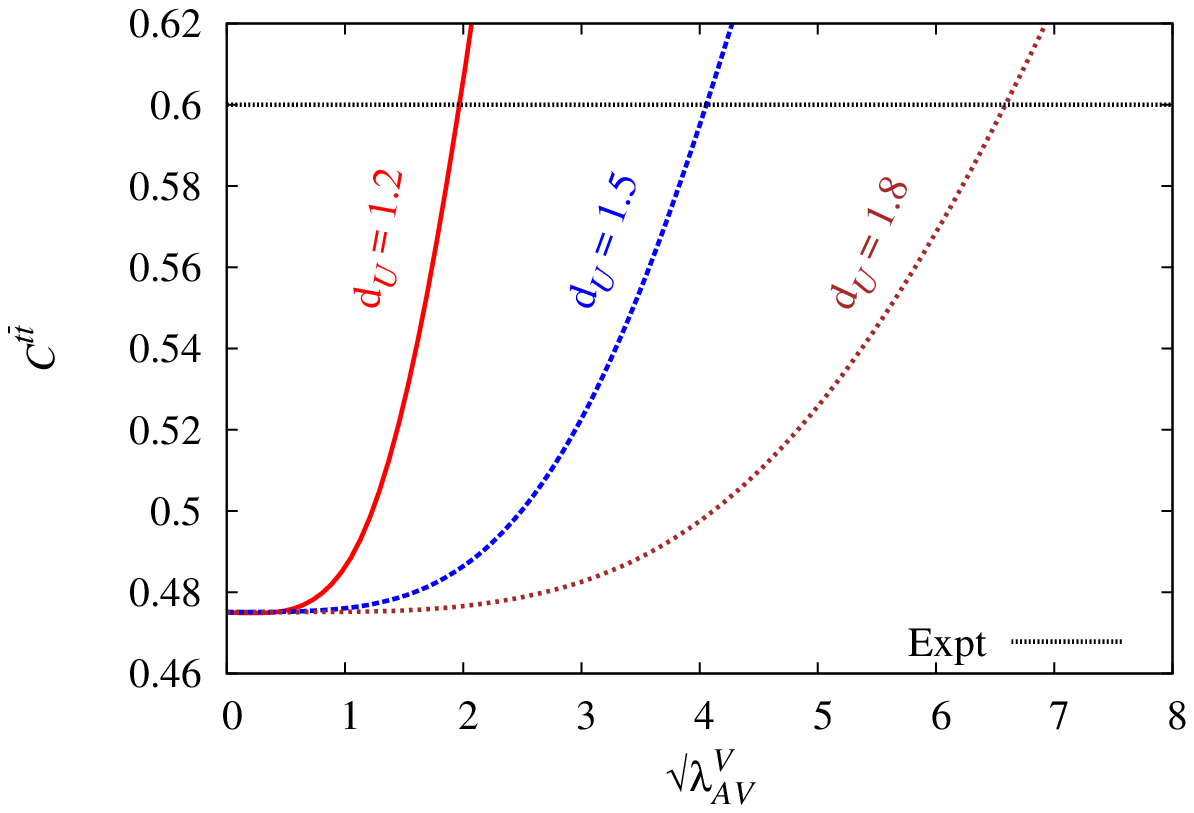}} \\
  \subfloat[$\lambda^V_{RR} \ne 0$ and $\lambda^V_{LR}=\lambda^V_{RL}=\lambda^V_{LL} =0 $]{\label{fig:FC_lambda_vec_oct_spcr-3-4-a}\includegraphics[width=0.5\textwidth]{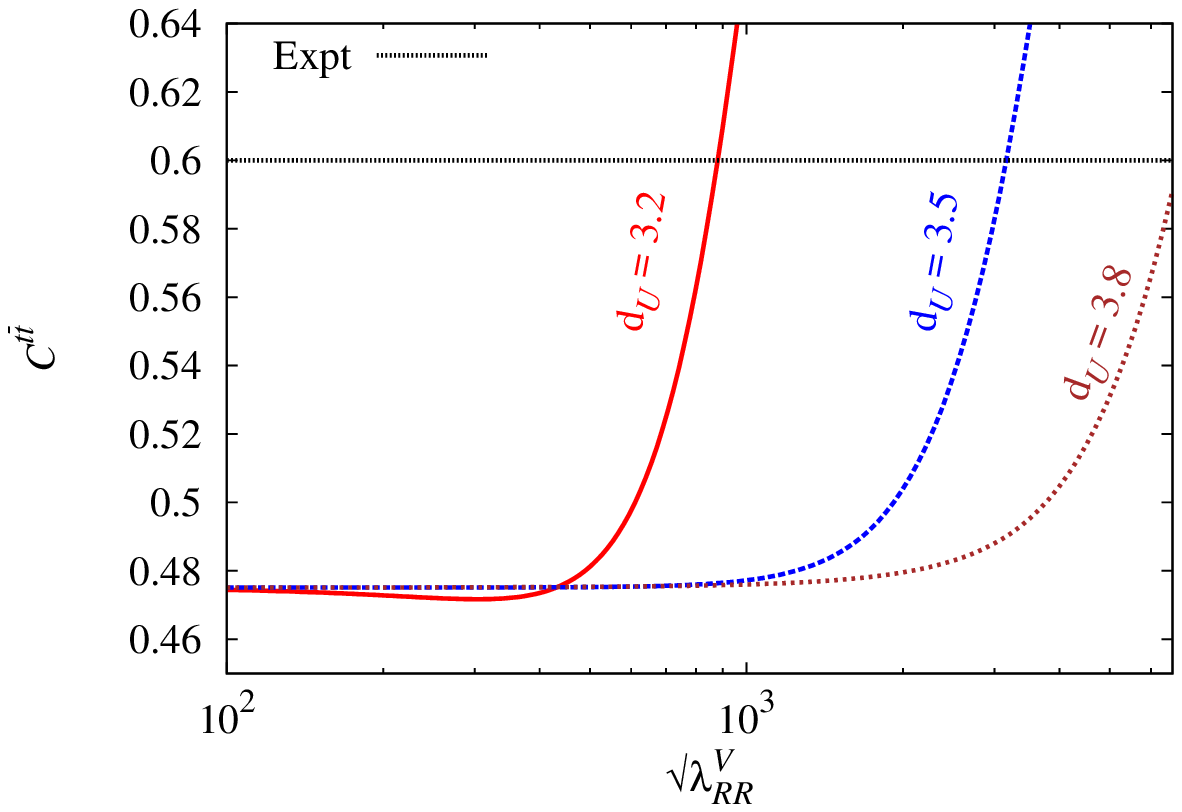}}
  \subfloat[$\lambda^V_{LL} = \lambda^V_{RR} = -\lambda^V_{RL}  = -\lambda^V_{LR} =\lambda^V_{AV}$]{\label{fig:FC_lambda_vec_oct_spcr-3-4-d}\includegraphics[width=0.5\textwidth]{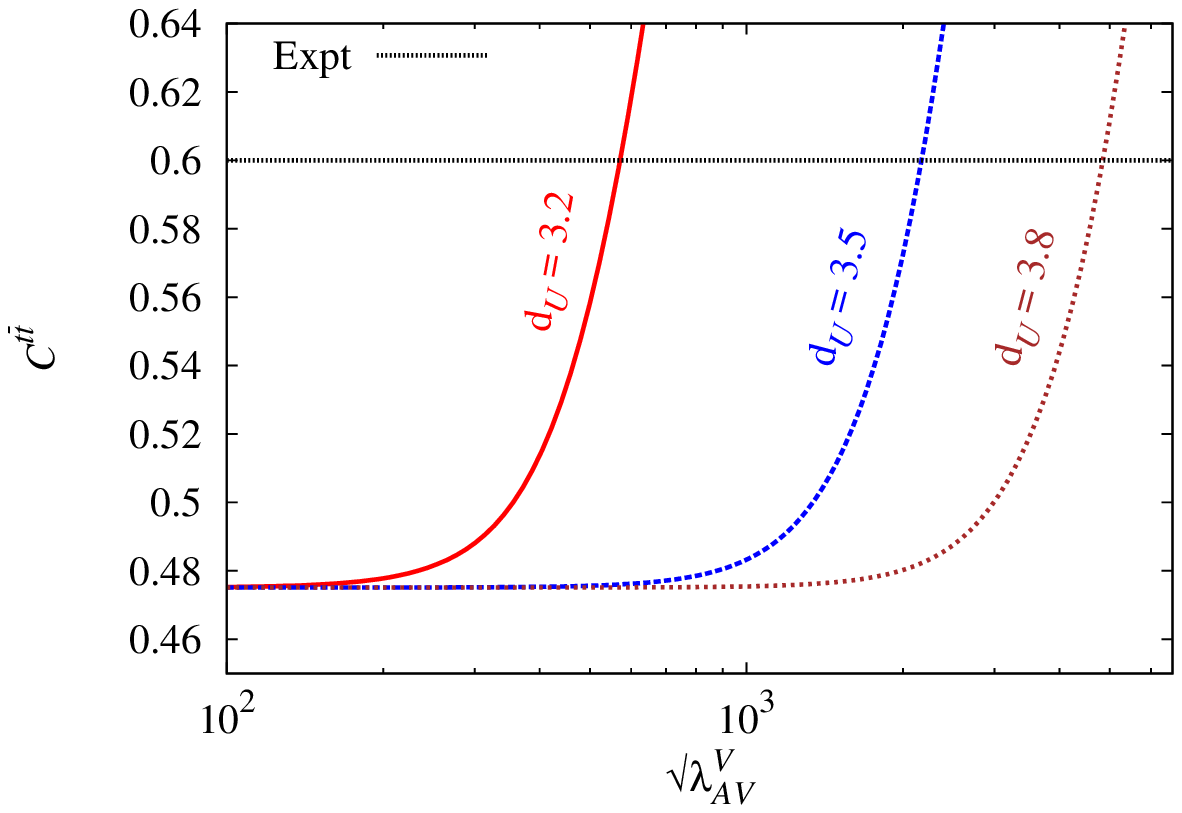}}
  \caption{\small \em{Variation of the spin-correlation coefficient \spincorr evaluated in helicity basis with  couplings $\surd \lambda^V_{i j}$ for color octet flavor conserving vector unparticles for various value of $\du$ at fixed $\Lambda_{\cal U}=1$ TeV and for different combinations of couplings. The experimental value is depicted with a dot-dashed line  at $0.60 \pm 0.50\ (stat) \pm 0.16\ (syst) $ ~\cite{Aaltonen:2010nz}.}}
  \label{fig:FC_vec_oct_spcr}
\end{figure*}
Next we consider the possibility of vector unparticles being color
octet with flavor conserving couplings. The variation of \crtt ,
$\afbt$ and \spincorr with couplings are shown in the
figures \ref{fig:FC_vec_oct_sigma},\ref{fig:FC_vec_oct_afb} and
\ref{fig:FC_vec_oct_spcr} respectively. 
The flavor conserving octet
unparticles do not interfere with flavor singlet electroweak
sector.  Thus the nature of these numbers can be explained
completely on the basis of the interplay of QCD and octet unparticle helicity amplitudes.
  From equations \eqref{def-Psq} and
\eqref{gprop-vec}, it may be seen that the interference terms given in
 equation \eqref{eq:mat_div}, can be rewritten as 
\begin{widetext}
\bea 
 \Bigl[2 {\cal M}^{\rm QCD} {\cal R}e({\cal M}_{_{\rm FCV}}^{\rm
  unp})\Bigr]_{\rm same\, hel} &=&  2 \, B_{_{\rm FCV}}^{\rm int}\,
(\lall +\llr +\lrl+\lrr) (1-\bt^2) \st^2 \label{eq:FCV_int_same} \\ 
{\rm and} \qquad \qquad \qquad \qquad \quad &&\nn\\ 
\Bigl[2 {\cal M}^{\rm
  QCD} {\cal R}e({\cal M}_{_{\rm FCV}}^{\rm unp} )\Bigr]_{\rm opp\, hel} &=&
2\, B_{_{\rm FCV}}^{\rm int}  \biggl[  (\lall +\lrr+\llr +\lrl) (1+\ct^2)\nn\\ 
&& \qquad \quad+\, 2 \,\beta_t \left(\lall +\lrr-\lrl-\llr \right)\ct \biggr]\label{eq:FCV_int_opp}\\
{\rm where} \qquad \qquad \qquad B_{_{\rm FCV}}^{\rm int}  &=&g_s^2\Bigl(
\frac{\hat s}{\Lambda_{\cal U}^2}\Bigr)^{(\du-1)} A_{\du} \biggl[-\frac{\cot{(\du\pi)}}{2}\biggr] \label{eq-int-matsq-FCV}
 \eea
\end{widetext}
The $\left\vert {\cal M}_{_{\rm FCV}}^{\rm unp}\right\vert^2_{\rm same\,\, hel}$ and $\left\vert M_{_{\rm FCV}}^{\rm opp}\right\vert^2_{\rm same\,\, hel}$ are same as given in equation (\ref{eq-matsq-FCV}).
Analysing these expressions we observe that 
\ben
\item  In the interference term, unlike in the spin correlation  only the opposite helicity amplitudes contributes to \afbt which is proportional to $4\,\bt \left(\lall +\lrr-\lrl-\llr\right)$. Hence \cfbt = \afbt.
\item
The helicity amplitudes for unparticles and QCD are left-right symmetric. Therefore  the variation of the observables with coupling products will be same for case (a) and case (b).
We show the variation for \crtt, \afbt
 and \spincorr for the case (a) in figures 
\ref{fig:FC_vec_oct_sigma-1-2-a}, \ref{fig:FC_vec_oct_afb-1-2-a} 
and \ref{fig:FC_lambda_vec_oct_spcr-1-2-a} for the phenomenologically interesting range of \du i.e. $1<\du<2$. The
 same variation for $3<\du<4$ (consistent with the conformally invariant
 sector) is shown in figures  \ref{fig:FC_vec_oct_sigma-3-4-a}, 
\ref{fig:FC_vec_oct_afb-3-4-a} and \ref{fig:FC_lambda_vec_oct_spcr-3-4-a}
 respectively. The case (c)
corresponds to pure vector couplings of unparticles with quarks which is similar to that of QCD. Hence negligible $\afbt$ is generated  only
 from the squared term of the EW neutral current. The case
(d) on the other hand generates appreciable $\afbt$ proportional to $16\bt \laa$ due to  the interference of vector (QCD)
with axial vector unparticle sector. The variation for \crtt, \afbt
 and \spincorr for case (d) is shown in the figures 
\ref{fig:FC_vec_oct_sigma-1-2-d}, \ref{fig:FC_vec_oct_afb-1-2-d} 
and \ref{fig:FC_lambda_vec_oct_spcr-1-2-d} for $1<\du<2$. The
 same variation for $3<\du<4$ (consistent with the conformally invariant
 sector) is shown in figures  \ref{fig:FC_vec_oct_sigma-3-4-d}, 
\ref{fig:FC_vec_oct_afb-3-4-d} and \ref{fig:FC_lambda_vec_oct_spcr-3-4-d}
 respectively.
\item The sign of interference term is
determined by the unparticle propagator which has a non-trivial phase
dependence upon $\du$.
We may write
\bea
\left\vert {\cal M}\right\vert^2 =  \left\vert {\cal M}^{\rm SM}\right\vert^2 + \delta\,\Bigl[ 2 \,{\cal M}^{\rm
  QCD}\bigl| {\cal R}e({\cal M}^{\rm unp})\bigr|\Bigr]
\eea
where $\delta$  is negative for $n <\du< (n+1/2)$, zero for $\du = (n+1/2)$  and positive for $ (n+1/2)<\du<(n+1)$ due to the presence of $\cot({\du\pi})$ ($n$ being an integer $\geq 1$). This effect is evident  in the figures~\ref{fig:FC_vec_oct_sigma}. We observe that due to the effect of the interference, the cross-section in presence of unparticles first decreases with increase in couplings and eventually increases with the onset of $\left\vert {\cal M}_{\rm unp}\right\vert^2$ for a rather large couplings.
\item This effect is also  well demonstrated for \afbt in figures \ref{fig:FC_vec_oct_afb-1-2-d} and \ref{fig:FC_vec_oct_afb-3-4-d} corresponding to case (d) where there is no contribution to \afbt  from $\left\vert {\cal M}_{\rm unp}\right\vert^2$. However for case (a), as seen from figures \ref{fig:FC_vec_oct_afb-1-2-a} and 
\ref{fig:FC_vec_oct_afb-3-4-a}, the unparticle squared term overtakes the interference term and drives the \afbt towards a   positive value with a gradual increase in the coupling strength. 
\par It is worthy to mention that for all axigluon models we require the light quark-axigluon and top-axigluon coupling  to be of opposite  sign  in order to generate positive asymmetry. This is not necessary with unparticles. However, if one chooses couplings in this non-universal way, then \afbt will pick up extra negative sign  as shown in  figure \ref{fig:FC_vec_oct_afb-1-2-e} for $1<\du<2$. Even for $\du>3$ one  obtains large positive asymmetry for $3<\du < 3.5$ for this choice of couplings.
\een 
\par Thus with octet vector unparticle sector having pure right or left
handed couplings or even with axial vector couplings, there is a
region in parameter space ({\it e.g.}  case(a) with $\du=1.2$ and case (d)  with $\du=1.8$ ) where it is possible to get appreciable
positive $\afbt$ keeping the \tt production cross-section and spin
correlation consistent with experimental measurement.  For
 higher \du consistent with the unitarity bound of completely conformally
 invariant sector, i.e. $\du>3$, not enough positive \afbt is obtained for
 any \du and coupling consistent with the \crtt. 
\par Chen et al~\cite{Chen:2010hm}
 have calculated $\afbt$ in the lab frame for the same case (for FC
 vector octet unparticles with pure right handed coupling). Using the
 definition of equation.~\eqref{eq:int-afb-def-cos} and  for an identical choice of  parameters
 our cross-section and \afbt are in agreement with them. 

\subsubsection{FC Color Singlet Tensor Unparticle}
As in the case of vector unparticles, the $s$ channel process through FC color singlet
tensor unparticles  interferes with the EW neutral current but not with the QCD.
Therefore pure
 unparticle amplitude (i.e. $\left\vert {\cal M}^{\rm unp}\right\vert^2$) determines the
 behavior of the observables. The left-right symmetry in helicity amplitudes results in almost identical behaviour  for the observables corresponding to the cases (a) and (b) which is similar to FC vector unparticles.
The same and opposite helicity contributions to $\left\vert {\cal M}^{\rm unp}\right\vert^2$   are given by
\begin{widetext}
\bea
 \left\vert {\cal M}_{_{\rm FCT}}^{\rm unp}\right\vert^2_{\rm same\,\, hel} &=& {\cal B}_{_{\rm FCT}}^{\rm unp}\Bigl[2 \bt^2 (1- \beta_t^2)
\left\{(\lrrt +\lrlt)^2 +(\lallt + \llrt)^2 \right\} s_\theta^2 \ct^2 \Bigr] \nn \\
 \left\vert {\cal M}_{_{\rm FCT}}^{\rm unp}\right\vert^2_{\rm opp\,\, hel}  &=&  \frac{1}{2}\bt^2 {\cal B}_{_{\rm FCT}}^{\rm unp}\biggl[\Bigl\{ (1+ \beta_t^2)
\left(\lrrts +\lallts +\lrlts + \llrts \right)  \nn\\
&&+ 2 (1- \beta_t^2)
\left(\lrrt \lrlt +\lallt  \llrt \right)\Bigr\} (1-3c_\theta^2 + 4c_\theta^4) \nn\\
&& - 4  \beta_t \left(\lrrts +\lallts -\lrlts - \llrts \right) (1-2c_\theta^2) c_\theta \biggr]
\label{tensor_unp_sq}
\\
{\rm where} \qquad 
{\cal B}_{_{\rm FCT}}^{\rm unp} &=& \frac{1}{16} {\cal C}_f^{{\cal U}^{\rm FC}_T}\biggl[g_s^2\bigl(
\frac{\hat s}{\Lambda_{\cal U}^2}\bigr)^{2\du}  \bigl[
\frac{A_{\du}}{2\sin{(\du \pi)}} \bigr]^2 (\du^2) (\du -1 )^2 \biggr]
\label{tensor_unp_sq_const}
\eea
\end{widetext}
 The variation of the observables cross section \crtt and spin correlation \spincorr  with $\lambda^T_{ij}$  for different cases is shown in figures \ref{fig:FC_ten_sing_sigma} and
\ref{fig:FC_ten_sing_spcr} respectively.   On comparing with pure vector case we observe that  an additional $\beta_t$ factor in the helicity amplitudes     suppresses the \afbt   for case (a) or  (b) in  the   range of parameters  which keeps the cross section and spin correlation \spincorr in agreement with the
allowed experimental values.  Cases (c) and (d) do not give any contribution to \afbt. Moreover, the order of couplings involved are much larger (keeping the \crtt within experimental limits) compared to vector unparticle case because the overall coupling in tensor case have extra factor of $\Lambda$ in the denominator. Unitarity bounds the \dut values for conformally invariant symmetric
 tensor to be $\dut>4$. However, for such values of \dut the unparticle
 effects are highly subdued as may be seen from figures 
\ref{fig:FC_ten_sing_sigma} and \ref{fig:FC_ten_sing_spcr}.
\begin{figure*}[!ht]
  \centering
  \subfloat[$\lambda^T_{RR} \ne 0$ and $\lambda^T_{LR}=\lambda^T_{RL}=\lambda^T_{LL} =0 $]{\label{fig:FC_ten_sing_sigma-1-2-a}\includegraphics[width=0.5\textwidth]{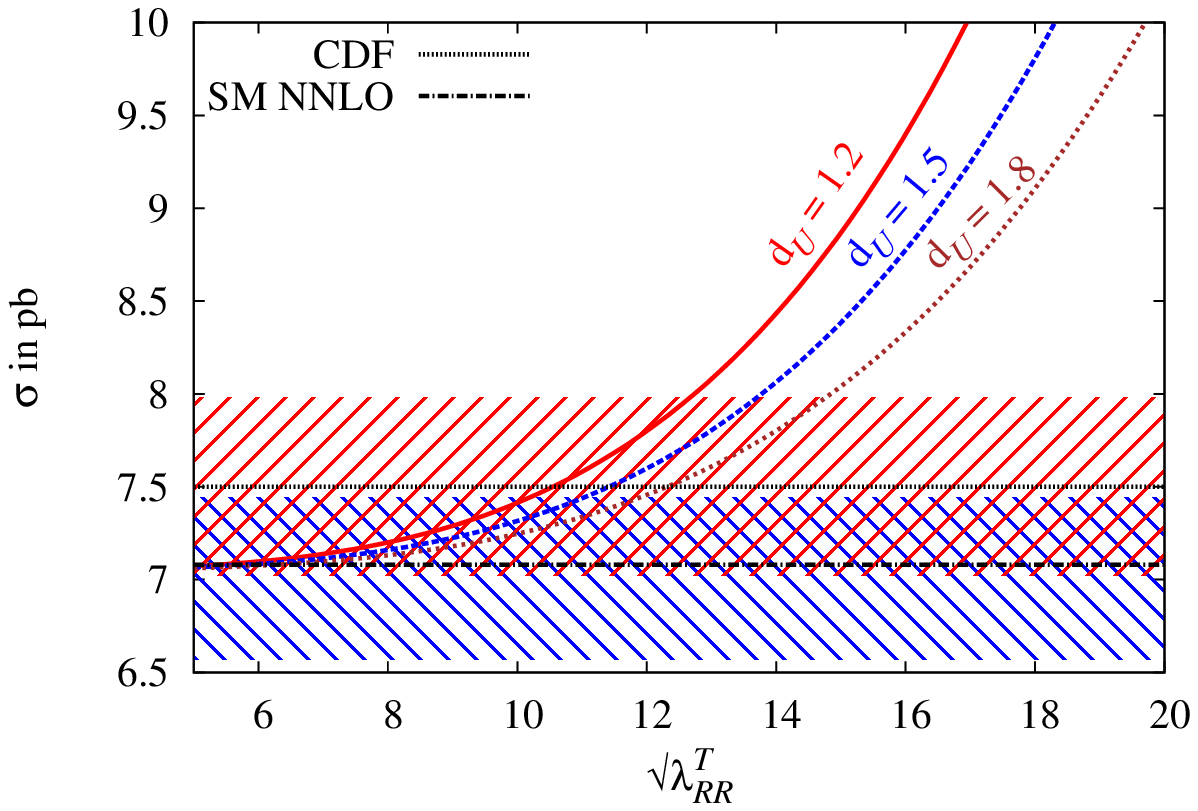}}
  \subfloat[$\lambda^T_{LL} = \lambda^T_{RR} = \lambda^T_{RL}  = \lambda^T_{LR} =\lambda^T_{TT}$]{\label{fig:FC_ten_sing_sigma-1-2-c}\includegraphics[width=0.5\textwidth]{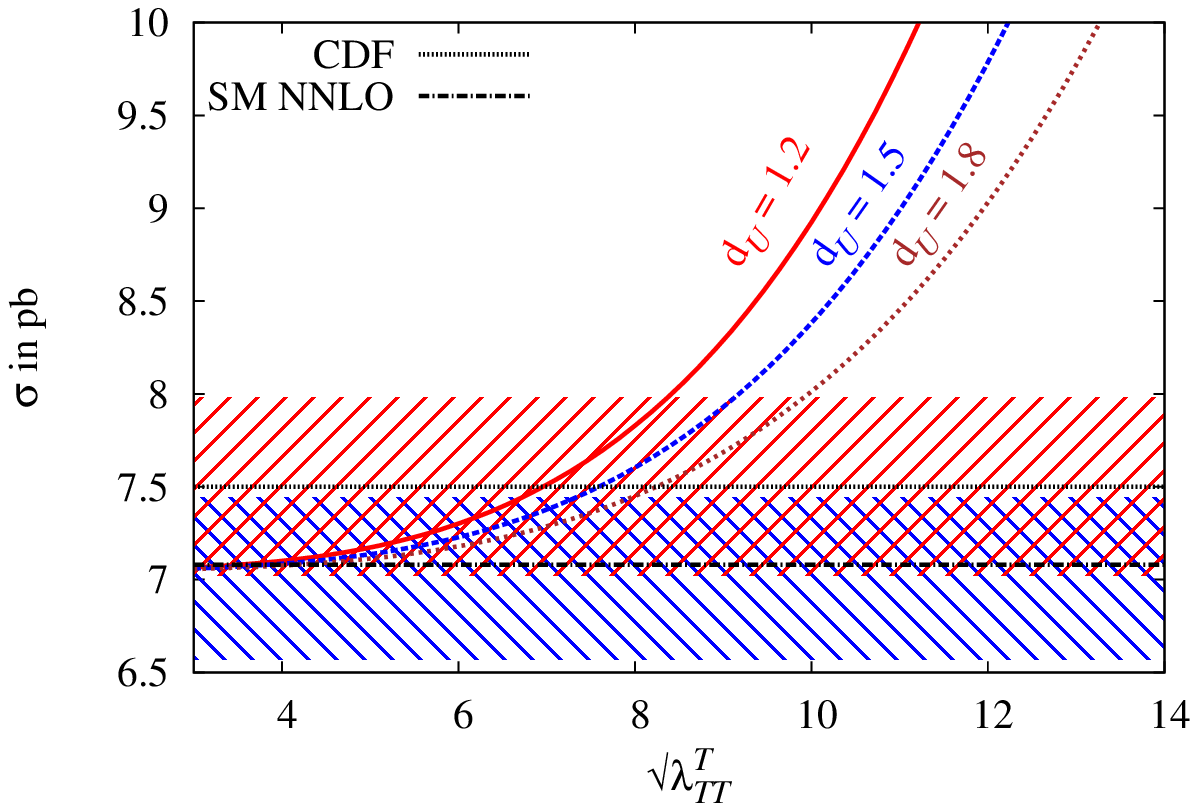}} \\
  \subfloat[$\lambda^T_{LL} = \lambda^T_{RR} = -\lambda^T_{RL}  = -\lambda^T_{LR} =\lambda^T_{AT}$]{\label{fig:FC_ten_sing_sigma-1-2-d}\includegraphics[width=0.5\textwidth]{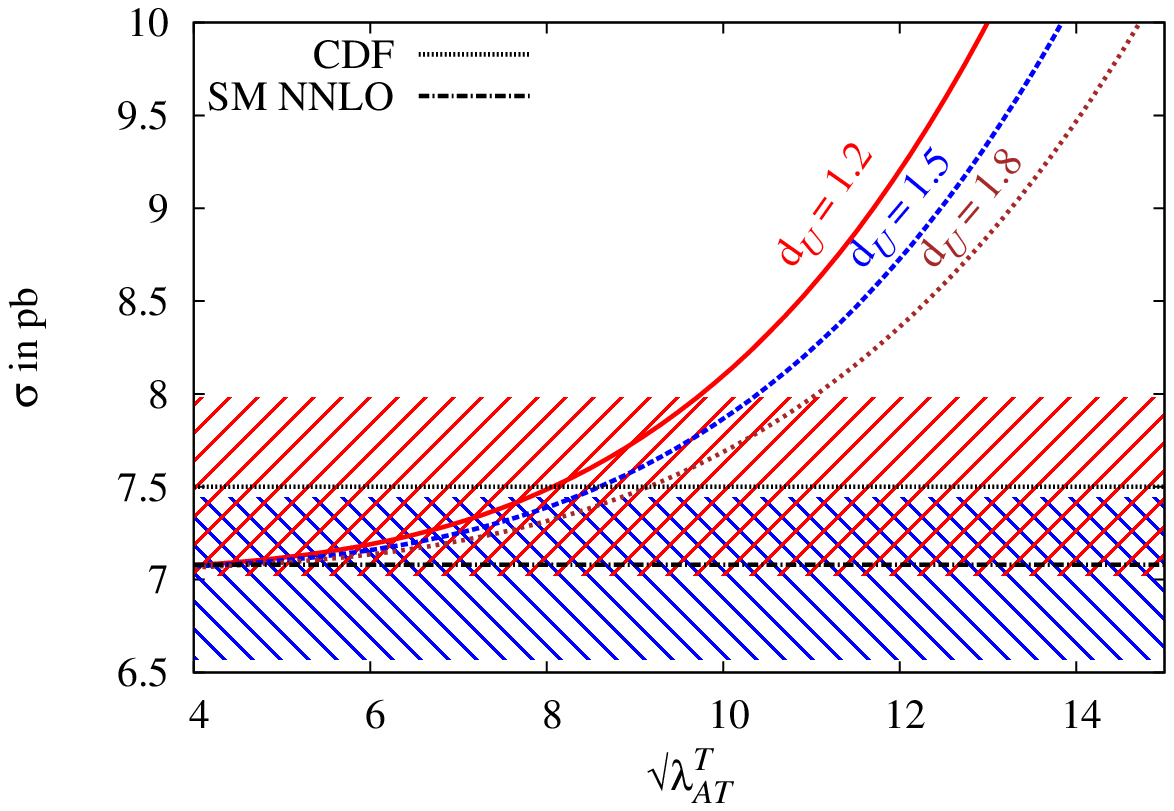}}
  \subfloat[$\lambda^T_{RR} \ne 0$ and $\lambda^T_{LR}=\lambda^T_{RL}=\lambda^T_{LL} =0 $]{\label{fig:FC_ten_sing_sigma-4-5-a}\includegraphics[width=0.5\textwidth]{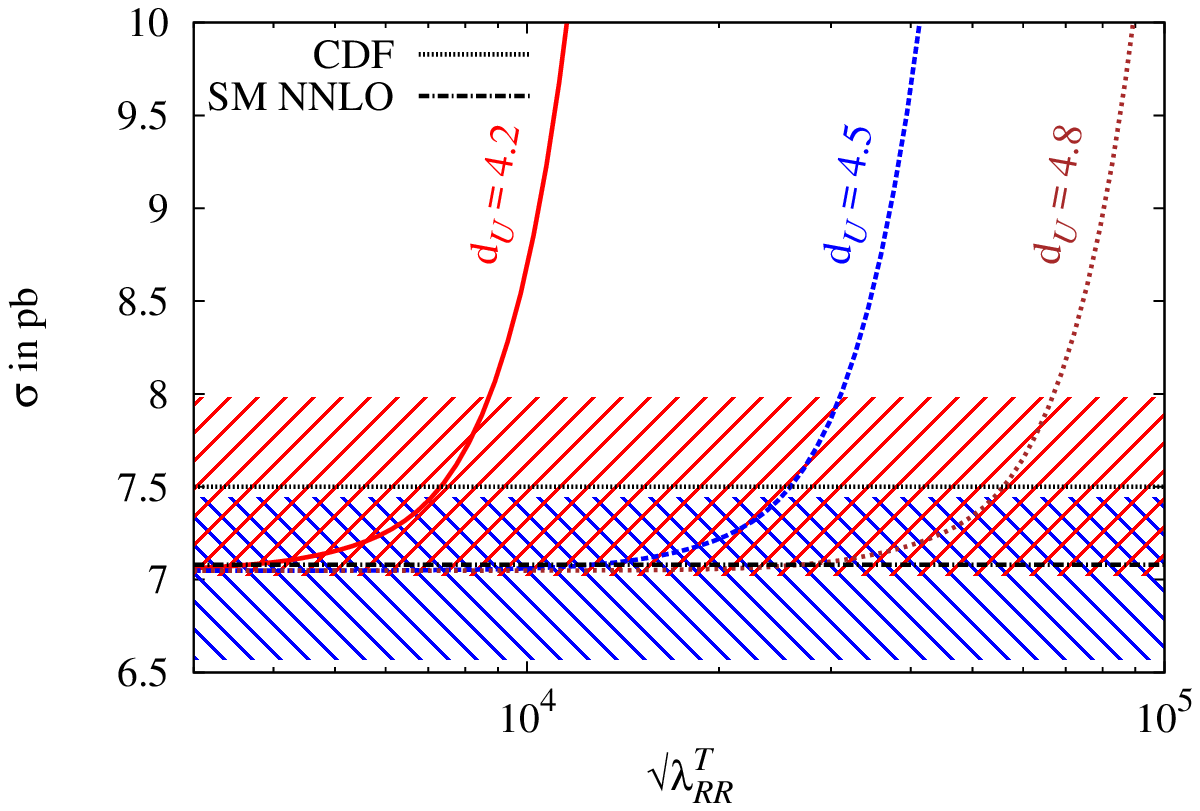}}\\
  \subfloat[$\lambda^T_{LL} = \lambda^T_{RR} = \lambda^T_{RL}  = \lambda^T_{LR} =\lambda^T_{TT}$]{\label{fig:FC_ten_sing_sigma-4-5-c}\includegraphics[width=0.5\textwidth]{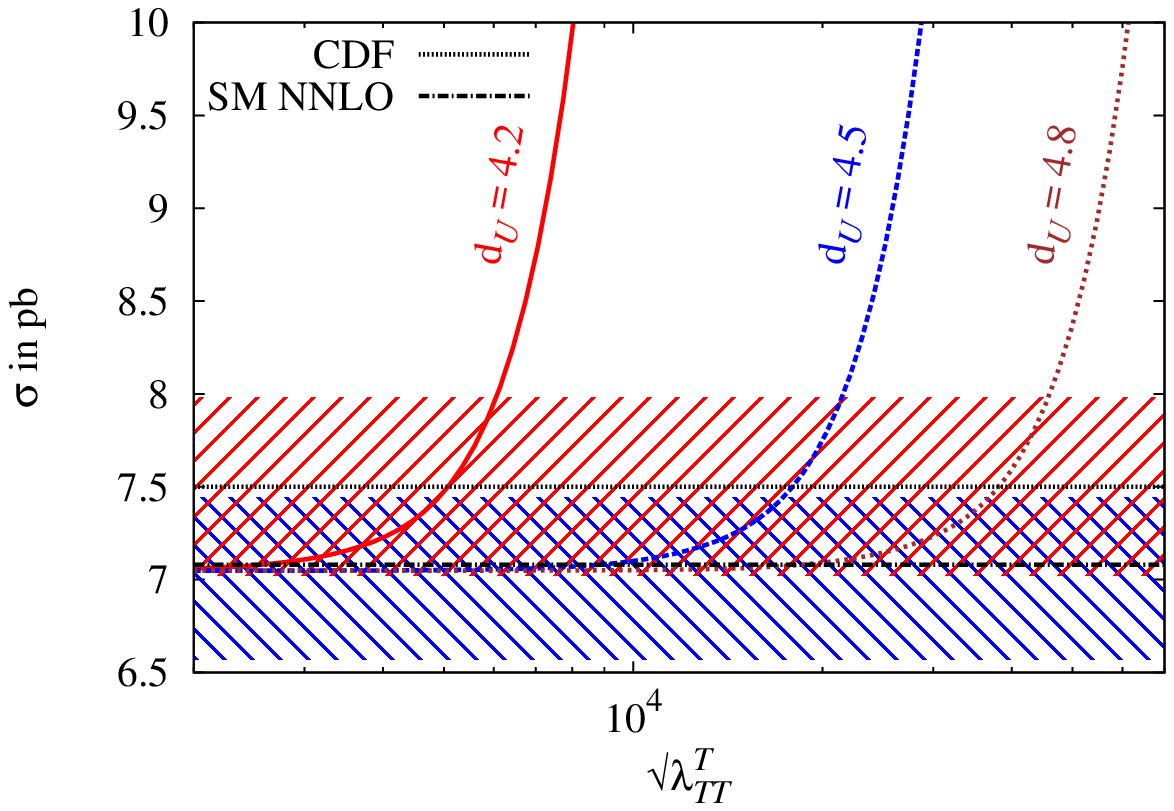}}
  \subfloat[$\lambda^T_{LL} = \lambda^T_{RR} = -\lambda^T_{RL}  = -\lambda^T_{LR} =\lambda_{AT}$]{\label{fig:FC_ten_sing_sigma-4-5-d}\includegraphics[width=0.5\textwidth]{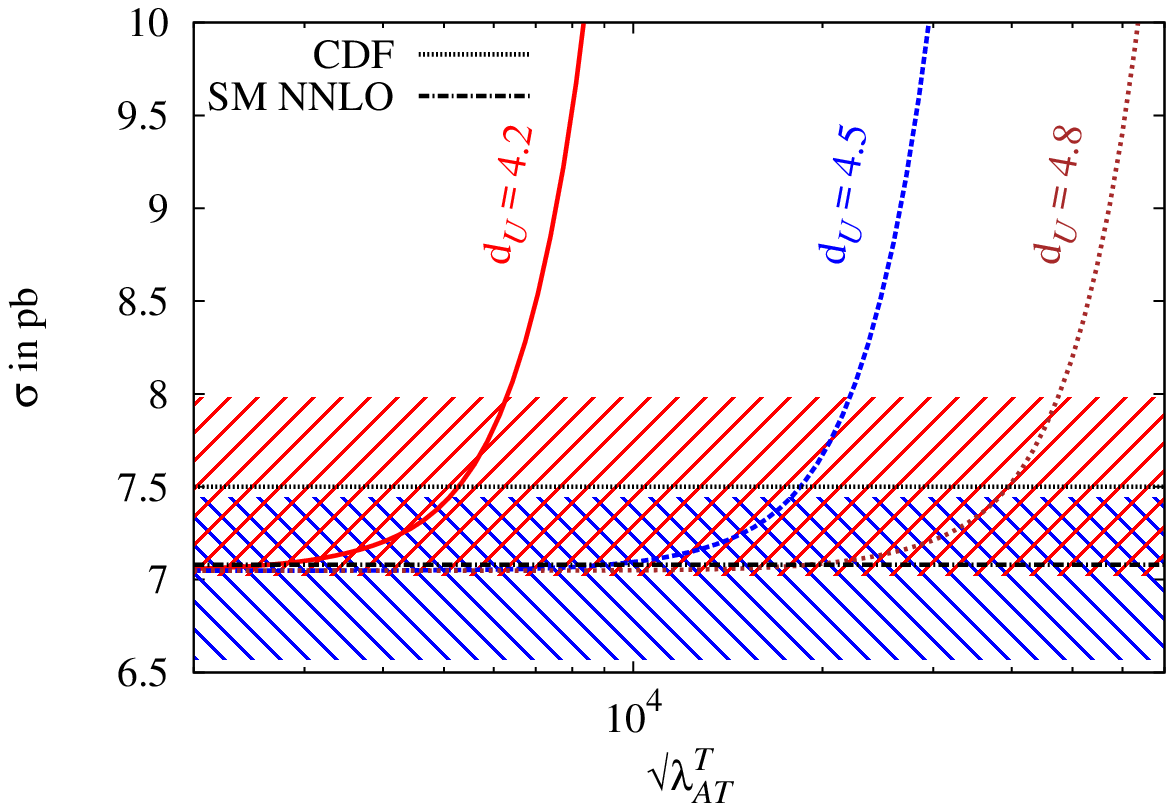}}
\caption{\small \em{Variation of the cross-section $\sigma \left( p\bar p \to t\bar t\right)$ with  couplings
$\surd\lambda^T_{i j}$ for
color singlet flavor conserving tensor unparticles corresponding to  different values of  $d_{\cal U}$ at fixed $\Lambda_{\cal U}=1$ TeV and for different combinations of couplings mentioned in the text . The  upper dotted line with a red band  depicts the  cross-section  $7.50 \pm 0.48$ pb from CDF (all channels) ~\cite{cdf-top-cross}, while the lower dot-dashed  line with a blue band show theoretical estimate $7.08\pm 0.36$ pb at NNLO  ~\cite{kidonakis-tcross}
.}} 
  \label{fig:FC_ten_sing_sigma}
\end{figure*}
\begin{figure*}[!ht]
  \centering
  \subfloat[$\lambda^T_{RR} \ne 0$ and $\lambda^T_{LR}=\lambda^T_{RL}=\lambda^T_{LL} =0 $]{\label{fig:FC_ten_sing_spcr-1-2-a}\includegraphics[width=0.5\textwidth]{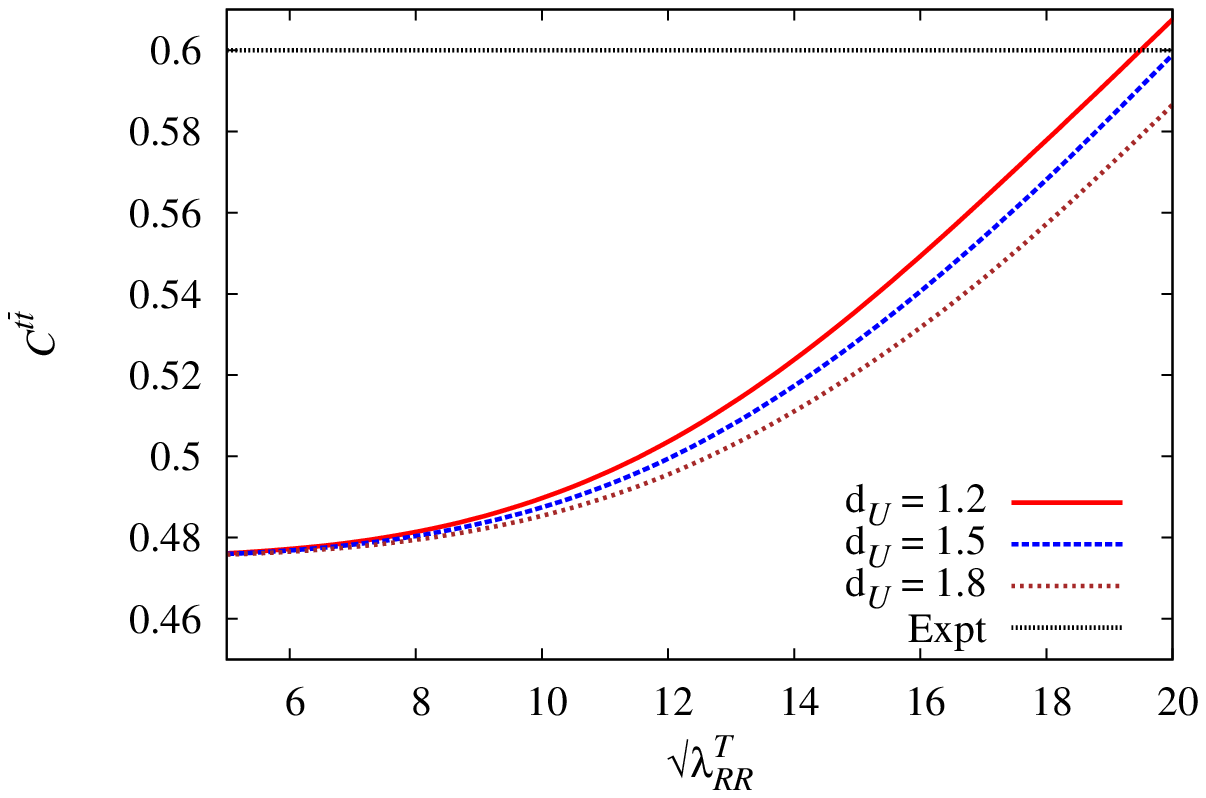}}
  \subfloat[$\lambda^T_{LL} = \lambda^T_{RR} = \lambda^T_{RL}  = \lambda^T_{LR} =\lambda^T_{TT}$]{\label{fig:FC_ten_sing_spcr-1-2-c}\includegraphics[width=0.5\textwidth]{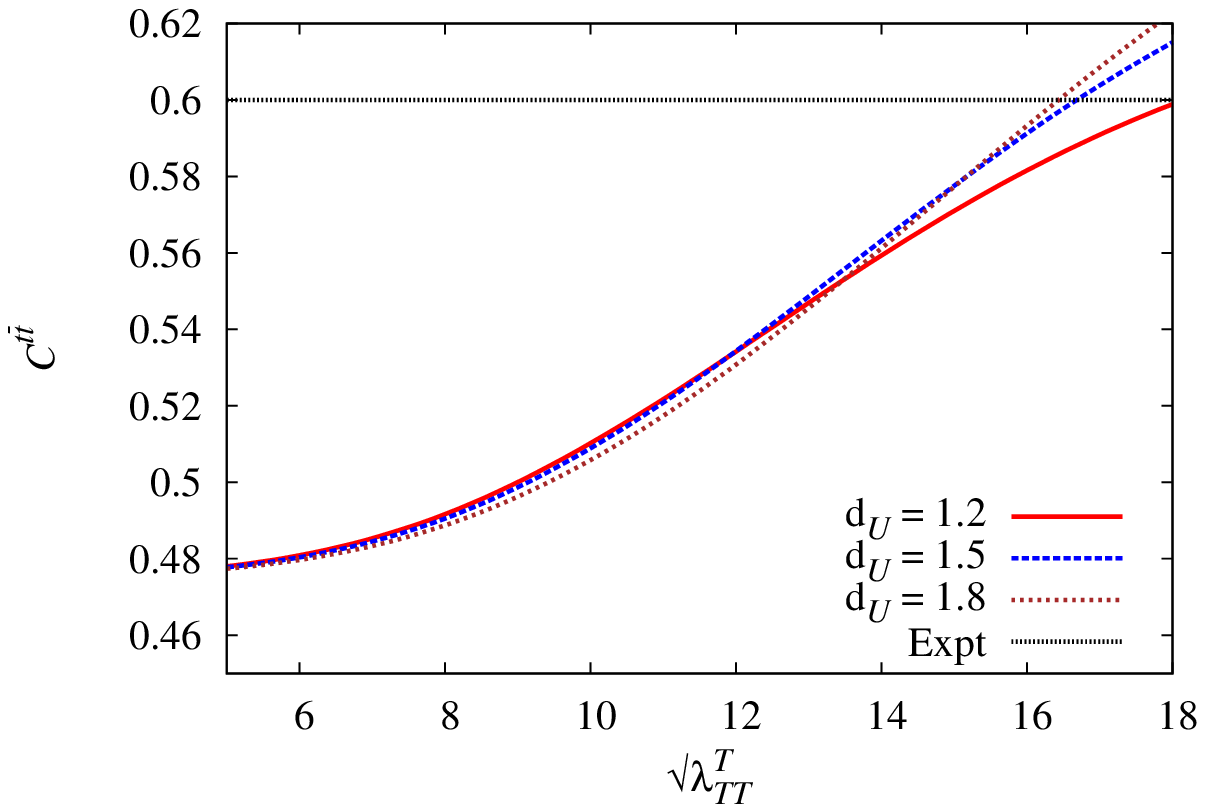}} \\
  \subfloat[$\lambda^T_{LL} = \lambda^T_{RR} = -\lambda^T_{RL}  = -\lambda^T_{LR} =\lambda_{AT}$]{\label{fig:FC_ten_sing_spcr-1-2-d}\includegraphics[width=0.5\textwidth]{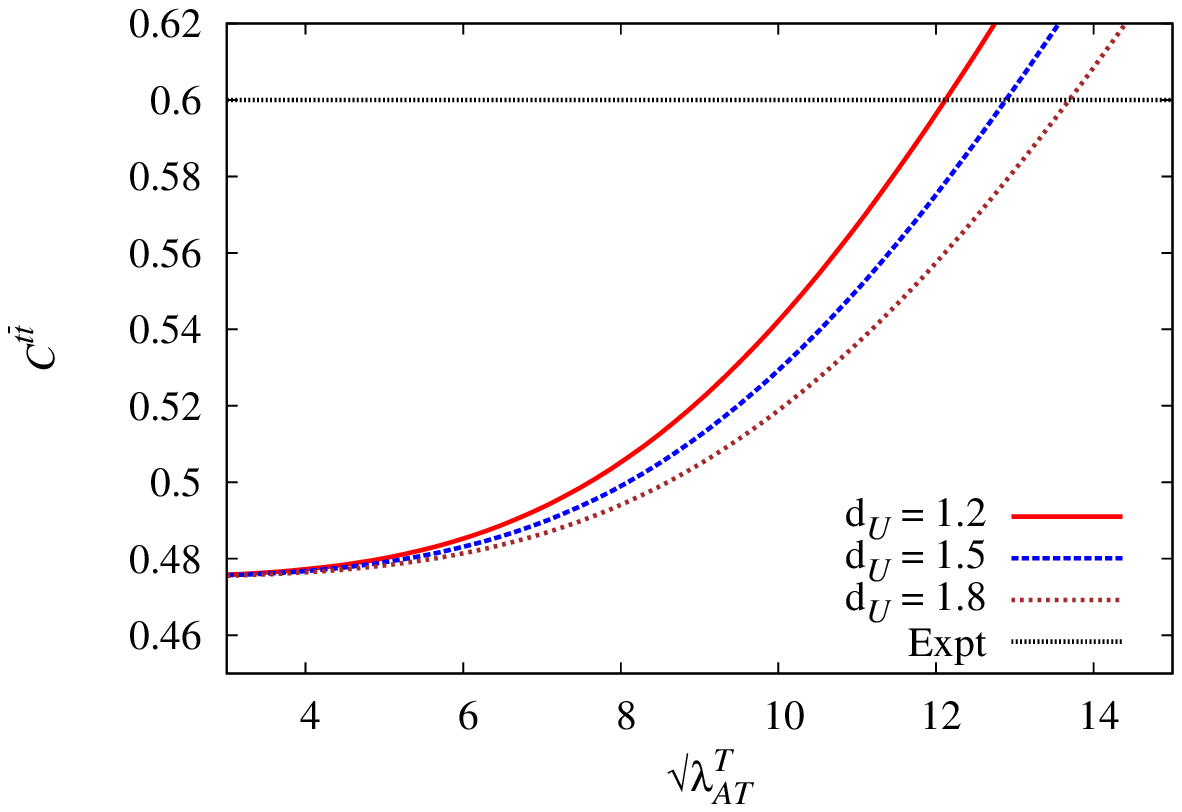}}
  \subfloat[$\lambda^T_{RR} \ne 0$ and $\lambda^T_{LR}=\lambda^T_{RL}=\lambda^T_{LL} =0 $]{\label{fig:FC_ten_sing_spcr-4-5-a}\includegraphics[width=0.5\textwidth]{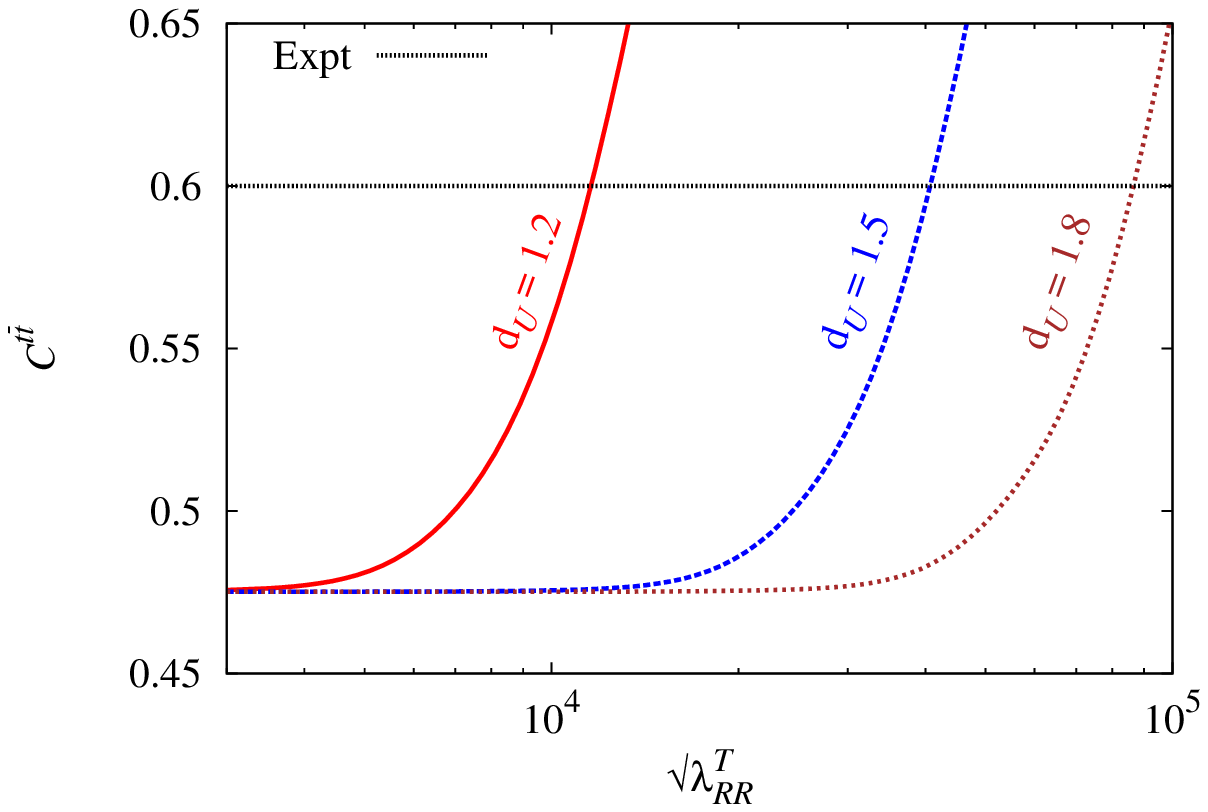}}\\
  \subfloat[$\lambda^T_{LL} = \lambda^T_{RR} = \lambda^T_{RL}  = \lambda^T_{LR} =\lambda^T_{TT}$]{\label{fig:FC_ten_sing_spcr-4-5-c}\includegraphics[width=0.5\textwidth]{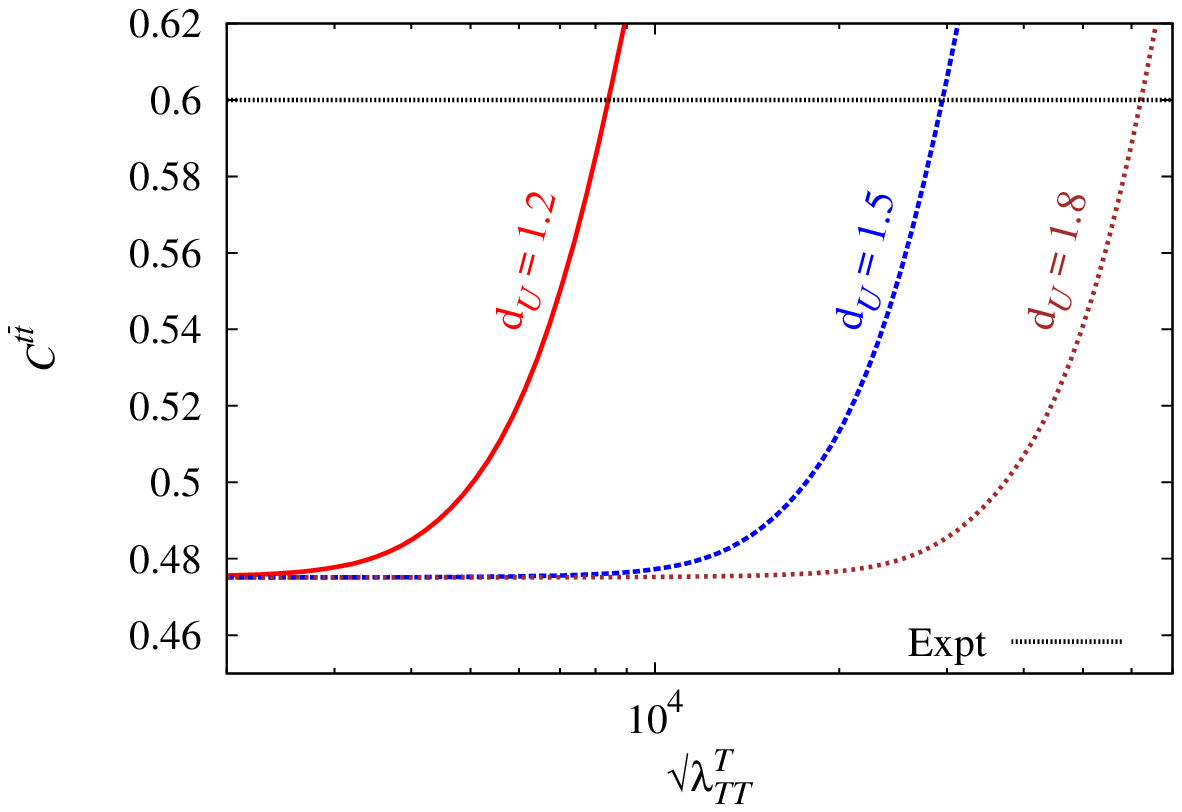}}
  \subfloat[$\lambda^T_{LL} = \lambda^T_{RR} = -\lambda^T_{RL}  = -\lambda^T_{LR} =\lambda_{AT}$]{\label{fig:FC_ten_sing_spcr-4-5-d}\includegraphics[width=0.5\textwidth]{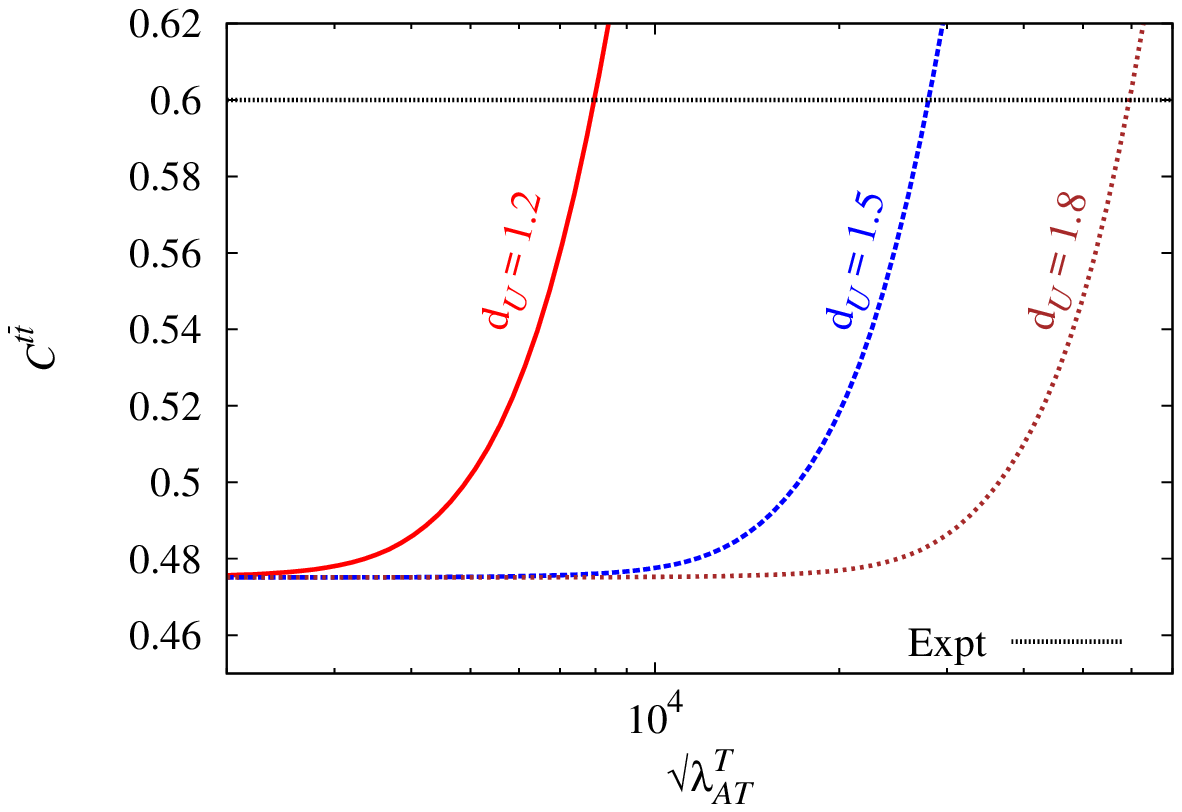}}
  \caption{\small \em{ Variation of the spin-correlation coefficient \spincorr evaluated in helicity basis with  couplings $\surd \lambda^T$ for color singlet flavor conserving tensor unparticles for various value of $\du$ at fixed $\Lambda_{\cal U}=1$ TeV and for different combinations of couplings. The experimental value is depicted with a dot-dashed line at $0.60 \pm 0.50\ (stat) \pm 0.16\ (syst) $ ~\cite{Aaltonen:2010nz}.} }
  \label{fig:FC_ten_sing_spcr}
\end{figure*}
\subsubsection{Color Octet Tensor Unparticle}
\begin{figure*}[!ht]
  \centering
  \subfloat[$\lambda^T_{RR} \ne 0$ and $\lambda^T_{LR}=\lambda^T_{RL}=\lambda^T_{LL} =0 $]{\label{fig:FC_ten_oct_sigma-1-2-a}\includegraphics[width=0.5\textwidth]{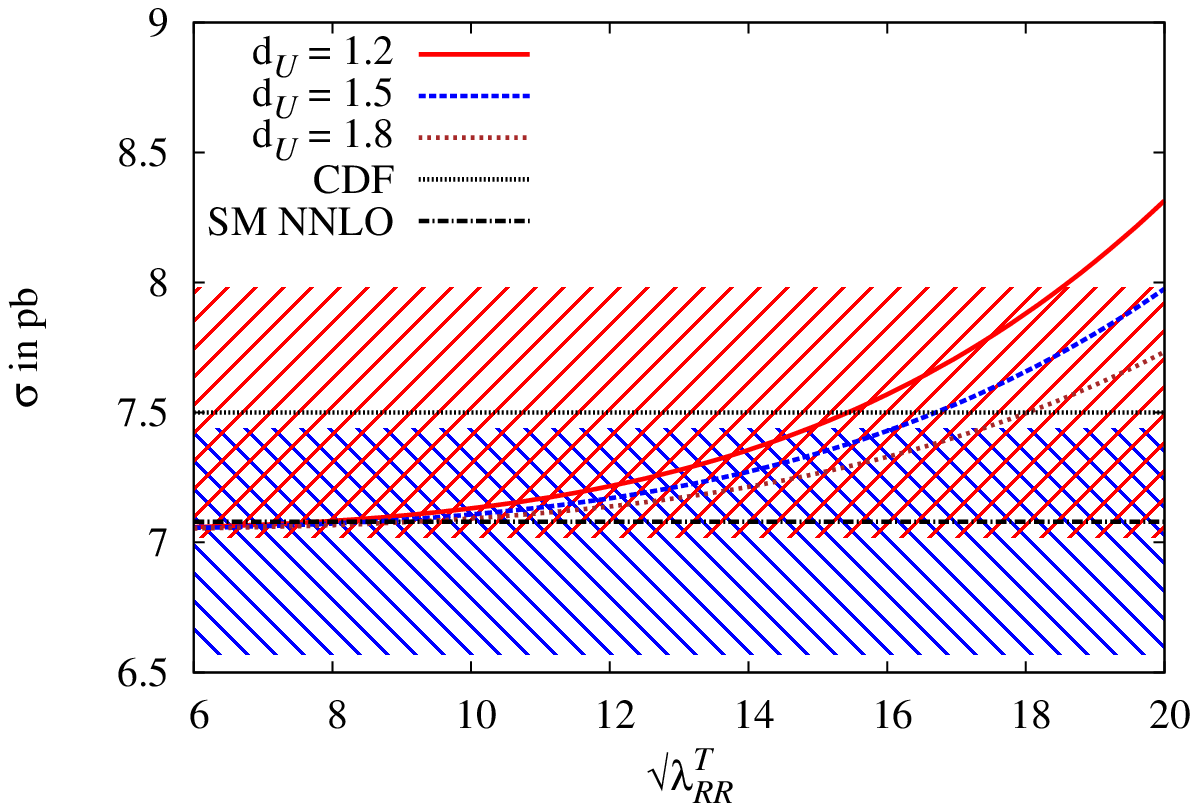}}
  \subfloat[$\lambda^T_{LL} = \lambda^T_{RR} = \lambda^T_{RL}  = \lambda^T_{LR} =\lambda^T_{TT}$]{\label{fig:FC_ten_oct_sigma-1-2-c}\includegraphics[width=0.5\textwidth]{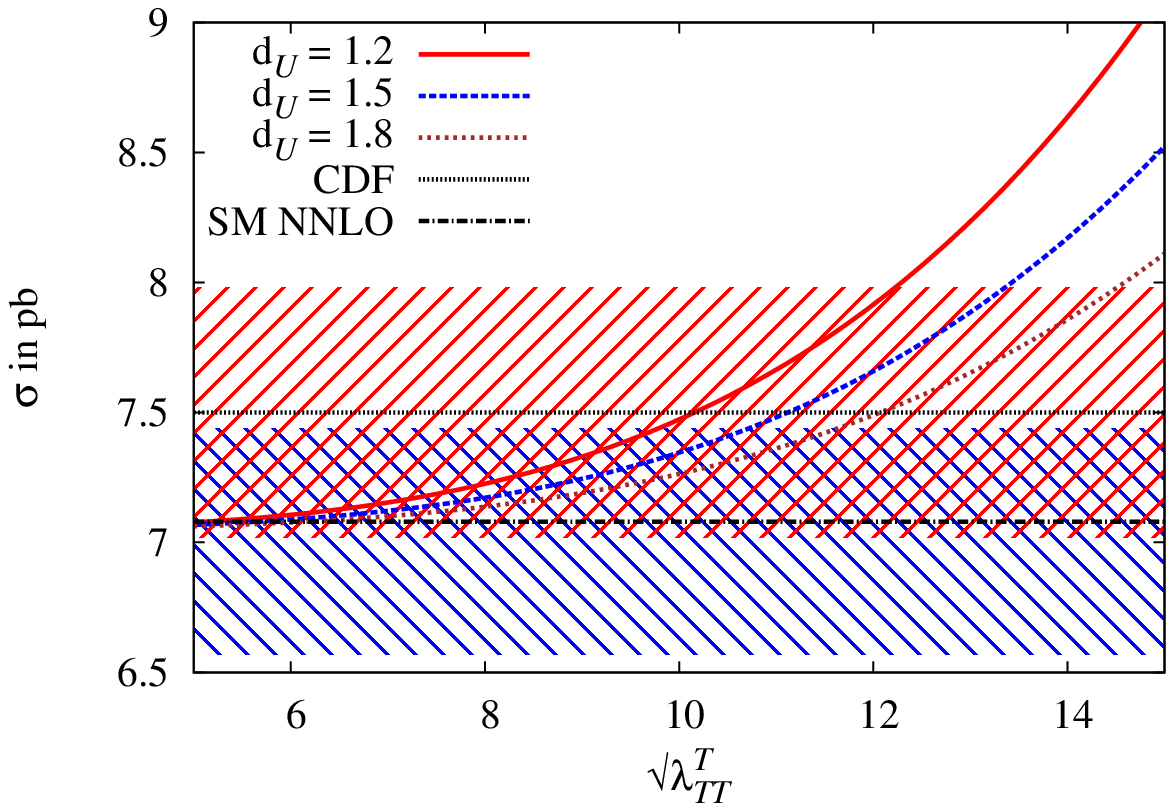}} \\
  \subfloat[$\lambda^T_{LL} = \lambda^T_{RR} = -\lambda^T_{RL}  = -\lambda^T_{LR} =\lambda_{AT}$]{\label{fig:FC_ten_oct_sigma-1-2-d}\includegraphics[width=0.5\textwidth]{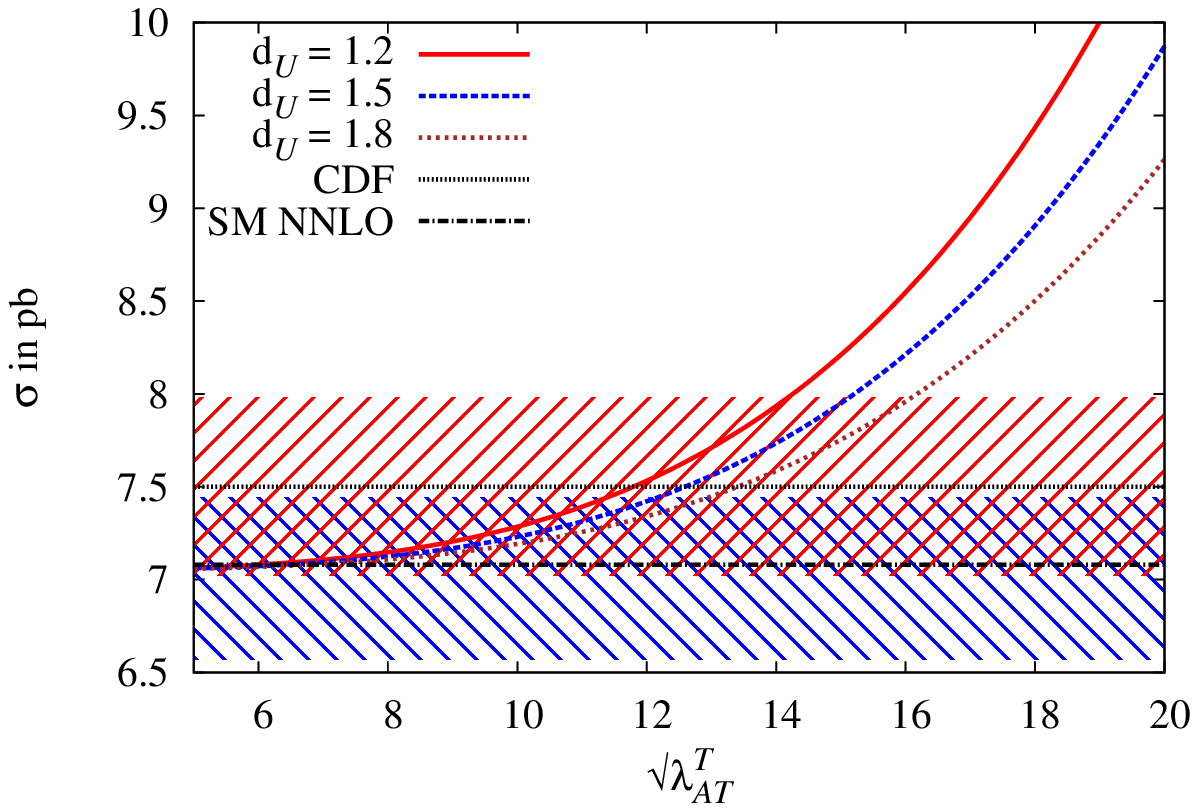}}
  \subfloat[$\lambda^T_{RR} \ne 0$ and $\lambda^T_{LR}=\lambda^T_{RL}=\lambda^T_{LL} =0 $]{\label{fig:FC_ten_oct_sigma-4-5-a}\includegraphics[width=0.5\textwidth]{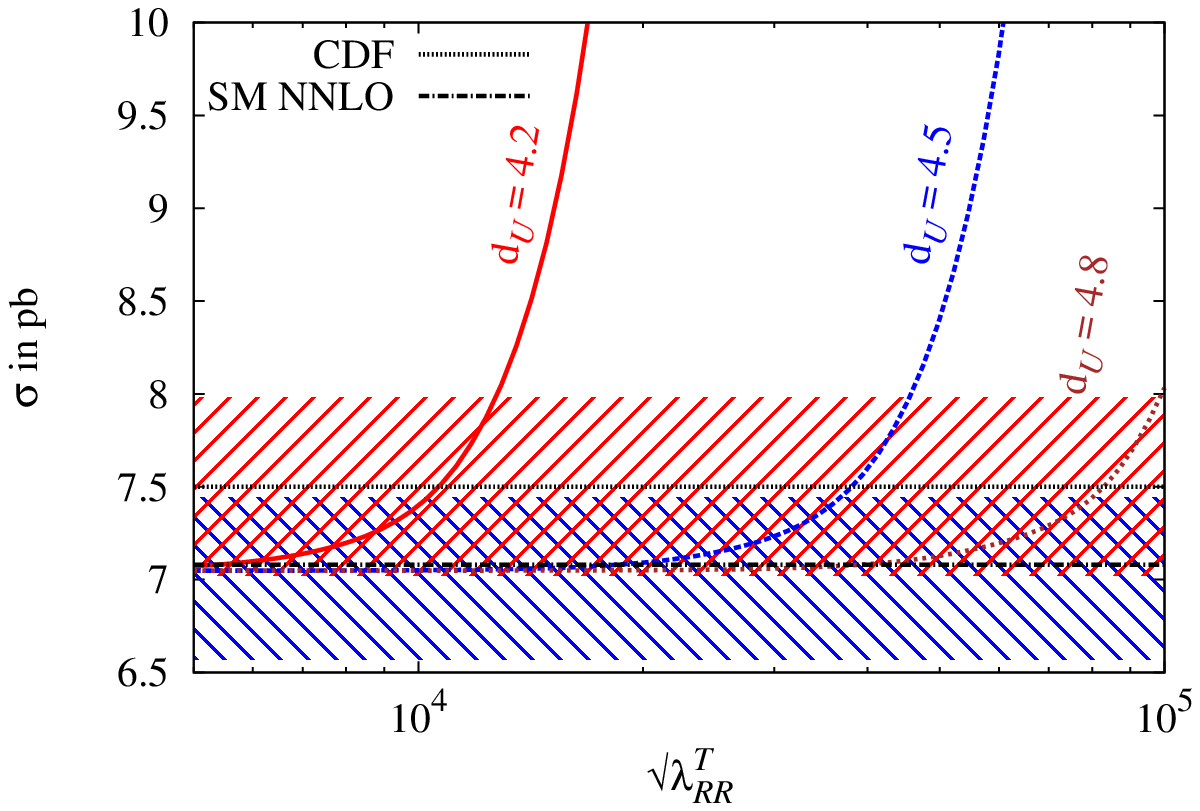}}\\
  \subfloat[$\lambda^T_{LL} = \lambda^T_{RR} = \lambda^T_{RL}  = \lambda^T_{LR} =\lambda^T_{TT}$]{\label{fig:FC_ten_oct_sigma-4-5-c}\includegraphics[width=0.5\textwidth]{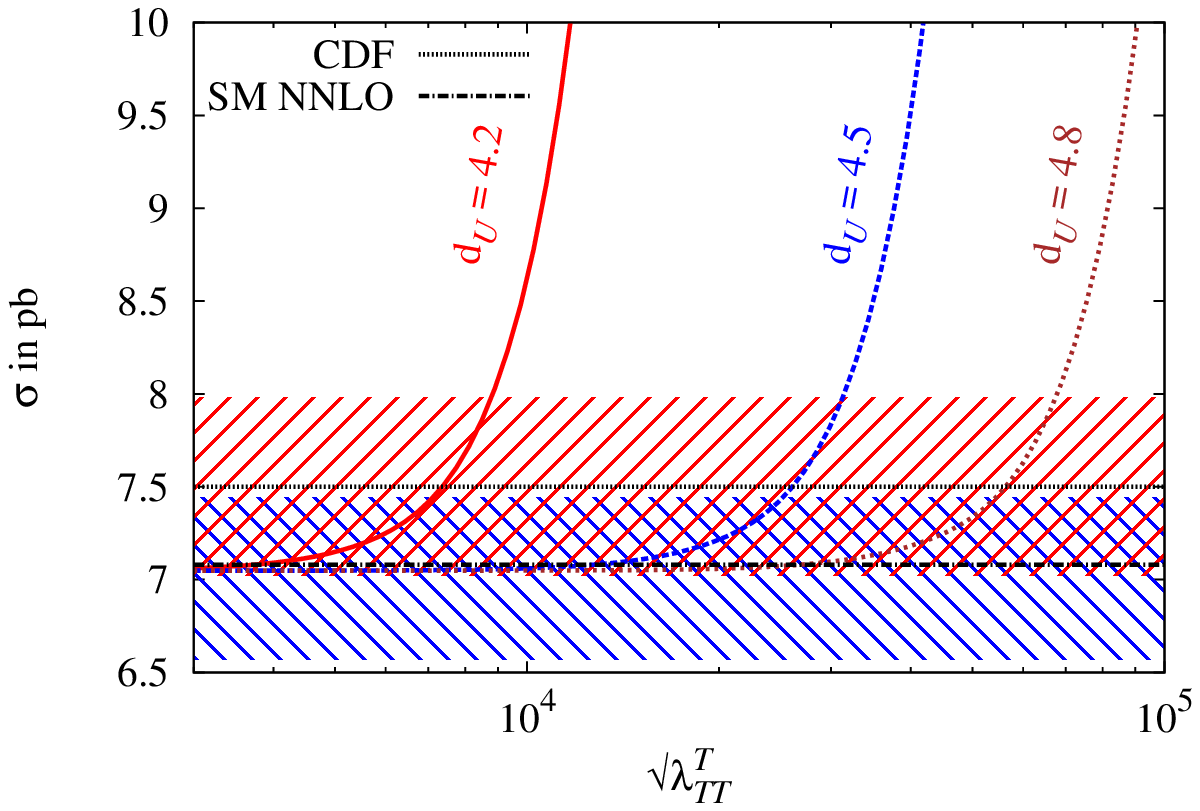}}
  \subfloat[$\lambda^T_{LL} = \lambda^T_{RR} = -\lambda^T_{RL}  = -\lambda^T_{LR} =\lambda_{AT}$]{\label{fig:FC_ten_oct_sigma-4-5-d}\includegraphics[width=0.5\textwidth]{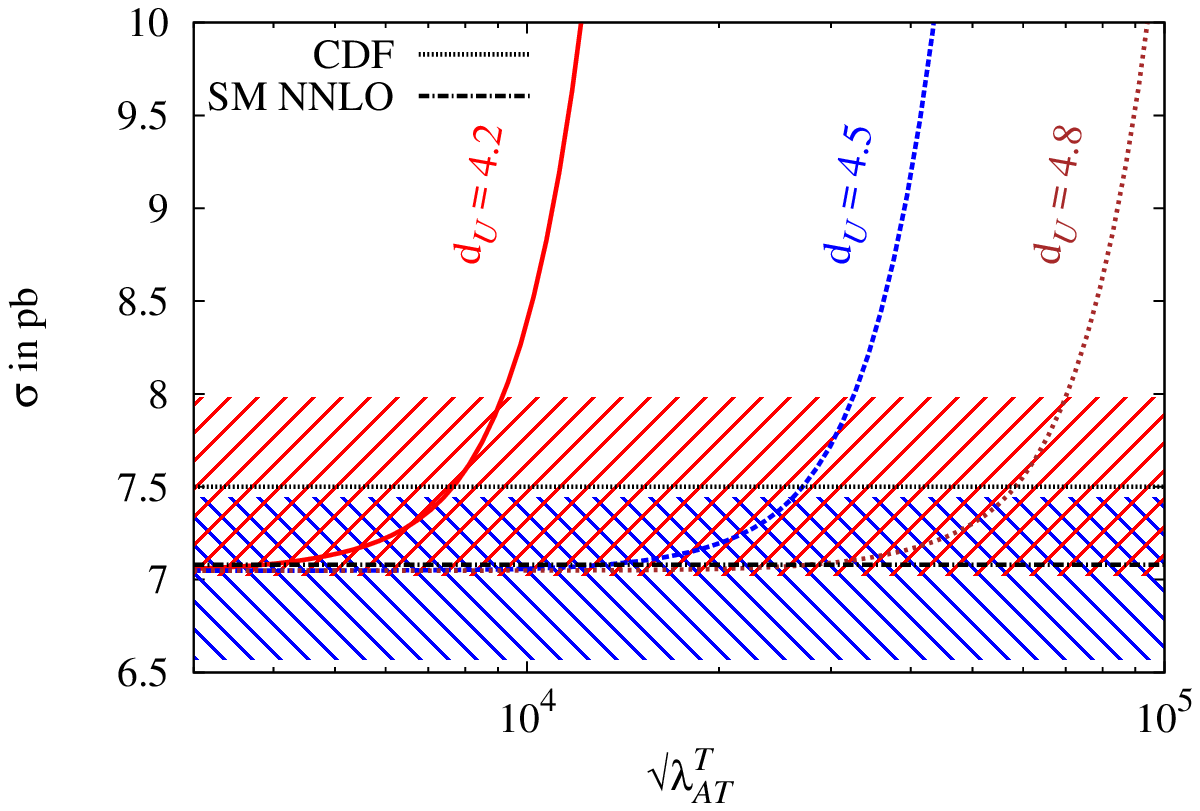}}
\caption{\small \em{Variation of the cross-section $\sigma \left( p\bar p \to t\bar t\right)$ with  couplings
$\surd\lambda^T$ for
color octet flavor conserving tensor unparticles corresponding to  different values of  $d_{\cal U}$ at fixed $\Lambda_{\cal U}=1$ TeV . The  upper dotted line with a red band  depicts the  cross-section  $7.50 \pm 0.48$ pb from CDF (all channels) ~\cite{cdf-top-cross}, while the lower dot-dashed  line with a blue band show theoretical estimate $7.08\pm 0.36$ pb at NNLO  ~\cite{kidonakis-tcross}.}}
  \label{fig:FC_ten_oct_sigma}
\end{figure*}

\begin{figure*}[!ht]
  \centering
  \subfloat[$\lambda^T_{RR} \ne 0$ and $\lambda^T_{LR}=\lambda^T_{RL}=\lambda^T_{LL} =0 $]{\label{fig:FC_ten_oct_afb-1-2-a}\includegraphics[width=0.5\textwidth]{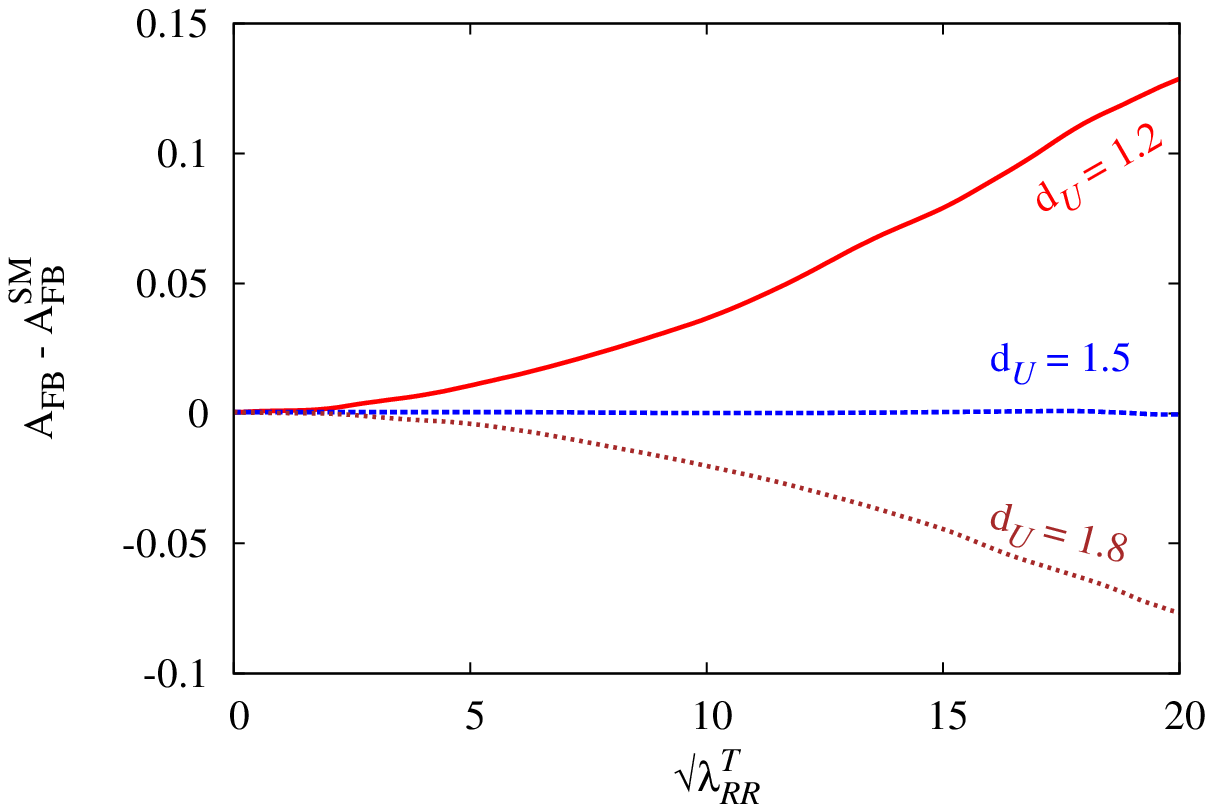}}
  \subfloat[$\lambda^T_{LL} = \lambda^T_{RR} = \lambda^T_{RL}  = \lambda^T_{LR} =\lambda^T_{TT}$]{\label{fig:FC_ten_oct_afb-1-2-c}\includegraphics[width=0.5\textwidth]{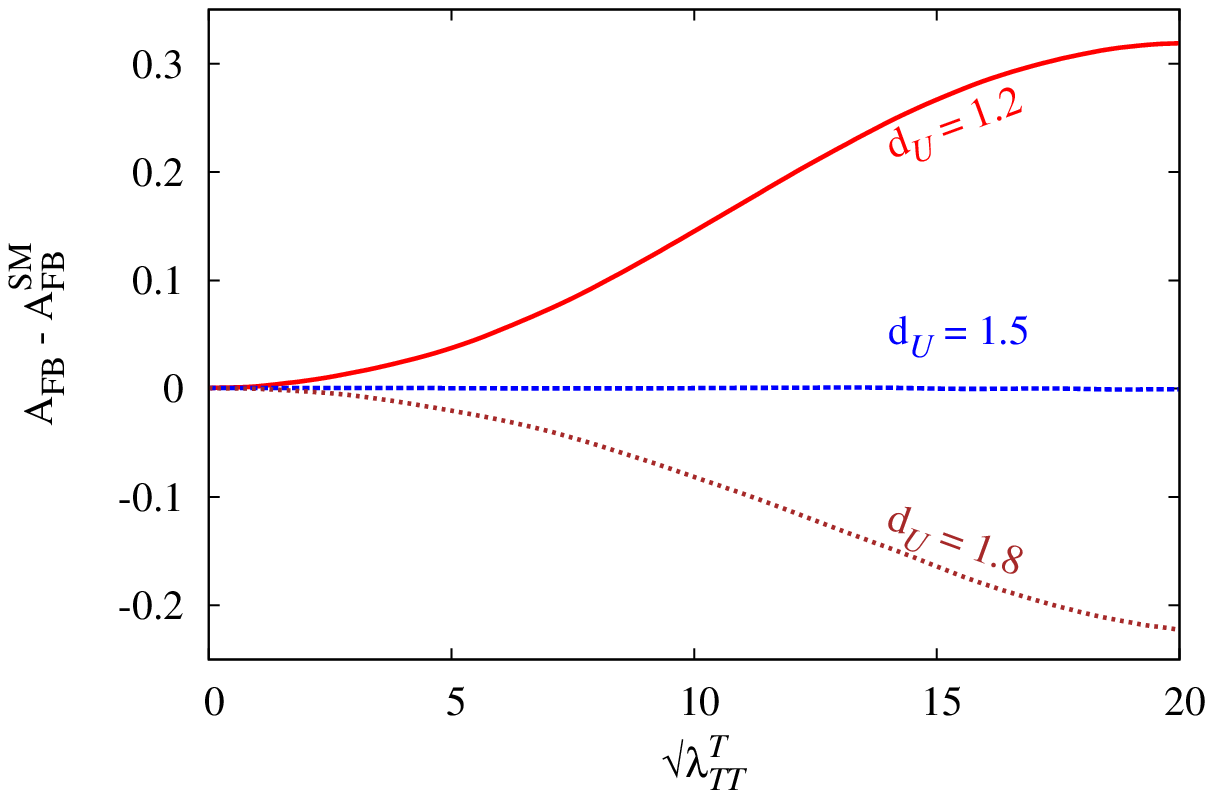}}
  \caption{\small \em{Variation of the $A_{FB} - A_{FB}^{SM}$ (the unparticle contribution to \afbt) with  couplings $\surd \lambda^T_{i j}$  for color octet flavor conserving tensor unparticles for various values of $\du$ for a fixed $\scaleu = 1$TeV. This is evaluated using the 1d differential distribution of rapidity in $t\bar t$ rest frame. The panels 'a' and 'b' correspond the cases (a) and (c) of the text.}}
  \label{fig:FC_ten_oct_afb}
\end{figure*}
The FC tensor octet unparticles do not interfere with electroweak neutral sector like  the vector unparticles and the behavior of
the plots (shown in Figs.~\ref{fig:FC_ten_oct_sigma},\ref{fig:FC_ten_oct_afb}
and \ref{fig:FC_ten_oct_spcr}) is almost
entirely determined by its interference with QCD. The same helicity and opposite  helicity contributions to interference with QCD are given by
\begin{figure*}[!ht]
  \centering
  \subfloat[$\lambda^T_{RR} \ne 0$ and $\lambda^T_{LR}=\lambda^T_{RL}=\lambda^T_{LL} =0 $]{\label{fig:FC_ten_oct_spcr-1-2-a}\includegraphics[width=0.5\textwidth]{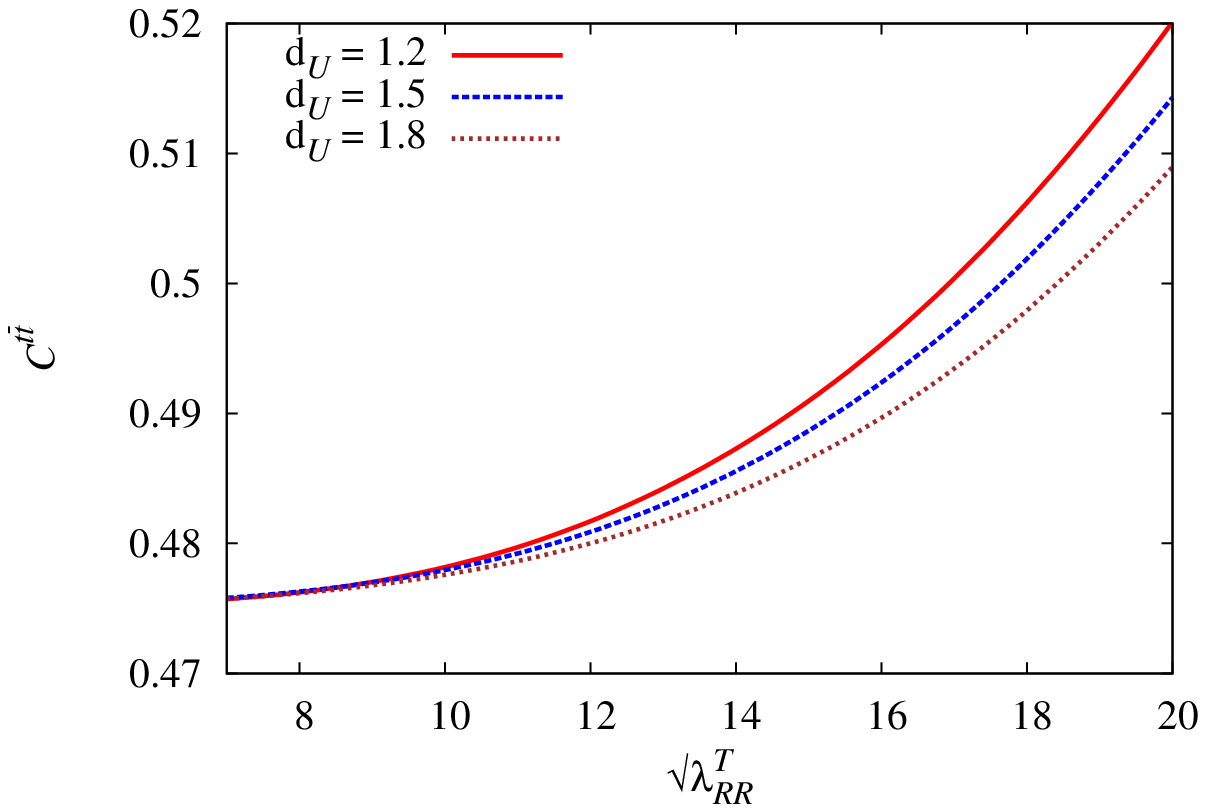}}
  \subfloat[$\lambda^T_{LL} = \lambda^T_{RR} = \lambda^T_{RL}  = \lambda^T_{LR} =\lambda^T_{TT}$]{\label{fig:FC_ten_oct_spcr-1-2-c}\includegraphics[width=0.5\textwidth]{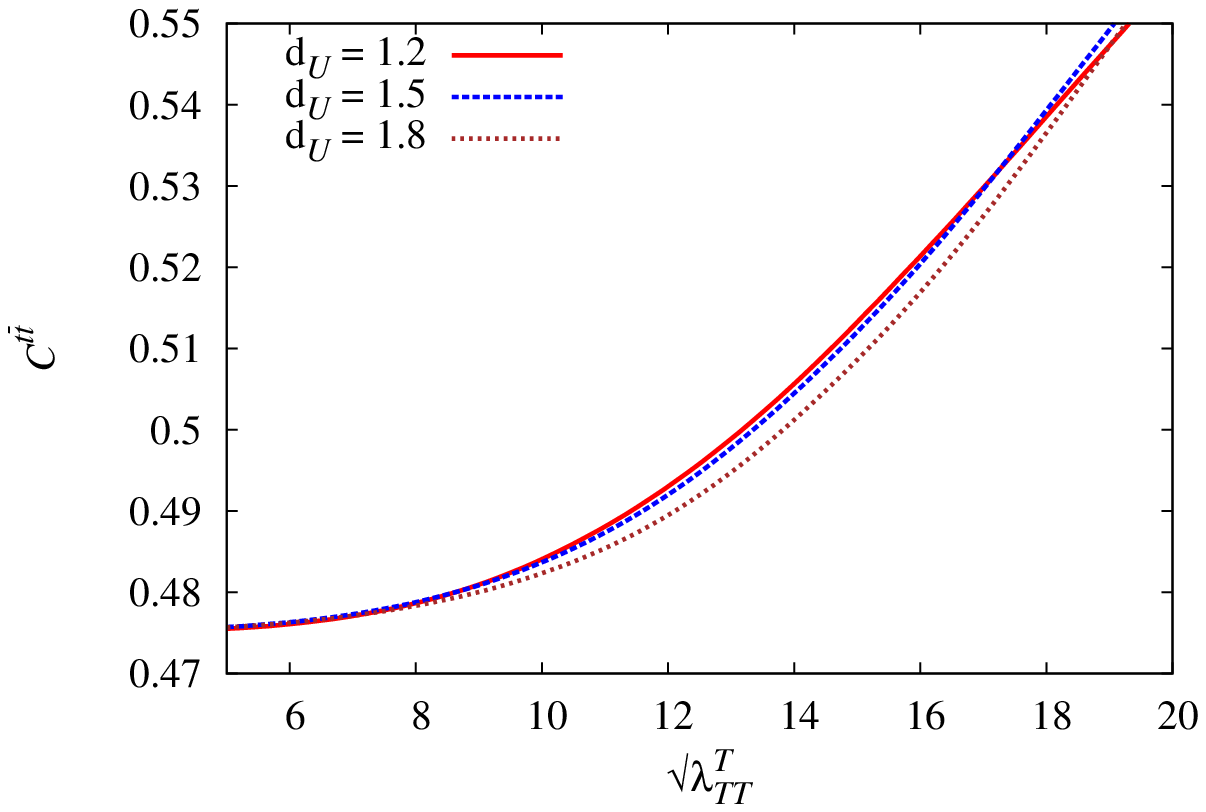}} \\
  \subfloat[$\lambda^T_{LL} = \lambda^T_{RR} = -\lambda^T_{RL}  = -\lambda^T_{LR} =\lambda_{AT}$]{\label{fig:FC_ten_oct_spcr-1-2-d}\includegraphics[width=0.5\textwidth]{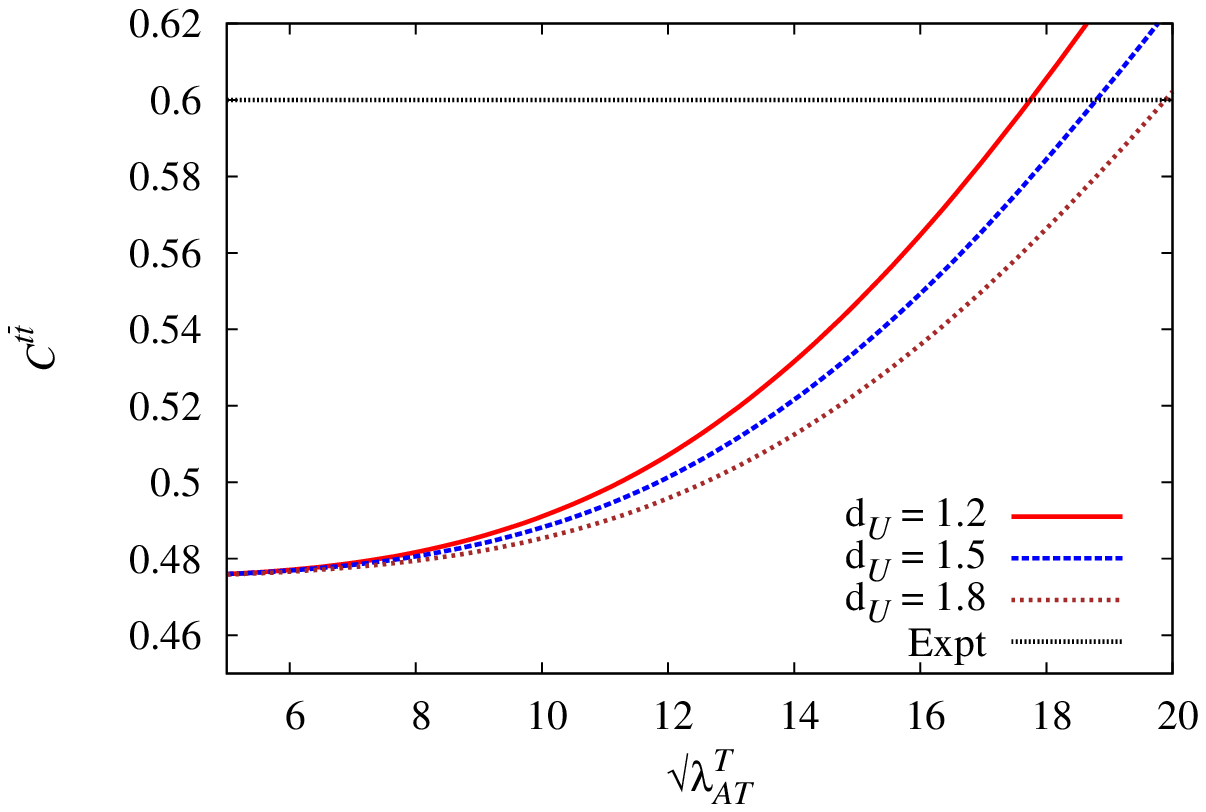}}
  \subfloat[$\lambda^T_{RR} \ne 0$ and $\lambda^T_{LR}=\lambda^T_{RL}=\lambda^T_{LL} =0 $]{\label{fig:FC_ten_oct_spcr-4-5-a}\includegraphics[width=0.5\textwidth]{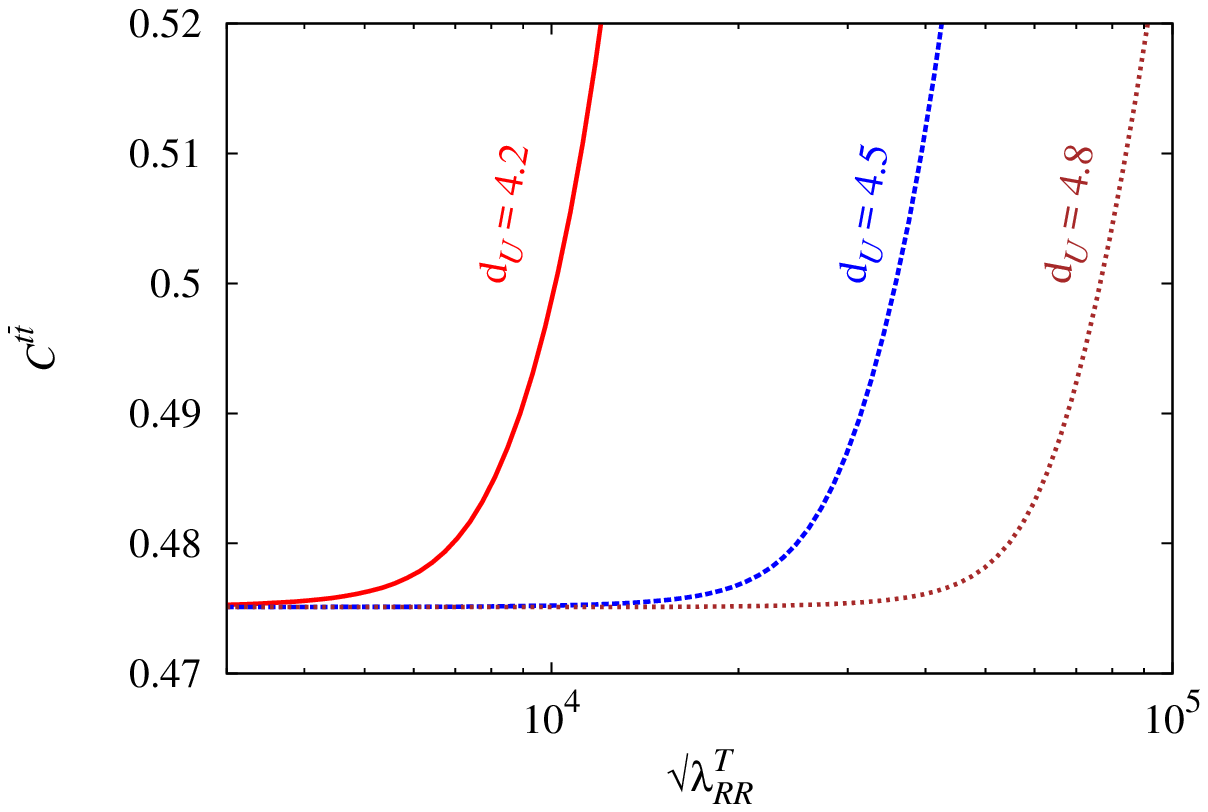}}\\
  \subfloat[$\lambda^T_{LL} = \lambda^T_{RR} = \lambda^T_{RL}  = \lambda^T_{LR} =\lambda^T_{TT}$]{\label{fig:FC_ten_oct_spcr-4-5-c}\includegraphics[width=0.5\textwidth]{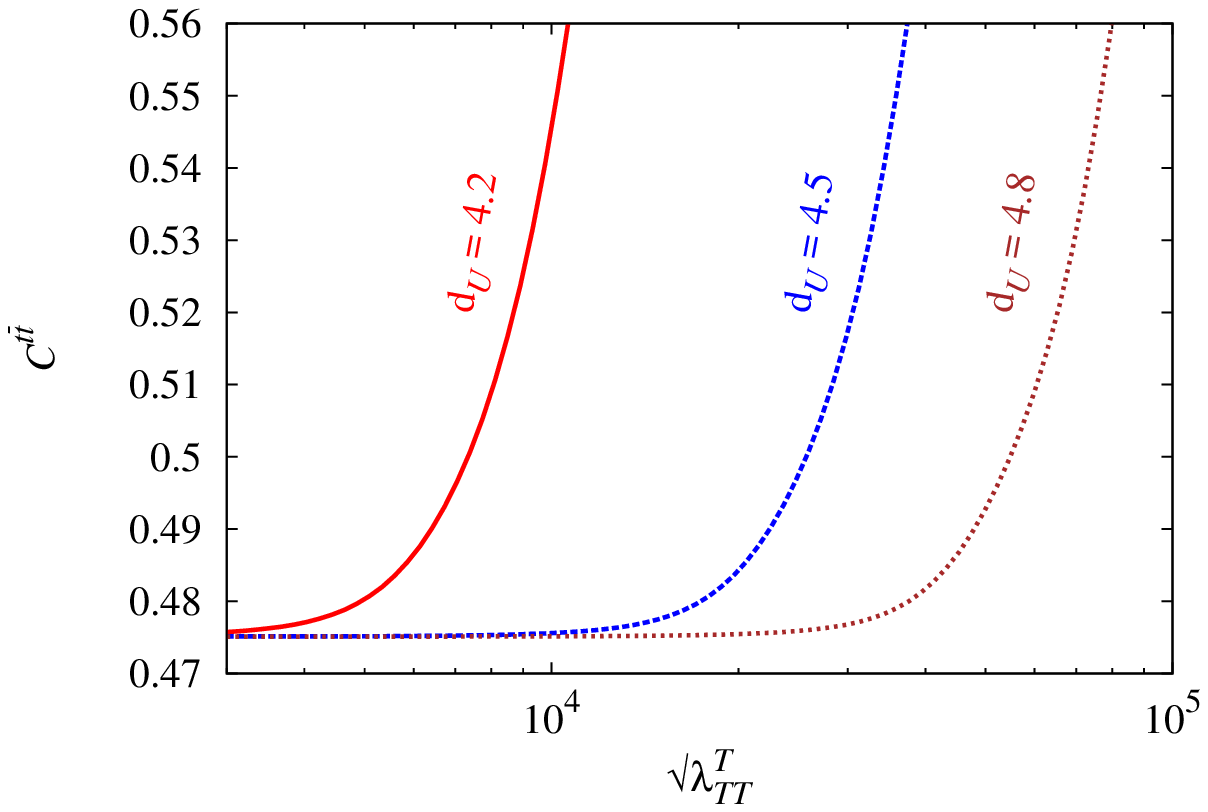}}
  \subfloat[$\lambda^T_{LL} = \lambda^T_{RR} = -\lambda^T_{RL}  = -\lambda^T_{LR} =\lambda_{AT}$]{\label{fig:FC_ten_oct_spcr-4-5-d}\includegraphics[width=0.5\textwidth]{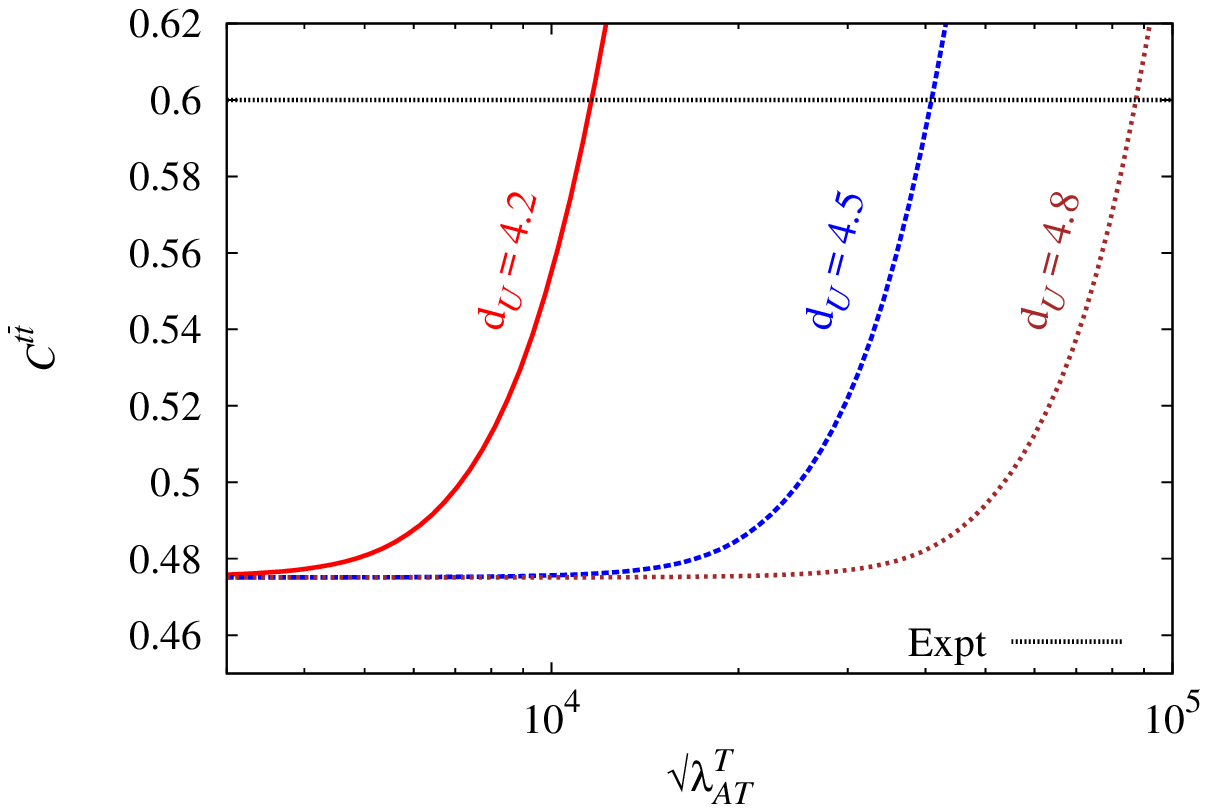}}
  \caption{\small \em{Variation of the spin-correlation coefficient \spincorr evaluated in helicity basis with  couplings $\surd \lambda^T_{i j}$  for color octet flavor conserving tensor unparticles for various values of $\du$ at fixed $\Lambda_{\cal U}=1$ TeV and for different coupling combinations as mentioned in the text. The experimental value is depicted with a dot-dashed line  at $0.60 \pm 0.50\ (stat) \pm 0.16\ (syst) $ ~\cite{Aaltonen:2010nz}.} }
  \label{fig:FC_ten_oct_spcr}
\end{figure*}
\begin{widetext} 
\bea 
 2 {\cal M}^{\rm QCD} {\cal R}({\cal M}^{\rm
  unp})\bigl\vert_{\rm same\, hel} &=& - 4 {\cal B}_T^{\rm int}
(\lallt +\llrt +\lrlt+\lrrt) \bt(1-\bt^2) \st^2 \ct
\label{tensor_unp_int_samehel} \\ 
2 {\cal M}^{\rm
  QCD} {\cal R}({\cal M}^{\rm unp})\bigl\vert_{\rm opp\, hel} &=&
- 2 {\cal B}_T^{\rm int} \bt \biggl[  \beta_t (\lallt +\lrrt-\llrt -\lrlt) (3\ct^2-1)\nn\\ 
&&  + 2 \left(\lallt +\lrrt+\lrlt+\llrt \right)\ct^3 \biggr] 
\label{tensor_unp_int_opphel}
\\
{\rm where} \qquad
{\cal B}_T^{\rm int} &=& {\cal C}_f^{{\rm int}^{\rm FC}_T} \Bigl( \frac{\du(\du -1)}{4} \Bigr)
\biggl[g_s^2\bigl(
\frac{\hat s}{\Lambda_{\cal U}^2}\bigr)^{\du} A_{\du} [\cot{(\du\pi)}] 
  \biggr] 
\label{tensor_unp_int_const}
\eea
\end{widetext}
Here ${\cal C}_f^{{\rm int}^{\rm FC}_T}= 2$.
Apart from an extra $\beta_t  $ suppression  in comparison to the vector unparticles, as  seen from equation \eqref{tensor_unp_int_samehel}, the
 other salient differences between FC octet tensor and vector are:
\ben
\item Since  the phase factor appearing in the tensor propagator is different to that of the vectors,  the interference term contains  an extra negative sign  leading to a positive \afbt for $n <\du<(n+1/2)$, zero for $\du = (n+1/2)$  and negative for $ (n+1/2)<\du<(n+1)$  interference with SM. Thus for FC tensor octet unparticle, \afbt is
 positive for $1<\du<1.5$ and negative for $1.5<\du<2$ which is just
 opposite to that of FC vector unparticle as depicted in
 figure \ref{fig:FC_ten_oct_afb}.
\item Unlike FC vector the same helicity amplitudes also contribute to \afbt for cases (a), (b) and (c) leading to $\cfbt \ne \afbt$. The contribution to \afbt from $\left\vert {\cal M}_{unp}\right\vert^2$ is zero for cases (c) and (d). The same helicity and the opposite helicity contribution to \afbt vanish for the interference term in case (d). 

The opposite helicity contribution vanishes for case (d) (axial
 tensor) while it is non-zero for case (c). Thus while appreciable \afbt
 contribution was coming in FC vector octet for axial vector couplings,
 the same is not true for FC tensor.
\een

\subsection{Flavor Violating Vector Unparticle}
\begin{figure*}[!ht]
  \centering
  \subfloat[$g^{u t}_L =0 \neq  g^{u t}_R $]{\label{fig:FV_vec_sing_sigma-1-3-a}\includegraphics[width=0.5\textwidth]{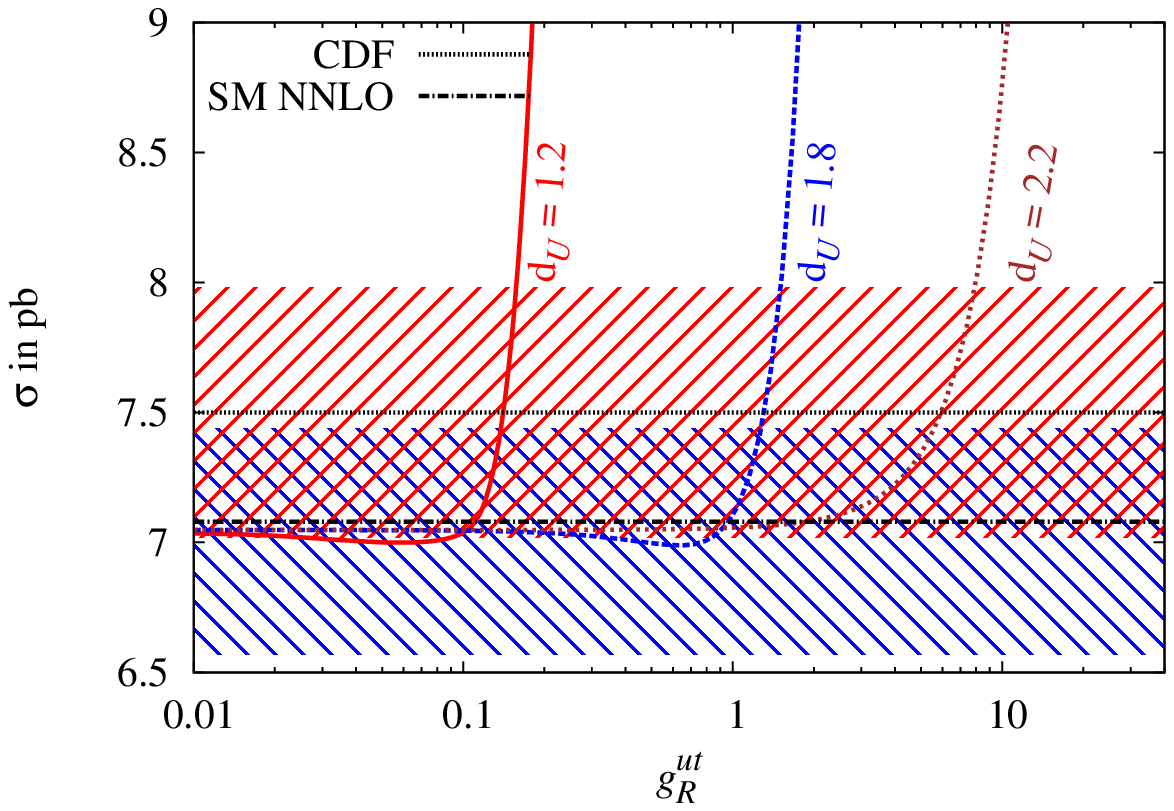}}
  \subfloat[$g^{u t}_R= -g^{u t}_L = g^{u t}_A$]{\label{fig:FV_vec_sing_sigma-1-3-d}\includegraphics[width=0.5\textwidth]{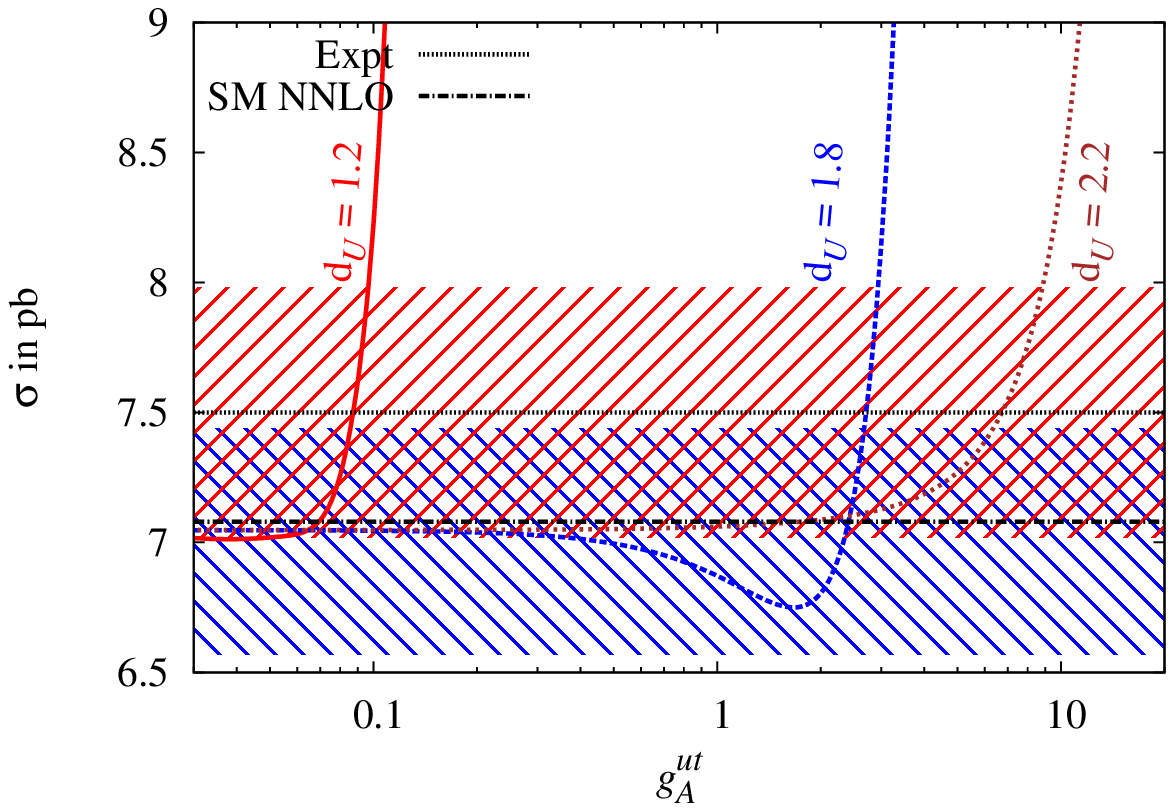}} 
\caption{\small \em{Variation of the cross-section $\sigma \left( p\bar p \to t\bar t\right)$ with  couplings
$g^{ut}_i$ for
color singlet flavor violating vector unparticles corresponding to  different values of  $d_{\cal U}$  in the range $1<\du<3$ at fixed $\Lambda_{\cal U}=1$ TeV . The  upper dotted line with a red band  depicts the  cross-section  $7.50 \pm 0.48$ pb from CDF (all channels) ~\cite{cdf-top-cross}, while the lower dot-dashed  line with a blue band show theoretical estimate $7.08\pm 0.36$ pb at NNLO  ~\cite{kidonakis-tcross}
.}}
 \label{fig:FV_vec_sing_sigma-1-3}
\end{figure*}
\begin{figure*}[!ht]
  \centering
  \subfloat[$g^{u t}_L =0 \neq  g^{u t}_R$]{\label{fig:FV_vec_sing_afb-1-3-a}\includegraphics[width=0.5\textwidth]{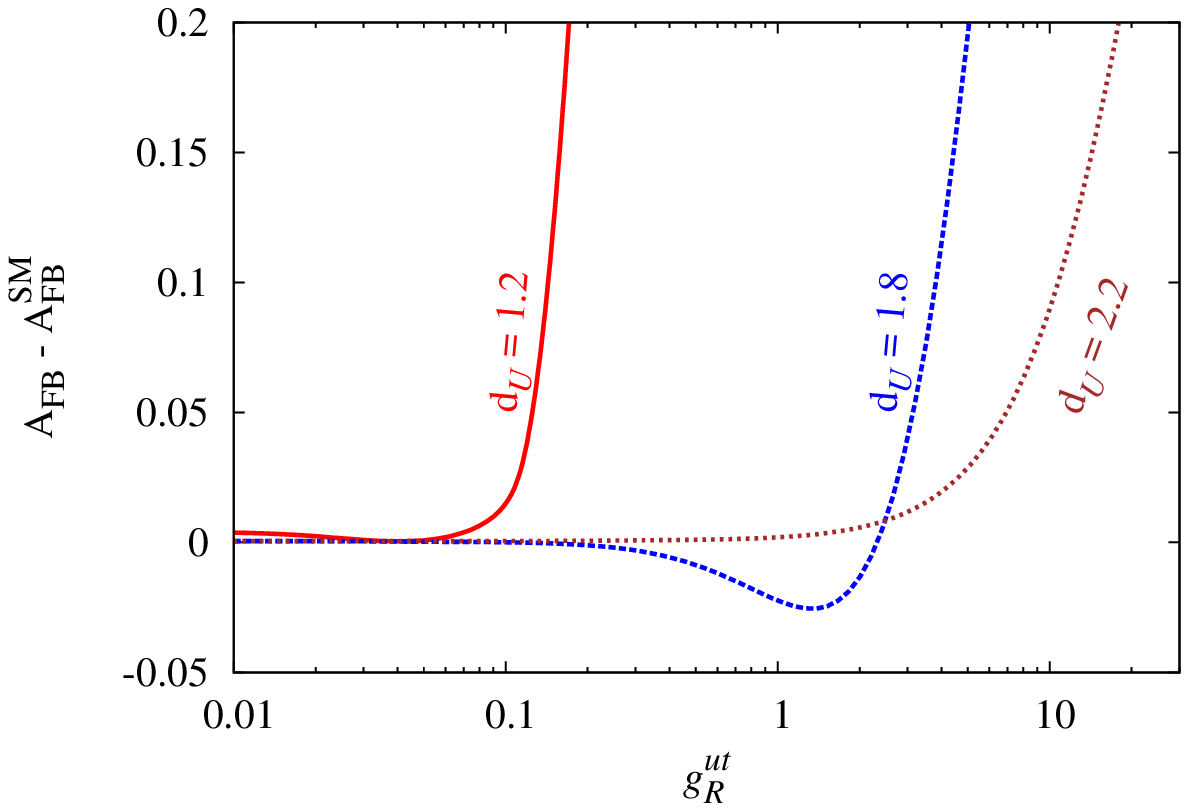}}
  \subfloat[$g^{u t}_R= -g^{u t}_L = g^{u t}_A$]{\label{fig:FV_vec_sing_afb-1-3-d}\includegraphics[width=0.5\textwidth]{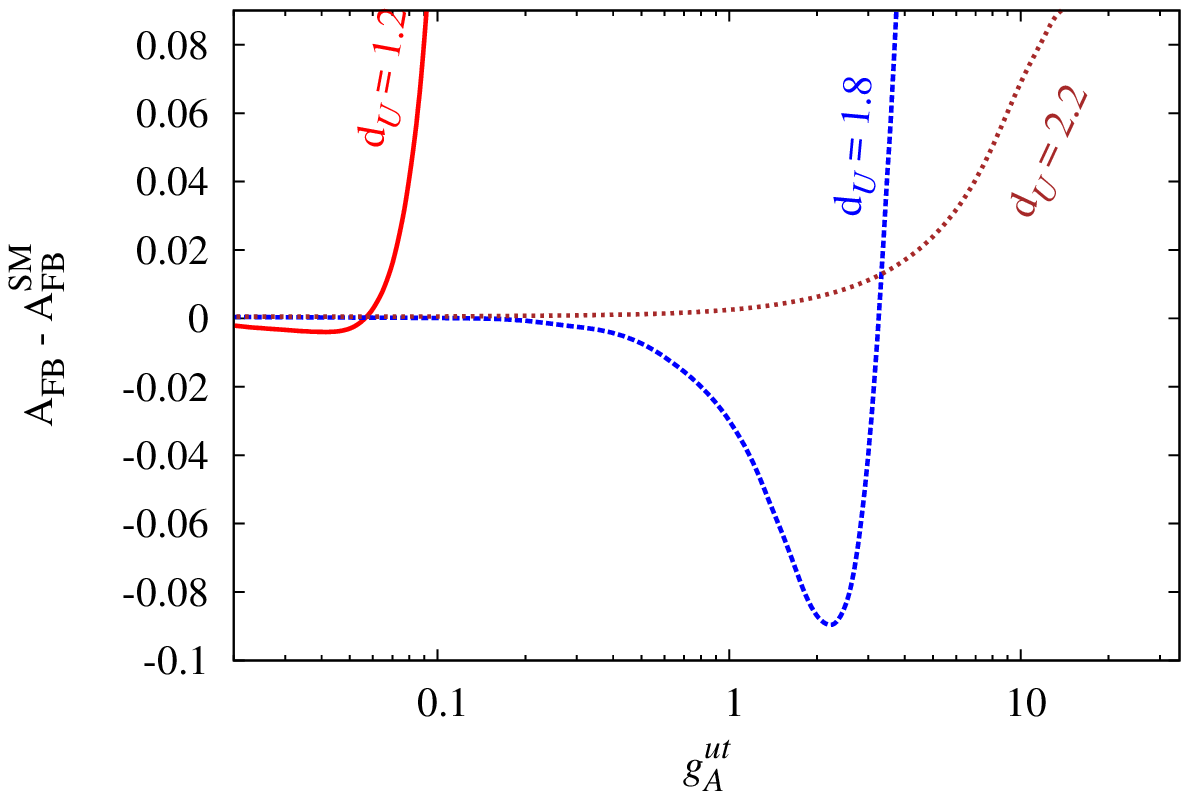}} 
  \caption{\small\em{ Variation of the unparticle contribution to charge asymmetry $A_{FB} -A_{FB}^{SM}$ with couplings $g^{ut}_i$  in the presence of flavor violating color singlet vector unparticles for various values of $\du$ in the range $1<\du<3$ at fixed $\Lambda_{\cal U}=1$ TeV. This is evaluated using the 1d differential distribution of rapidity in $t\bar t$ rest frame.}}
  \label{fig:FV_vec_sing_afb-1-3}
\end{figure*}
\begin{figure*}[!ht]
  \centering
  \subfloat[$g^{u t}_L =0 \neq  g^{u t}_R$]{\label{fig:FV_vec_sing_spcr-1-3-a}\includegraphics[width=0.5\textwidth]{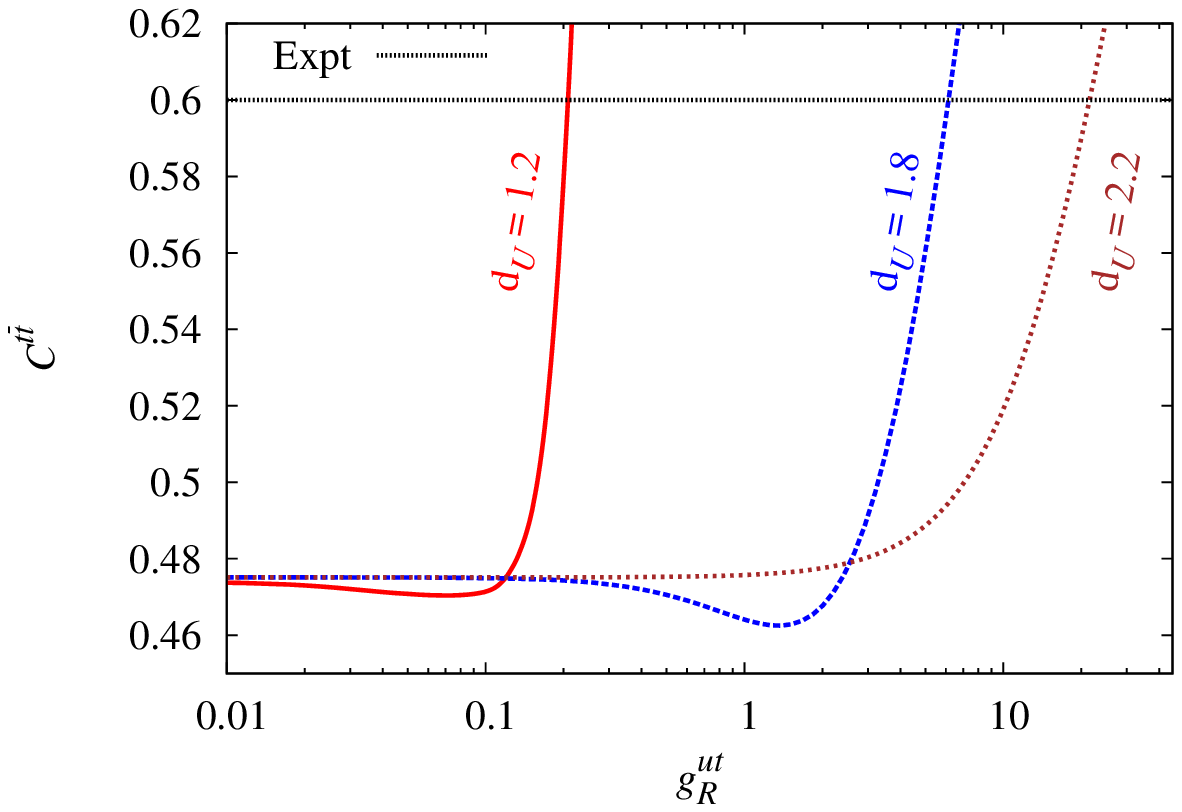}}
  \subfloat[$g^{u t}_R= -g^{u t}_L=g^{u t}_A$]{\label{fig:FV_vec_sing_spcr-1-3-d}\includegraphics[width=0.5\textwidth]{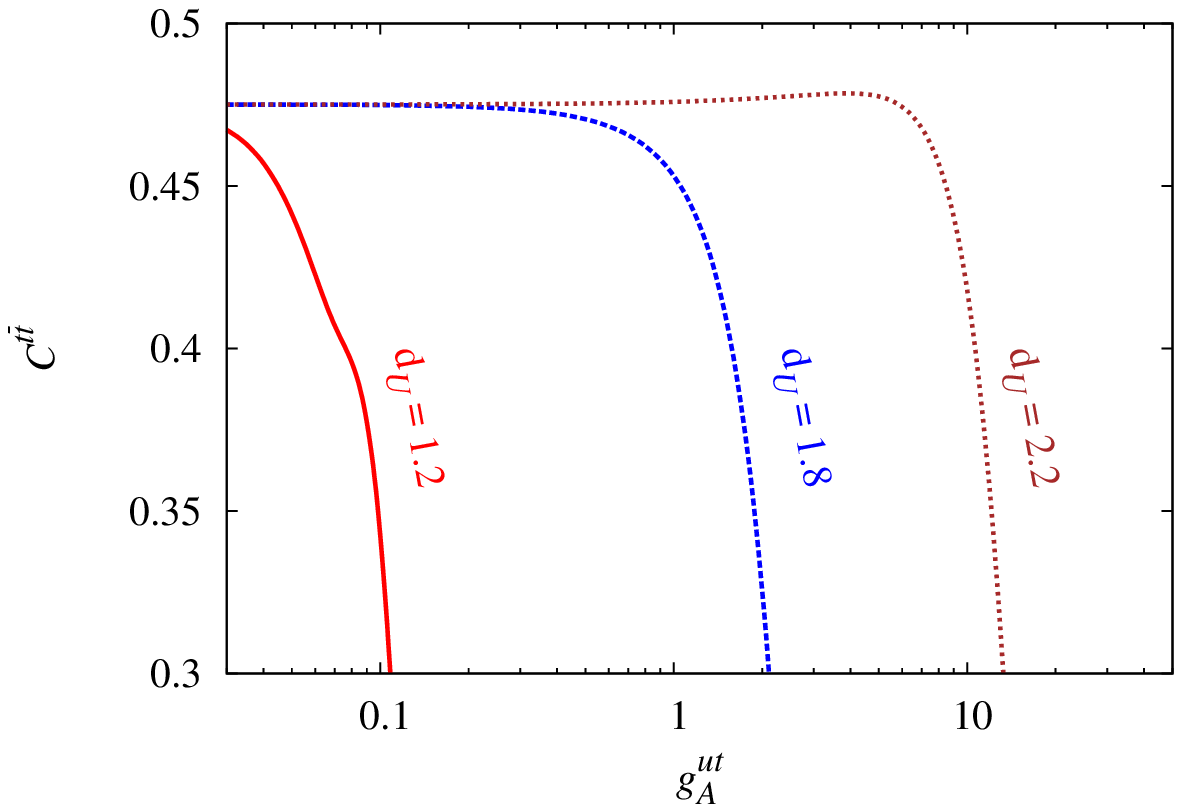}} 
  \caption{\small\em{ Variation of the spin correlation coefficient \spincorr with  couplings $g^{ut}_i$ in the presence of flavor violating color singlet vector unparticles for various values of $\du$ in the range $1<\du<3$ at fixed $\Lambda_{\cal U}=1$ TeV. The experimental value is depicted with a dot-dashed line at $0.60 \pm 0.50\ (stat) \pm 0.16\ (syst) $ ~\cite{Aaltonen:2010nz}. }}
  \label{fig:FV_vec_sing_spcr-1-3}
\end{figure*}
\begin{figure*}[!ht]
  \centering
  \subfloat[$g^{ut}_{i} $ variation of \crtt for cases (a) $g^{u t}_L =0 \neq  g^{u t}_R$ and (d) $g^{u t}_R=-g^{u t}_L=g^{u t}_A$ with $\du = 3.2 $]{\label{fig:FV_vec_sing_sigma-3.2}\includegraphics[width=0.5\textwidth]{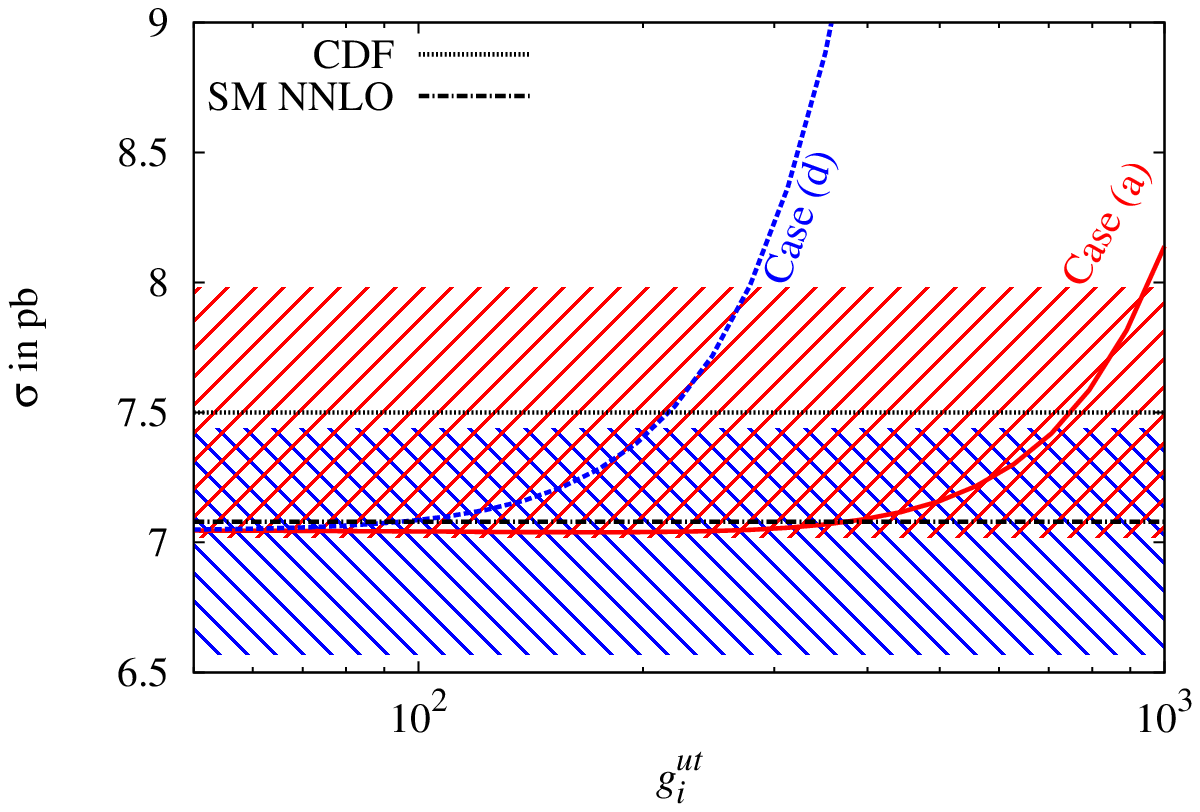}}
  \subfloat[$g^{u t}_i $ variation of $A_{FB}-A_{FB}^{SM}$ for cases (a) $g^{u t}_L = 0 \neq  g^{u t}_R$ and (d) $g^{u t}_R=-g^{u t}_L=g^{u t}_A$ with $\du = 3.2 $]{\label{fig:FV_vec_sing_afb-3.2}\includegraphics[width=0.5\textwidth]{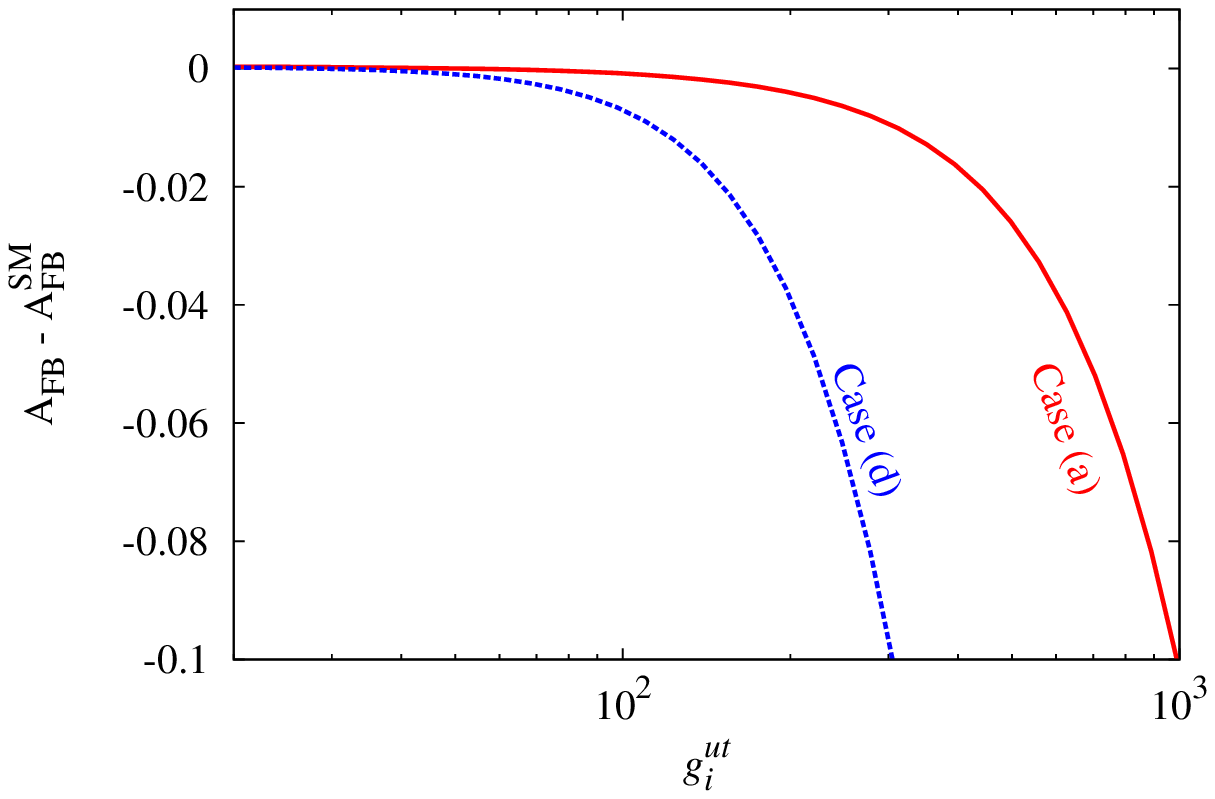}} \\
  \subfloat[$g^{ut} $ variation of \spincorr for cases (a) and (d) with $\du = 3.2 $]{\label{fig:FV_vec_sing_spcr-3.2}\includegraphics[width=0.5\textwidth]{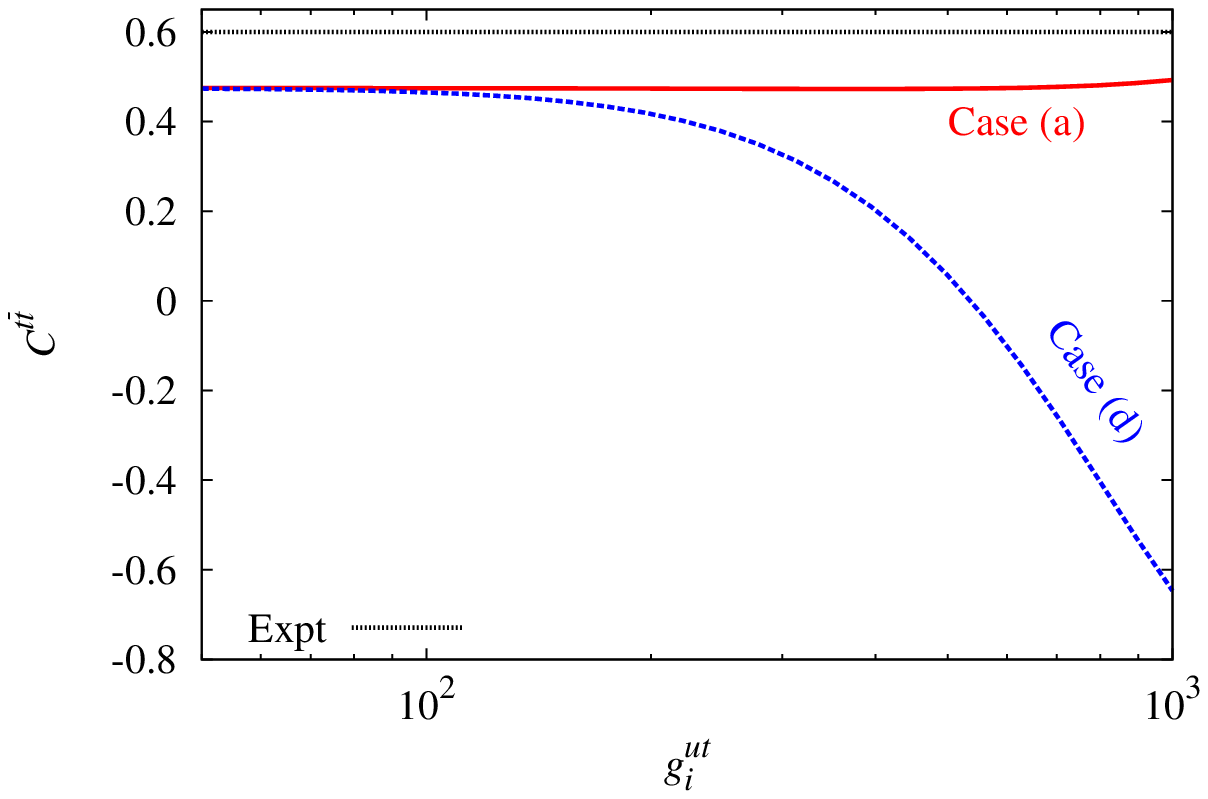}}
\caption{\small \em{Variation of the cross-section $\sigma \left( p\bar p \to t\bar t\right)$, the unparticle contribution to charge asymmetry  $A_{FB}-A_{FB}^{SM}$ and spin correlation coefficient \spincorr  with  couplings
$g^{ut}_i$ for
color singlet flavor violating vector unparticles  at fixed  $d_{\cal U}=3.2$  and $\Lambda_{\cal U}=1$ TeV corresponding
 to cases (a) and (d) mentioned in the text. In  plot (a) the  upper dotted line with a red band  depicts the  cross-section  $7.50 \pm 0.48$ pb from CDF (all channels) ~\cite{cdf-top-cross}, while the lower dot-dashed  line with a blue band show theoretical estimate $7.08\pm 0.36$ pb at NNLO  ~\cite{kidonakis-tcross}. Plot (b) is evaluated using the 1d differential distribution of rapidity in $t\bar t$ rest frame.
In plot (c) the expertimental value is depicted with a dot-dashed line at $0.60 \pm 0.50\ (stat) \pm 0.16\ (syst) $ ~\cite{Aaltonen:2010nz}.}}
  \label{fig:FV_vec_sing-3.2}
\end{figure*}
\begin{figure*}[!ht]
  \centering
  \subfloat[$g^{u t}_L =0 \neq  g^{u t}_R$]{\label{fig:FV_vec_oct_sigma-1-3-a}\includegraphics[width=0.5\textwidth]{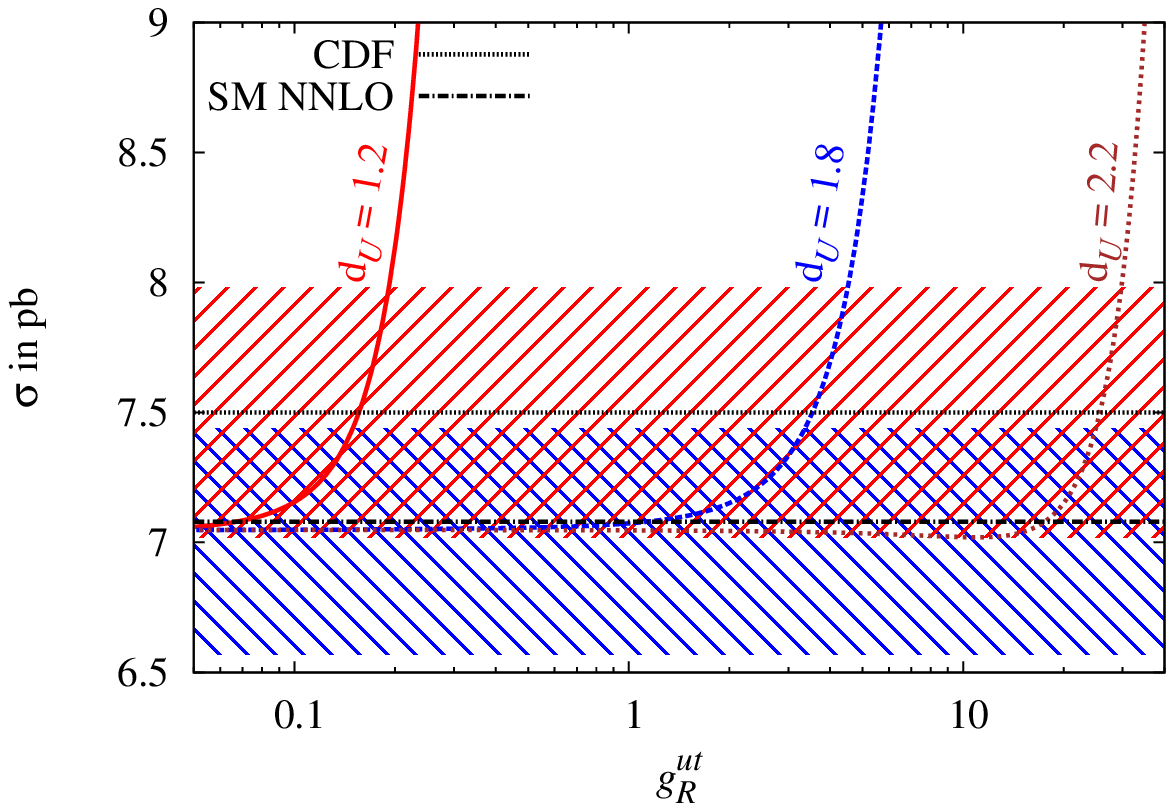}}
  \subfloat[$g^{ut}_R=-g^{ut}_L=g^{ut}_A$]{\label{fig:FV_vec_oct_sigma-1-3-d}\includegraphics[width=0.5\textwidth]{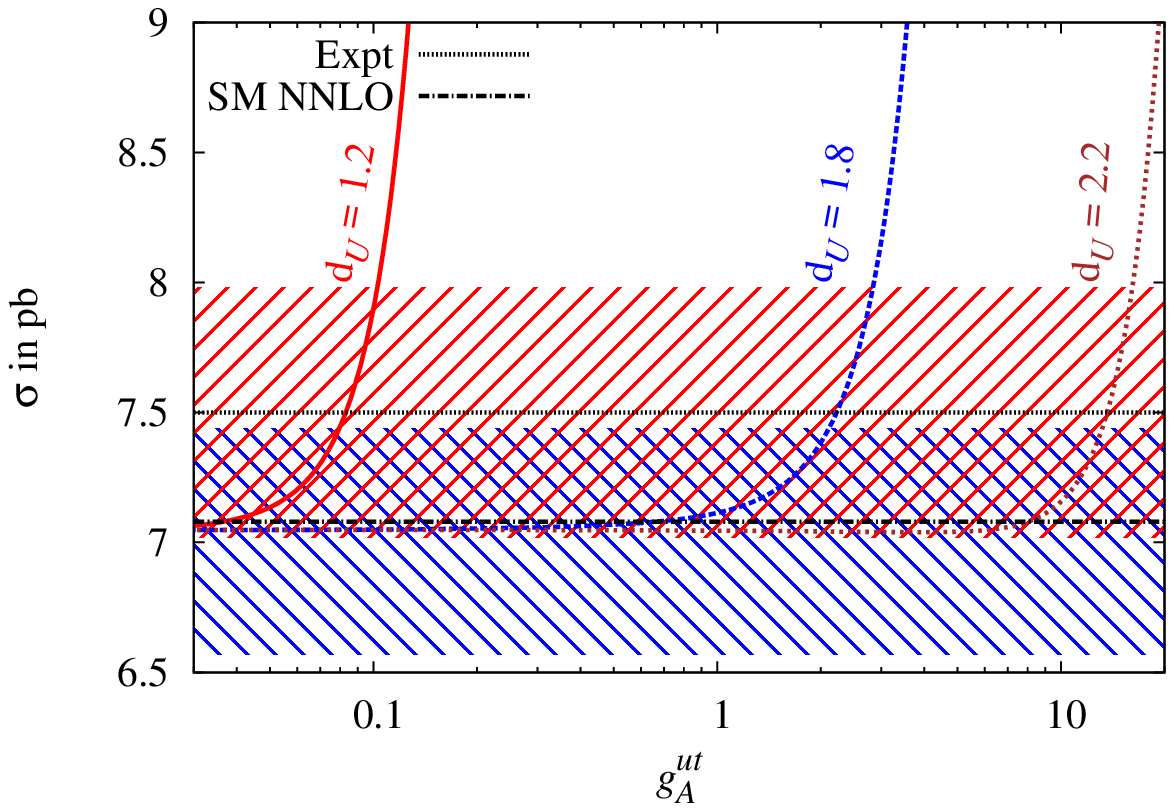}} 
\caption{\small \em{Variation of the cross-section $\sigma \left( p\bar p \to t\bar t\right)$ with  couplings
$g^{ut}_i$ for
color octet flavor violating vector unparticles for various values of \du in the range  $1 <d_{\cal U}<3$ at fixed $\Lambda_{\cal U}=1$ TeV and for cases (a) (or (b)) and (d) (or (c)) mentioned in the text. The  upper dotted line with a red band  depicts the  cross-section  $7.50 \pm 0.48$ pb from CDF (all channels) ~\cite{cdf-top-cross}, while the lower dot-dashed  line with a blue band show theoretical estimate $7.08\pm 0.36$ pb at NNLO  ~\cite{kidonakis-tcross}.}} 
  \label{fig:FV_vec_oct_sigma-1-3}
\end{figure*}
\begin{figure*}[!ht]
  \centering
  \subfloat[$g^{u t}_L =0 \neq  g^{u t}_R$]{\label{fig:FV_vec_oct_afb-1-3-a}\includegraphics[width=0.5\textwidth]{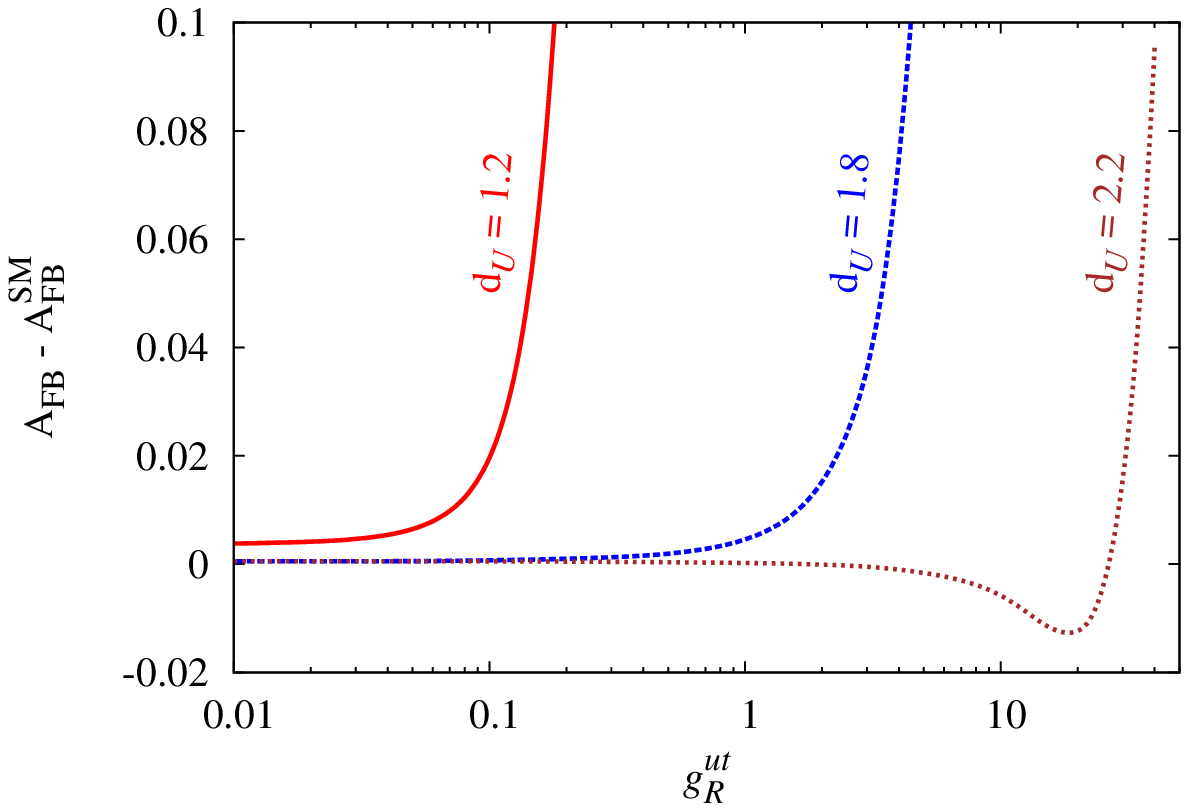}}
  \subfloat[$g^{ut}_R=-g^{ut}_L=g^{ut}_A$]{\label{fig:FV_vec_oct_afb-1-3-d}\includegraphics[width=0.5\textwidth]{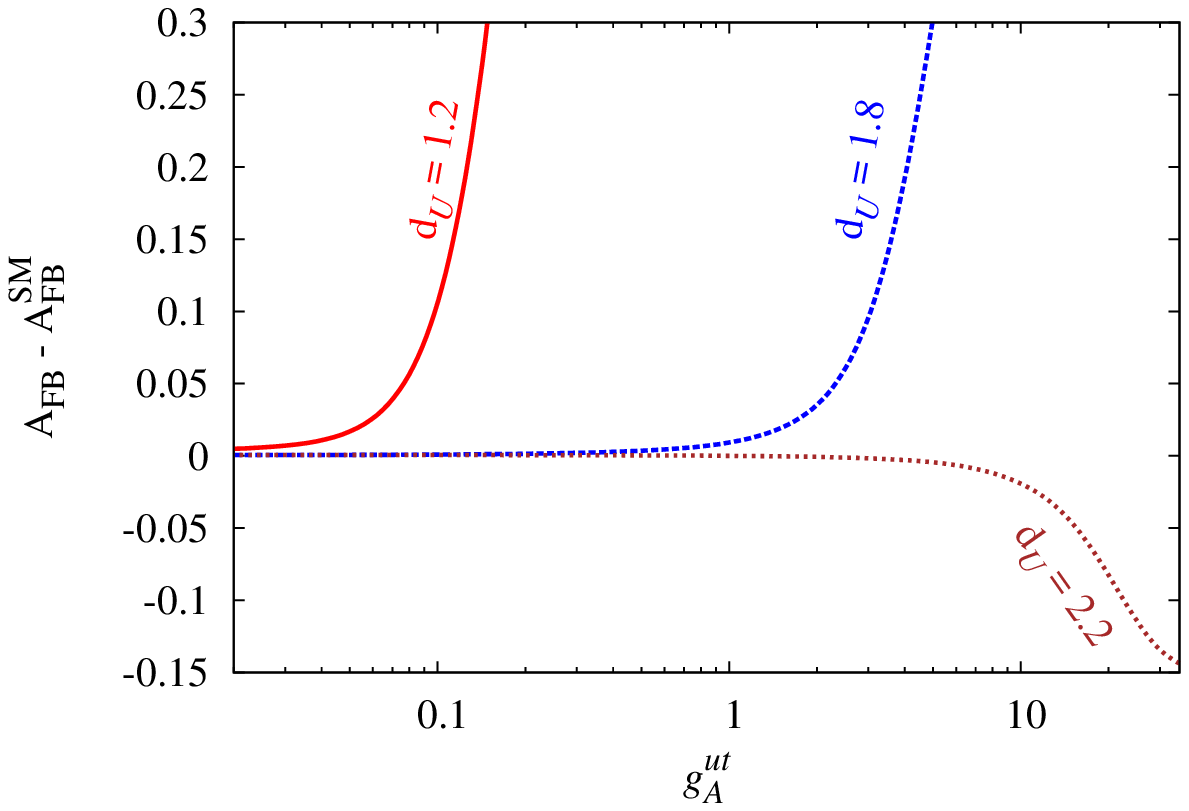}} 
  \caption{\small\em{ Variation of the unparticle contribution to charge asymmetry $A_{FB}^{SM}$ with couplings $g^{ut}_i$  in the presence of flavor violating color octet vector unparticles for various values of $\du$ in the range $1<\du<3$ at fixed $\Lambda_{\cal U}=1$ TeV and for the cases (a)(or (b)) and (d) (or ((C)). This is evaluated using the 1d differential distribution of rapidity in $t\bar t$ rest frame.}}
  \label{fig:FV_vec_oct_afb-1-3}
\end{figure*}
\begin{figure*}[!ht]
  \centering
  \subfloat[$g^{u t}_L =0 \neq  g^{u t}_R$]{\label{fig:FV_vec_oct_spcr-1-3-a}\includegraphics[width=0.5\textwidth]{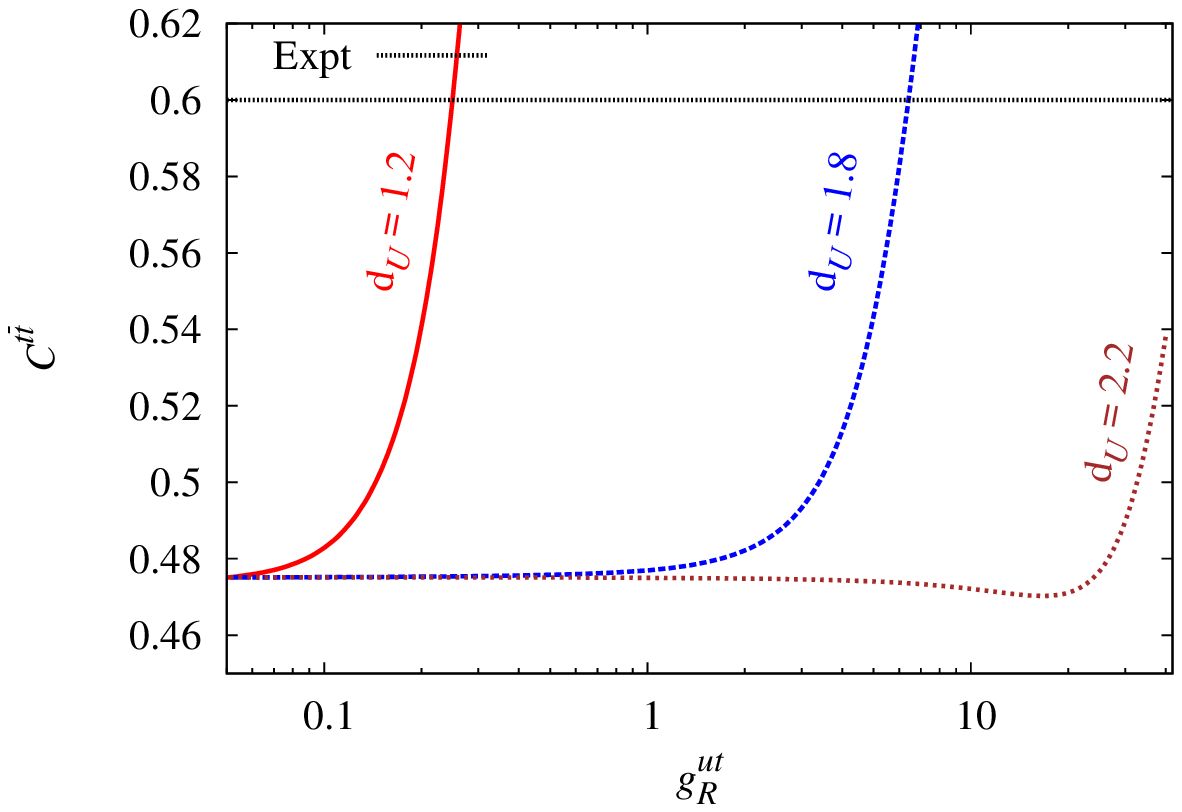}}
  \subfloat[$g^{ut}_R=-g^{ut}_L =g^{ut}_A$]{\label{fig:FV_vec_oct_spcr-1-3-d}\includegraphics[width=0.5\textwidth]{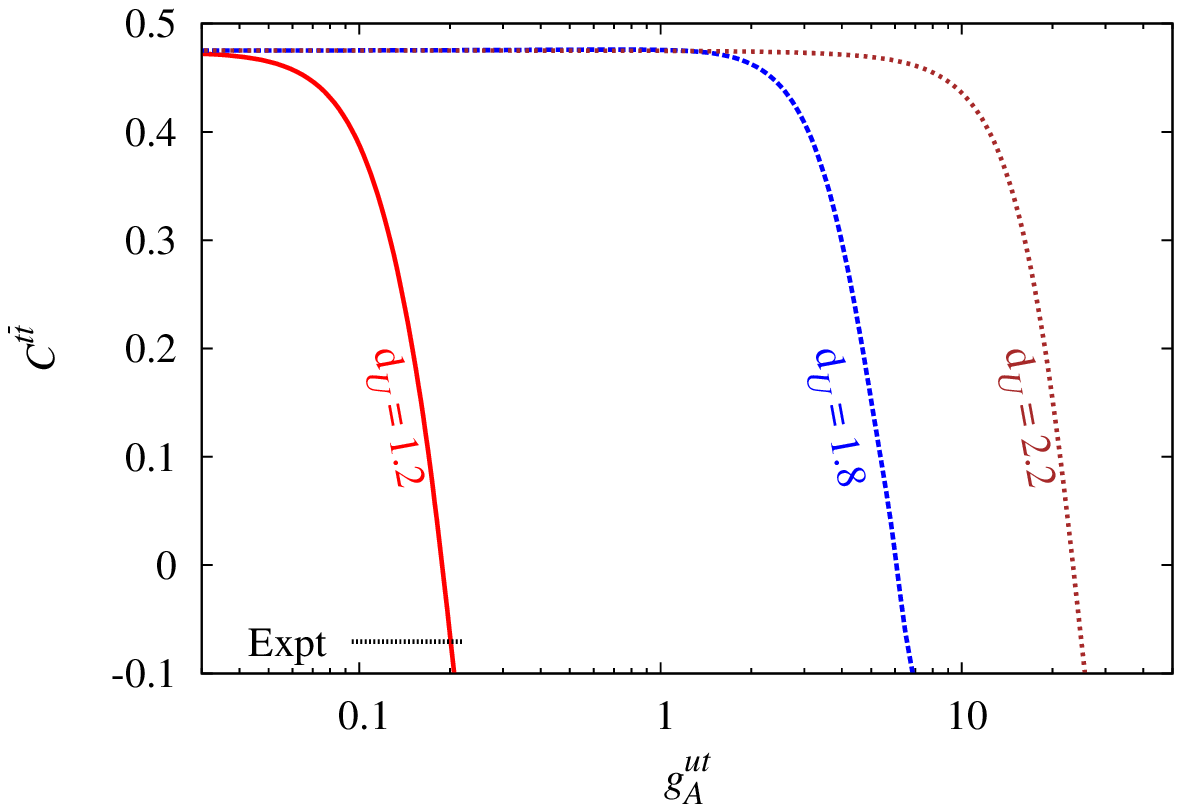}} 
  \caption{\small\em{ Variation of the spin correlation coefficient \spincorr with  couplings $g^{ut}_i$  in the presence of flavor violating color octet vector unparticles for various values of $\du$ in the range $1<\du<3$ at fixed $\Lambda_{\cal U}=1$ TeV and for cases (a)(or(b)) and (d)(or (c)) mentioned in the text. The experimental value is depicted with a dot-dashed line  at $0.60 \pm 0.50\ (stat) \pm 0.16\ (syst) $ ~\cite{Aaltonen:2010nz}.} }
  \label{fig:FV_vec_oct_spcr-1-3}
\end{figure*}
\begin{figure*}[!ht]
  \centering
  \subfloat[$g^{ut}_i$ variation of \crtt for cases (a) $g^{u t}_L =0 \neq  g^{u t}_R$ and (d) $g^{ut}_R=-g^{ut}_L=g^{ut}_A$ with $\du = 3.2 $]{\label{fig:FV_vec_oct_sigma-3.2}\includegraphics[width=0.5\textwidth]{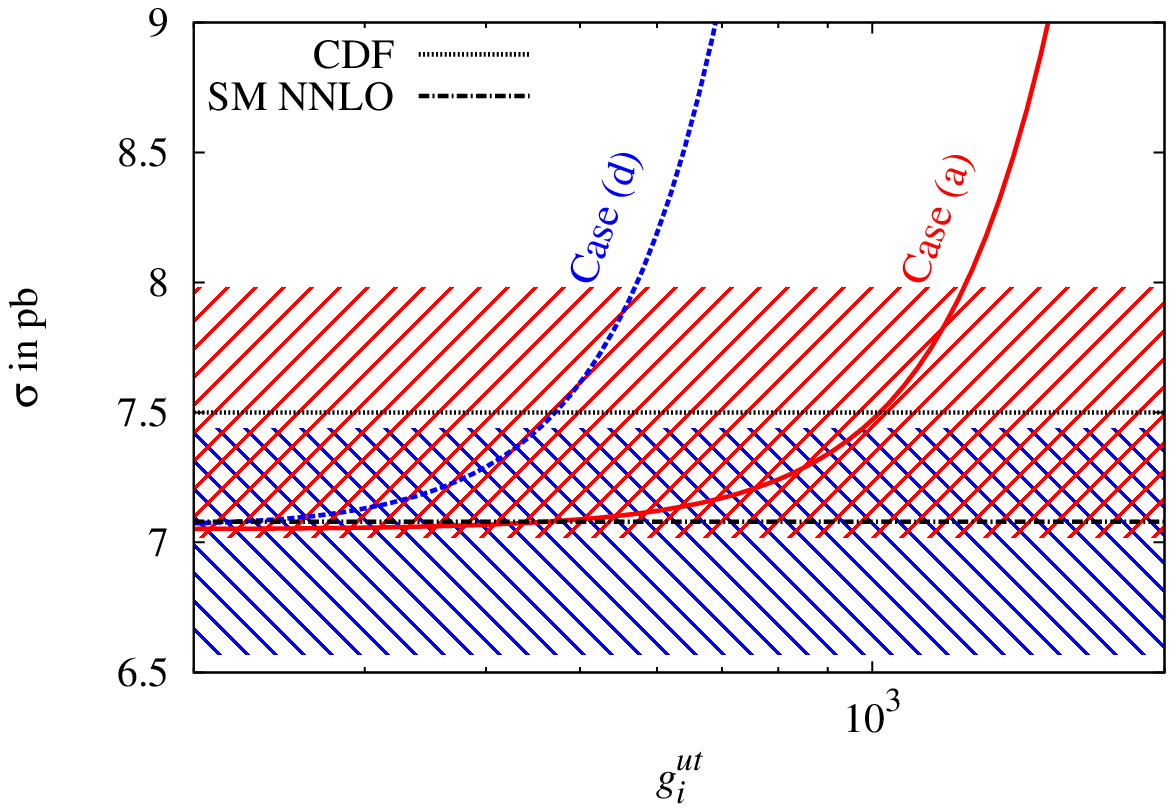}}
  \subfloat[$g^{ut}_i$ variation of $A_{FB}-A^{SM}_{FB}$  for cases (a) $g^{u t}_L =0 \neq  g^{u t}_R$ and (d) $g^{ut}_R=-g^{ut}_L=g^{ut}_A$  with $\du = 3.2 $]{\label{fig:FV_vec_oct_afb-3.2}\includegraphics[width=0.5\textwidth]{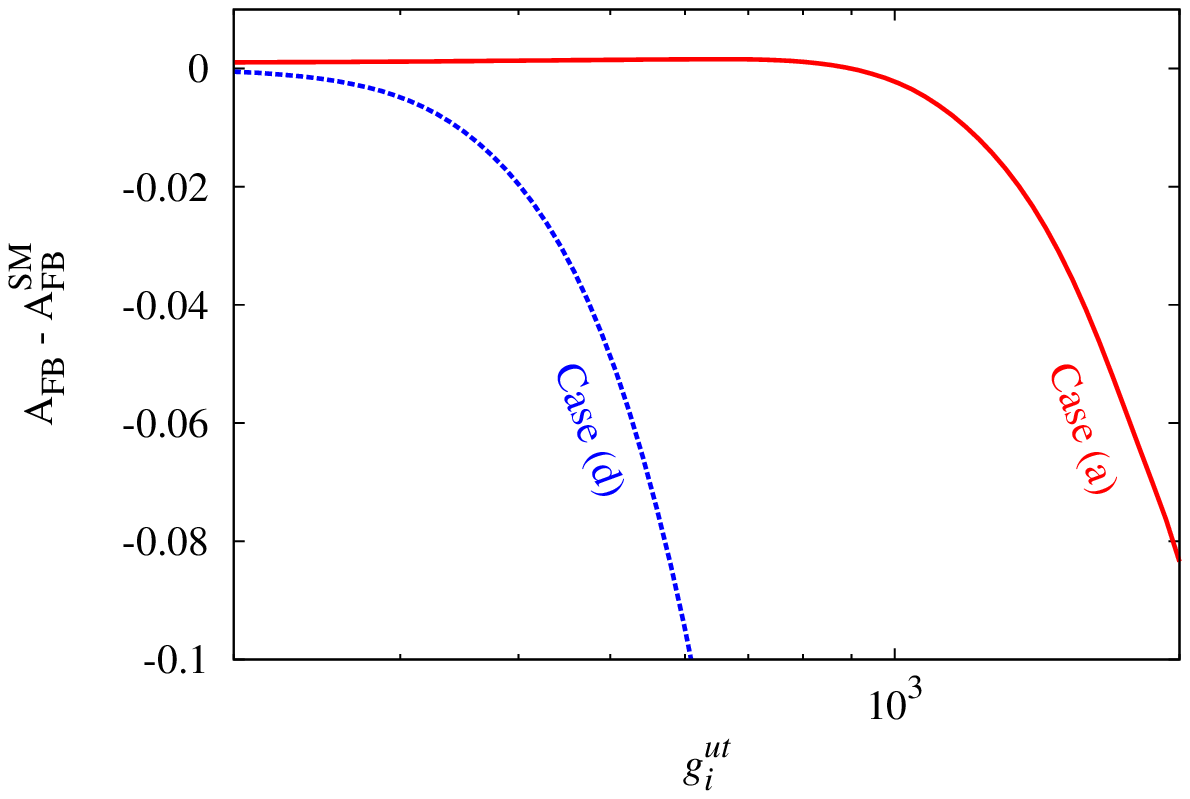}} \\
  \subfloat[$g^{ut}_i $ variation of \spincorr for cases  (a) $g^{u t}_L =0 \neq  g^{u t}_R$ and (d) $g^{ut}_R=-g^{ut}_L=g^{ut}_A$ with $\du = 3.2 $]{\label{fig:FV_vec_oct_spcr-3.2}\includegraphics[width=0.5\textwidth]{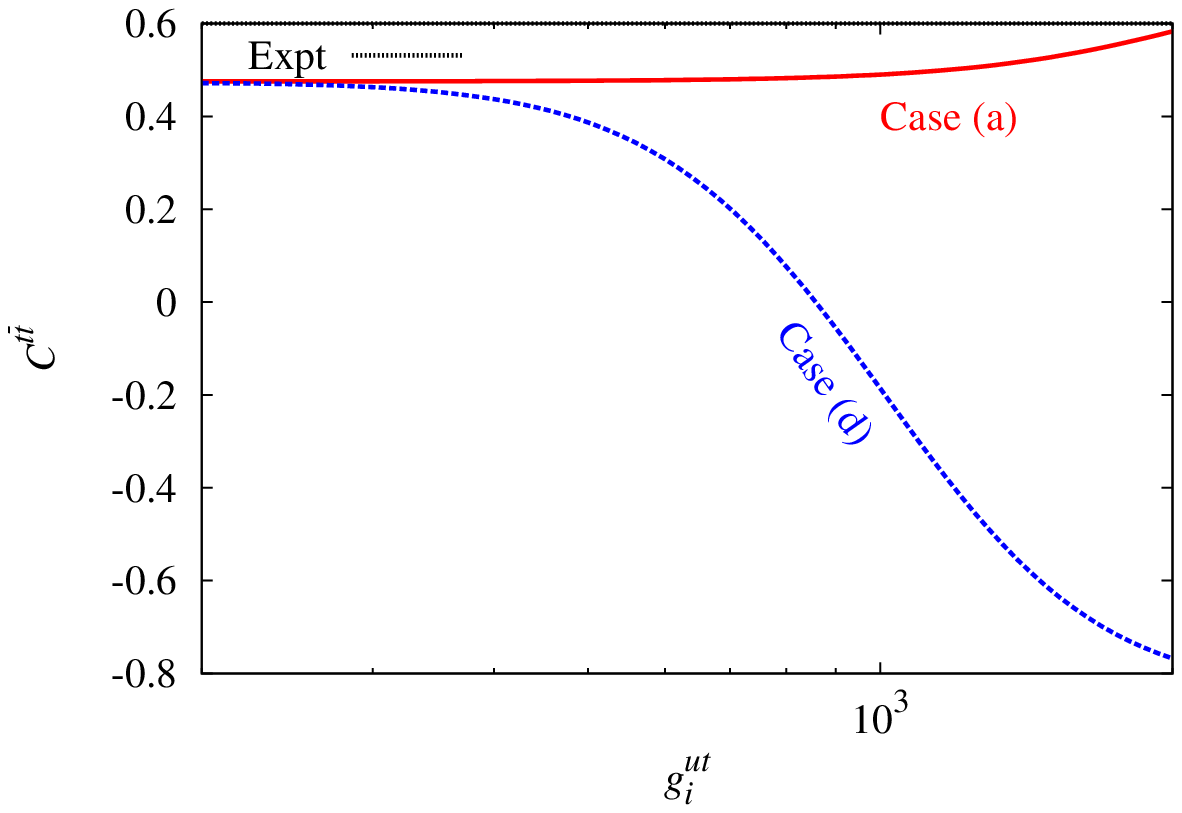}}
\caption{\small \em{Variation of the cross-section $\sigma \left( p\bar p \to t\bar t\right)$,  the unparticle contribution to charge asymmetry $A_{FB}-A^{SM}_{FB}$
and the spin correlation coefficient \spincorr  with  couplings
$g^{ut}_i$ for 
color octet flavor violating vector unparticles at fixed  $d_{\cal U}=3.2$ and $\Lambda_{\cal U}=1$ TeV corresponding
 to cases (a) and (d) mentioned in the text. In  plot (a) the  upper dotted line with a red band  depicts the  cross-section  $7.50 \pm 0.48$ pb from CDF (all channels) ~\cite{cdf-top-cross}, while the lower dot-dashed  line with a blue band show theoretical estimate $7.08\pm 0.36$ pb at NNLO  ~\cite{kidonakis-tcross}. Plot (b) is evaluated using the 1d differential distribution of rapidity in $t\bar t$ rest frame.
In plot (c) the experimental value is depicted with a dot-dashed line  at $0.60 \pm 0.50\ (stat) \pm 0.16\ (syst) $ ~\cite{Aaltonen:2010nz}.} }
\label{fig:FV_vec_oct-3.2}
\end{figure*}
The unparticle theory
 being an effective theory it is possible to have flavor violating
 couplings of unparticles with quarks.
The Flavor violating couplings involving first two generations of quarks and  vector unparticles are tightly constrained.
However, here we explore the vector unparticles which couples to the first and third generation quarks only.
The top pair production mediated by these unparticles are realized through
 $t$ channel processes. Hence the second term in the propagator 
\eqref{gprop-vec} also contributes. 
\par 
It is to be noted that the vector
 unparticle propagator given in equation (\ref{prop-vec-tensor}) has a
 pole for $\du=2$ at $\hat t=0$, and hence constraints the $\du$ to be
 greater
 than 2 \cite{Choudhury:2007cq, Choudhury:2007js} for consistency of the
 theory.  However in our case $\left \vert \hat t \right\vert$ is quite large and we do not encounter this problem. But this kind of flavor violating couplings also initiates the  top decay to unparticles which is studied in section \ref{SST}. This decay is only possible for \du $> 2$. We can overcome this constraint by introducing a  small infrared cut off  which is discussed in section \ref{sec:unp} and later on in subsection \ref{mass_gap}. The implication of this broken scale invariance on the cross-section, \afbt and spin-correlation is negligibly small as long as $ \mu \ll \left\vert \hat t\right\vert $. This enables us to make  our study for the top pair production  in the region $\du > 1$. 
\par Following the notation introduced in the previous subsection, the four possible coupling  for a given $\du$ \& $\Lambda_{\cal U}$ are (a) $g^{u t}_L = 0 \neq  g^{u t}_R$, (b) $g^{u t}_R \neq 0 =g^{u t}_L$ (c) $g_L^{ut} = g_R^{ut}=g^{ut}_V$   and (d) $g_L^{ut} = -g_R^{ut}=g_A^{ut}$. 
\par The $t$-channel process involving flavor violating singlet/octet  unparticles
 interferes  with  QCD and negligibly small electroweak sector.
  Hence the results can be explained
by the flavor violating unparticle new physics sector and its
interference with QCD.  Writing separately the same and opposite helicity
 contributions to these terms,
\begin{widetext}
\begin{eqnarray}
\left\vert {\cal M}^{\rm unp}\right\vert^2_{\rm same\,\, hel} &=&  c_f^{sq} g_s^2  {\cal B}_{_{FVV}}^2 \Biggl[ \frac{1}{2} 
\left\{\lprrsq +\lpllsq \right\}
 (1- \beta_t^2) (1+A_t)^2 \st^2 + 4\left(\lplr\right)^2 \times\nn \\
&& \,\,\left\{ (1+\bt^2) (1+A_t^2) 
+(1+\bt^2) A_t^2\ct^2  + 4 \bt A_t  + 2\, (1+\bt^2 +2\bt A_t ) A_t\ct \right\}\Biggr] \label{eqn_FV_vec_unp2_same}\\
\left\vert {\cal M}^{\rm unp}\right\vert^2_{\rm opp\,\, hel}  &=&  c_f^{sq}  g_s^2  {\cal B}_{_{FVV}}^2 \Biggl[ 4\lprlsq 
(1- \beta_t^2) A_t^2 \st^2 + 2  \biggl\{\lprrsq +\lpllsq \biggr\}\times\nn\\
&& \,\,\biggl\{ (1+A_t)^2\bt c_\theta+\,\, \frac{1}{4}  (1+ \beta_t^2) (1+A_t)^2
(1+\ct^2) \biggl\} \Biggr]\nn\\
\label{eqn_FV_vec_unp2_opp}\\
2 {\cal M}^{\rm QCD} {\cal R}e\, ({\cal M}^{\rm
  unp})\bigl\vert_{\rm same\, hel} &=& c_f^{int} g_s^2  {\cal B}_{_{\rm FVV}}
\biggl[\biggl\{\lpll +\lprr \biggr\} (1-\bt^2) (1+A_t)\st^2 \biggr] 
\label{eqn_FV_vec_int_samehel} \\
{\rm and }\quad
2 {\cal M}^{\rm
  QCD} {\cal R}e\, ({\cal M}^{\rm unp})\bigl\vert_{\rm opp\, hel} &=&
2\,c_f^{int}  g_s^2 {\cal B}_{_{\rm FVV}} \biggl[ \biggl\{\lpll +\lprr \biggr\} 
 (1+A_t)(1+\ct^2 +2 \bt \ct) \biggr]
\label{eqn_FV_vec_int_opphel}\\
{\rm with} \qquad {\cal B}_{_{\rm FVV}} &=& \bigl(\frac{1}{\Lambda_{\cal U}^2}
 \bigr)^{(\du -1)} \left[\frac{A_{\du}}{2\sin{(\du\pi)}}  \right]\, ({-\hat t})^{(\du -2)} 
{\hat s}; \,\,\,\,A_t = \frac{a \,m_t^2}{4\hat t}\,\,\, {\rm and}\,\, a = \frac{2(\du -2)}{(\du -1 )}.\nonumber
\end{eqnarray}
\end{widetext}
 Here $c_f^{sq} =9 (2)$ for FV singlet
 (octet) while   $c_f^{int} = 4 (-2/3)$ for FV singlet (octet).
\par 
The inherent symmetry of the helicity amplitudes  makes the contribution  to the observables
identical for cases (a) and (b) and similarly for cases (c) and (d).
The figures \ref{fig:FV_vec_sing_sigma-1-3}-\ref{fig:FV_vec_sing-3.2} 
give the variation of cross-section \crtt, charge asymmetry \afbt and 
spin correlation coefficient \spincorr corresponding to the color singlet FV coupling,
 while the same variation for flavor violating
 color octet are given in figures  
\ref{fig:FV_vec_oct_sigma-1-3}--\ref{fig:FV_vec_oct-3.2}. The salient features of these flavor violating unparticles are as follows:
\begin{enumerate}
\item Unlike the FC vector, the explicit and implicit dependence of $\hat t$  in the matrix element and in $A_t$ respectively  restrains the straight forward simplification of  the matrix element squared   as a polynomial in $\ct$. 
 \par The contribution to \afbt as well as \crtt 
 comes from both, same  and opposite helicity amplitudes. Therefore  \cfb
 is not same as \afbt when the flavor violating couplings are present.
\item  $\hat
t = m_t^2-( \hat s/2) (1-\bt\ct)$ is always negative and $\sin{(\du
  \pi)}$ is negative for $(2n-1)<\du<(2n)$  while positive for 
$(2n)<\du<(2n+1)$. And finally the sign of
 color factor decide the sign of contribution of interference
term of FV vector unparticle with QCD. Thus the interference term will be
 negative (positive ) for FV singlet (octet) for $1<\du<2$  while the sign
 will be reversed for $2<\du<3$ . However the negative values for singlet
 are visible only for
very small couplings as  with large  couplings  $\left\vert {\cal M}^{\rm unp}\right\vert^2$
contribution overshadows the interference term and hence the
cross section, \afbt and spin correlation for FV singlet first
decrease then increase.
\par Since $ -1<m_t^2/(4\,\hat t) \leq 0$, it implies $\left\vert A_t\right\vert \leq 1$, the factor $(1+A_t) $ is always positive. Thus even though $a$ flips sign from positive for $1<\du<2$ to negative for $2<\du<3$, it does not affect the sign of interference term for the $\du$ values considered. 
 \end{enumerate}
\par Thus for coupling to be within the range that gives   \crtt consistent with the CDF value, FV singlet vector unparticle gives favorable \afbt only for \du close to 1. On the other hand octet gives favorable values for the entire range $1<\du<2$. For $2<\du <3$, octet gives negative \afbt for allowed values of couplings while singlet gives very low values. 
\section{FV couplings: Same Sign Tops and Top Decay Width }
\label{SST}
\par As we have introduced the FV neutral current {\it via} non diagonal coupling for the fermion-fermion unparticle involving first and third generation, it is likely to have its bearings on the same sign top quark production and top decay width~\cite{AguilarSaavedra:2011zy}.
\subsection{Same Sign Top Production}
\begin{figure*}[!ht]
  \centering
  \subfloat[Flavor Violating Singlet Vector Unparticle]{\label{fig:contour_sing}\includegraphics[width=0.5\textwidth]{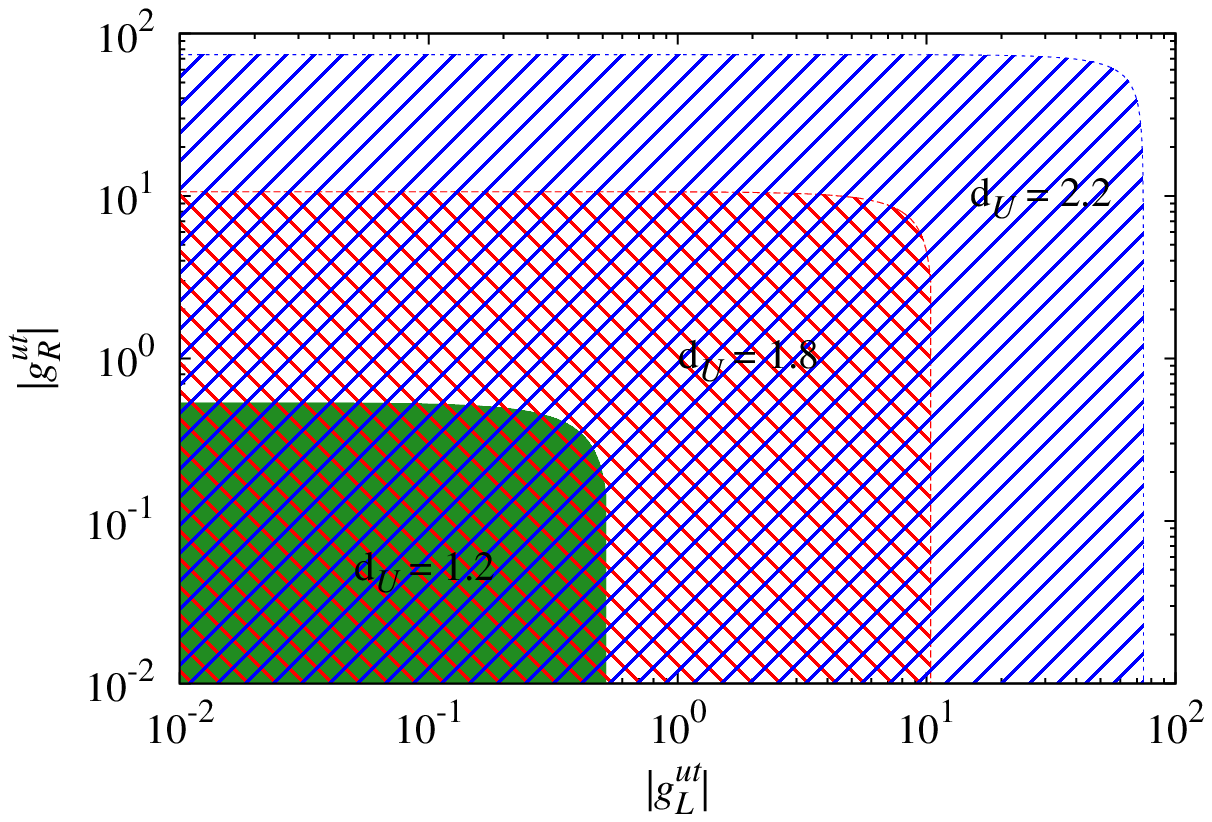}}
  \subfloat[Flavor Violating Octet Vector Unparticle]{\label{fig:contour_oct}\includegraphics[width=0.5\textwidth]{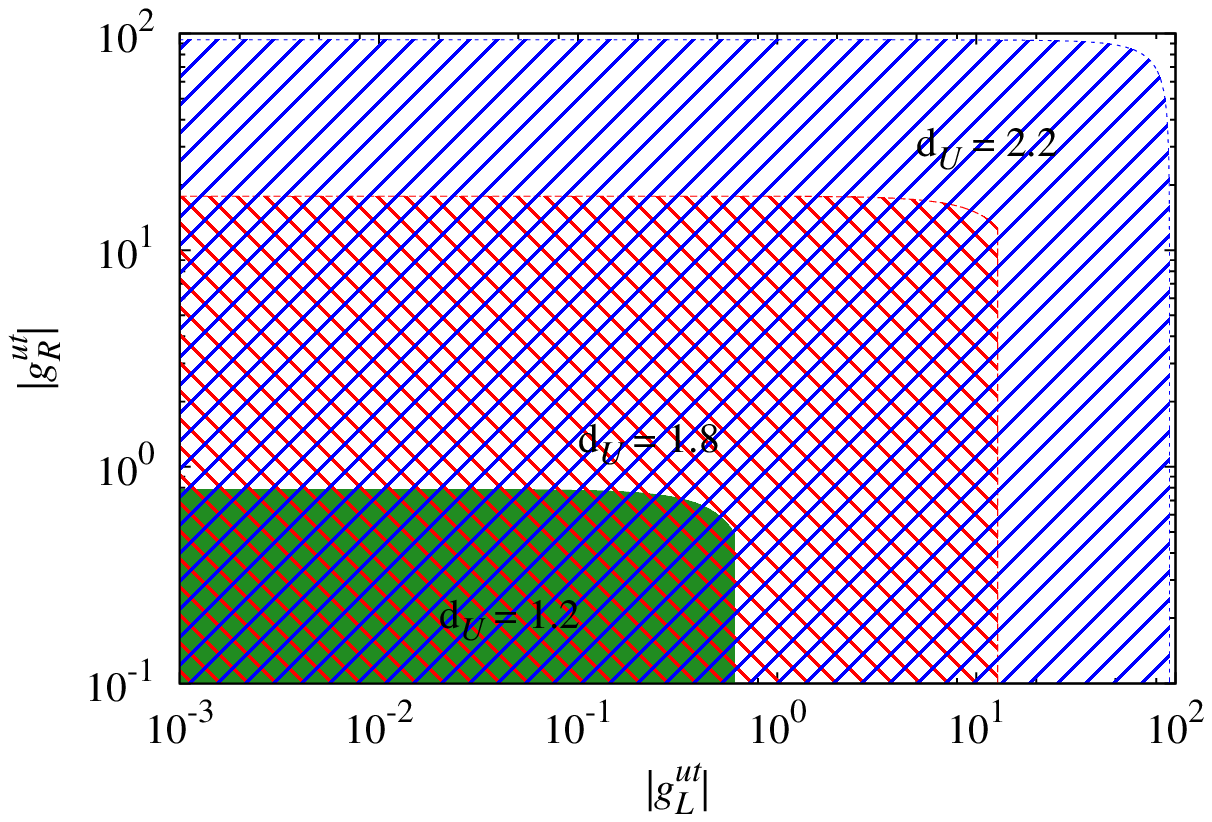}}
  \caption{\small \em{Exclusion contours at  95\% CL corresponding to  the cross-section $\sigma\left(p\bar p\to tt+\bar t\bar t\right)=54 $ fb \cite{Chatrchyan:2011dk} on the $\left\vert g_L^{ut}\right\vert - \left\vert g_R^{ut} \right\vert $ plane for a given \du and $\Lambda_{\cal U}=1$ TeV. The value of the couplings bounded by the enclosed region (for a specific \du value) are allowed.}}
  \label{fig:contour}
\end{figure*}
Recently an inclusive search of same sign tops at CMS dis-favored the FCNC solution in $Z^\prime $ model for \afbt anomaly at TeVatron \cite{Chatrchyan:2011dk}. The SM contribution is highly suppressed and any affirmative signal would indicate a presence of new physics. The unparticle theory through their FV couplings also generates the same sign tops $p\bar p\to tt$ {\it via} $t$ and $u$ channel Feynman diagrams. We compute the parton level helicity amplitudes for $uu\to tt$ and are given in the Appendix \ref{app:same_sign_hel_amp}.
\par As mentioned earlier, we have performed our analysis with $\Lambda_{\cal U} =1$ TeV to  probe the physics at TeVatron, where the partonic c.m.  energy $<< $ 1 TeV.  However, to examine the physics at LHC and validate the same model, we need to enhance this scale at least to $\Lambda_{\cal U}=10$ TeV. In this light at present we only see the effect of the unparticle physics at TeVatron and check their cross-validity among various processes. Same sign top production was also constrained by TeVatron \cite{sstops:cdf}. The CDF data was based on same sign dilepton search and their observations  are summarized in the table II of reference \cite{sstops:cdf}. They predicted the  upper limits at 95\% CL, on the production cross-section $\sigma \left(p\bar p\to tt +\bar t\bar t\right)$ times branching ratio BR$(W\to l\nu)^2$ for all distinct chirality modes: left-left $(LL)$ to be 54 fb, right-right $(RR)$ to be 51 fb and left-right $(LR)$ to be 51 fb, assuming only one non-zero mode at one time. We present the 95 \% CL contours for varying \du  on $\left\vert g_L^{ut}\right\vert  - \left\vert g_R^{ut}\right\vert $ plane using the result from $LL$ mode, assuming that there is remarkable difference in efficiency or shape among the contribution from $LL$, $LR$ or $RR$ modes. The contours corresponding to color singlets and octets are shown in figures \ref{fig:contour_sing} and \ref{fig:contour_oct} respectively.
\par We expect that with the change of the scale $\Lambda_{\cal U}$, we will be able to probe the same parameter region for LHC. This  work is currently under progress.
\subsection{Top Decay Width}
\begin{figure*}[!ht]
  \centering
 \subfloat[\du =1.2 ] {\includegraphics[width=0.5\textwidth]{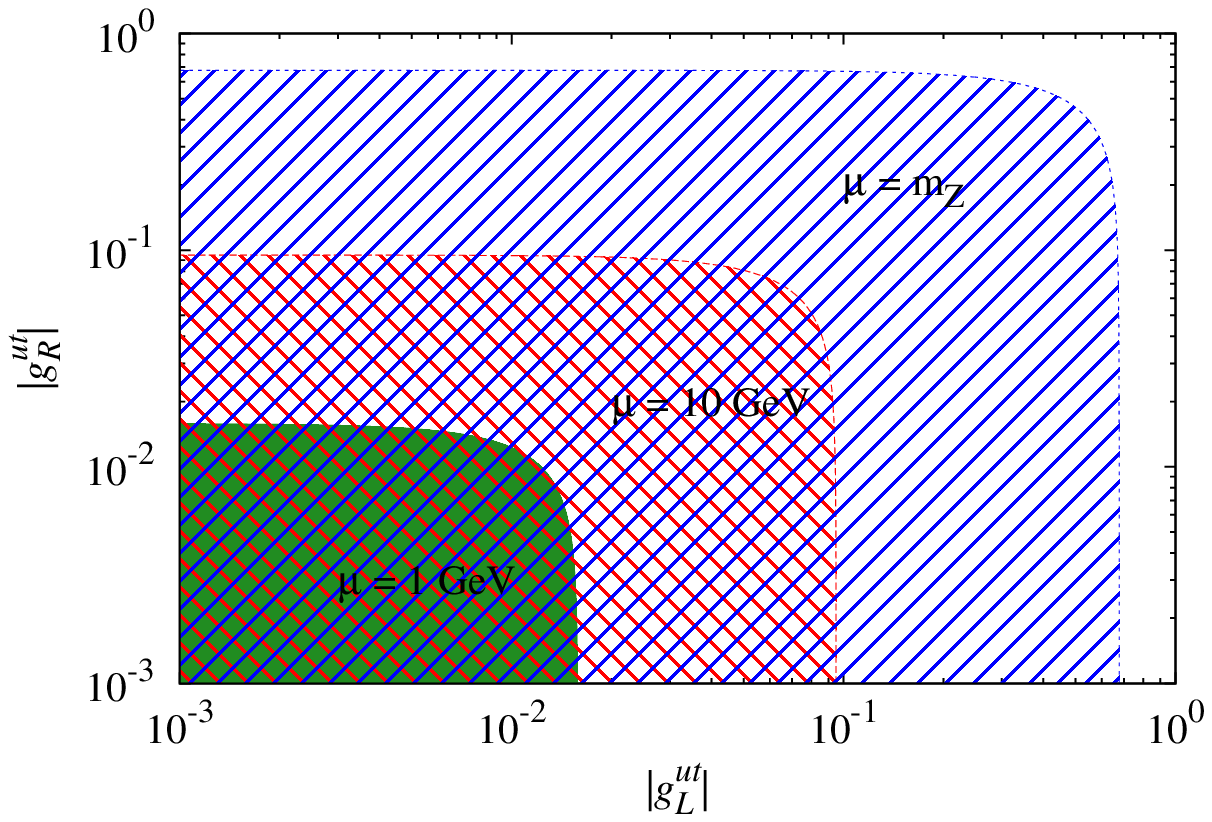}}
\subfloat[\du =1.8 ] {\includegraphics[width=0.5\textwidth]{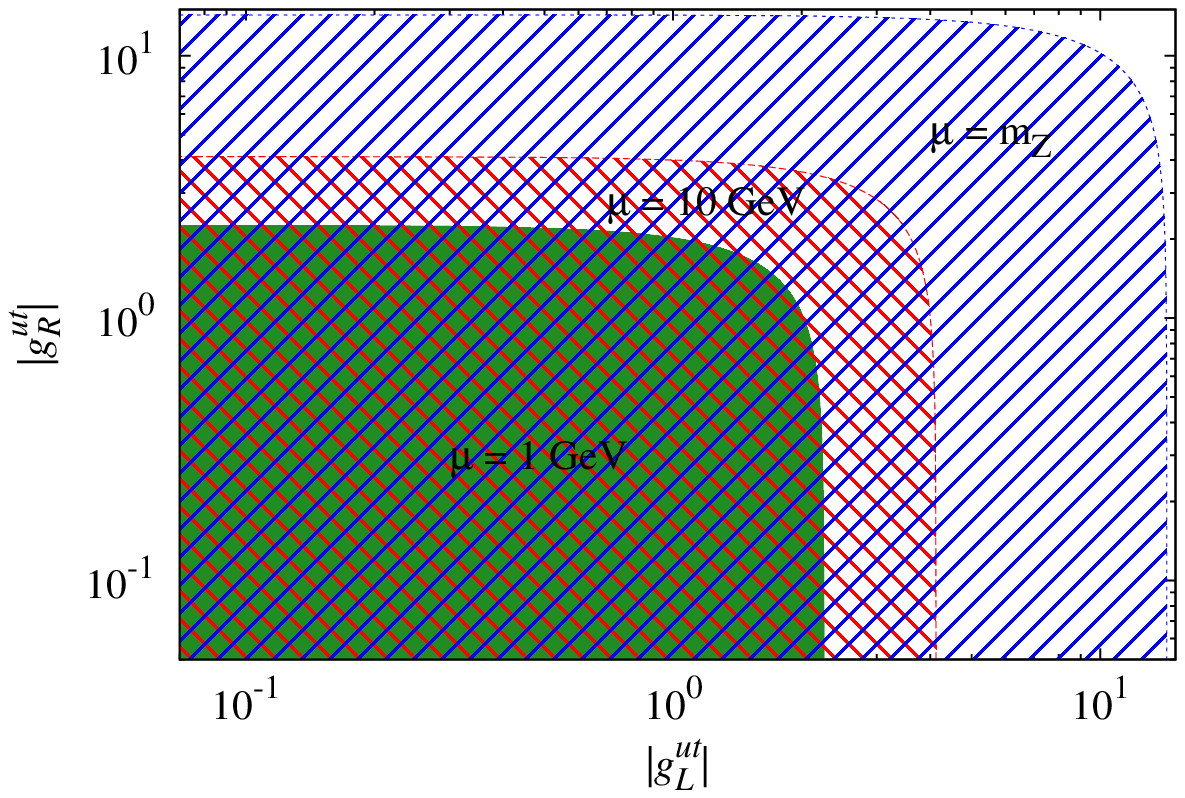}}\\
\subfloat[\du =2.2 ] {\includegraphics[width=0.5\textwidth]{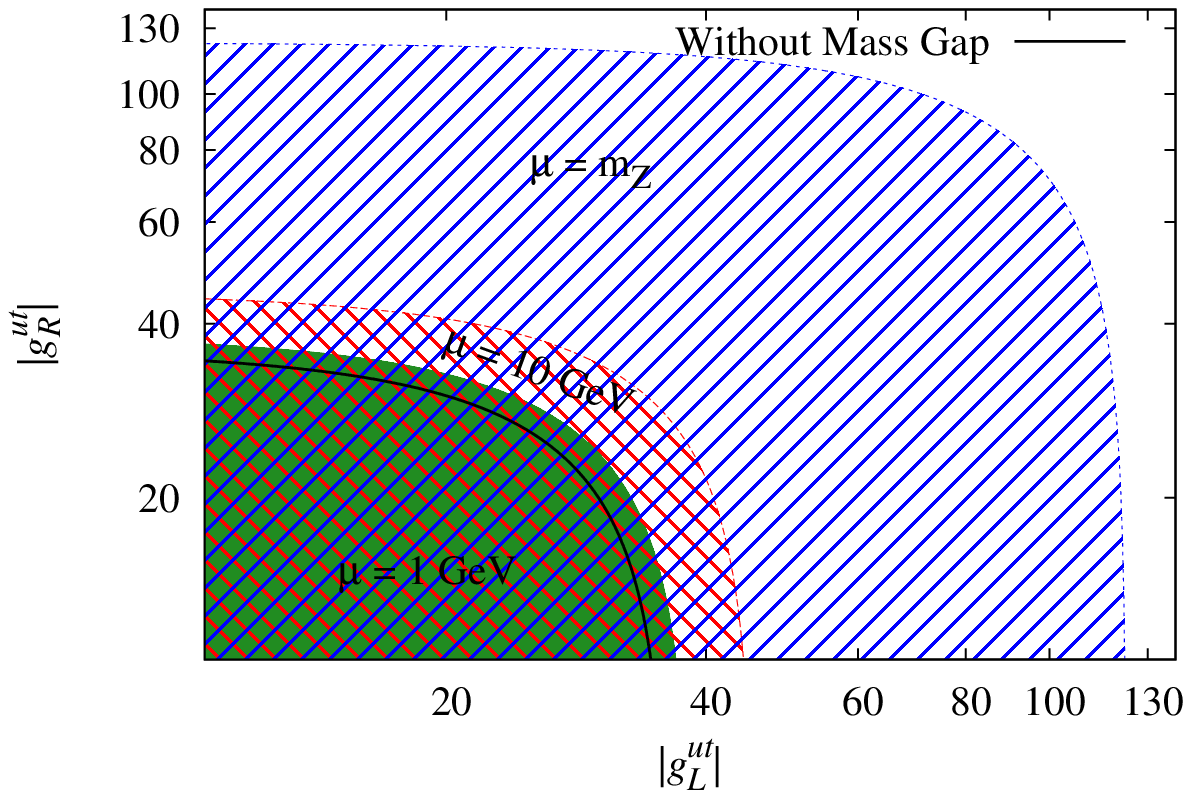}}
\caption{\small \em{ Exclusion contours at  95\% CL corresponding $\Gamma_t=1.99^{+0.69}_{-0.55}$ GeV  \cite{Abazov:2010tm} on the $\left\vert g_L^{ut}\right\vert - \left\vert g_R^{ut} \right\vert $ plane for  flavor Violating octet vector Unparticle with  a given mass gap $\mu $ and $\Lambda_{\cal U}=1$ TeV.  The value of the couplings bounded by the enclosed region (for a specific mass $\mu $ value) are allowed. In figure (c) the black contour line corresponds to  $\mu =$ 0 GeV.}}
\label{fig:dky_contour_oct}
\end{figure*}
The total decay width of top quark is one of the  fundamental property of top physics.
It is measured with precision from the partial decay width $\Gamma (t\to W\,b)$ in the $t$ channel of the single top quark production and from $t\bar t$ events.  Recently the top total width is measured to be $\Gamma_t=1.99^{+0.69}_{-0.55}$ GeV corresponding to 2.3 fb$^{-1}$ data by D\O\, collaboration \cite{Abazov:2010tm}. The SM contribution to top decay width at NLO  in $\alpha_s$ is given as 
\begin{widetext}
\begin{eqnarray}
\Gamma_{SM} (t\to W\,b)=\frac{G_Fm_t^3}{8\pi\sqrt{2}}\left\vert V_{tb}\right\vert^2 \left(1-\frac{M^2_W}{m_t^2}\right)^2 \left(1+2 \, \frac{M^2_W}{m_t^2}\right)\left[1- \frac{2\alpha_s}{3\pi}\left(\frac{2\pi^2}{3}-\frac{5}{2}\right)\right]
\end{eqnarray}
\end{widetext}
Computing with   $\alpha_s (M_Z) = 0.118$, $G_F = 1.16637 \times 10^{-5}$ GeV$^{-2}$,
$M_W = 80.399$ GeV, $\left\vert V_{tb}\right\vert = 1$, and $m_t = 173$ GeV, we find $\Gamma_{SM} =\left(t\to W\, b \right)_{SM} = 1.34 $ GeV.
 \par In unparticle sector, this was first mooted by Georgi \cite{Georgi:2007ek} where he considered the flavor violating derivative coupling of the scalar unparticles given as $i \lambda\,\Lambda^{-\du} \,\bar u \gamma_\mu\,\left(1-\gamma_5\right) t\,\,\partial^\mu{\cal O}_{\cal U} +h.c.\, $. The FV vector unparticle initiates a new channel  for top decaying to unparticle and a lighter quark $q$. We compute the partial decay width $\Gamma_{\cal U} (t\to q\,{\cal U}^{V}) $ as given in the Appendix \ref{app:decaywidth}. This decay width diverges for \du $<2$ while for \du $> 2$ it is given as 
\begin{widetext}
\bea
\Gamma_t &=&\Gamma_{SM} +  \frac{N_C}{6} \frac{A_{\du}}{4\pi^2} g_s^2 \left[\left(g^{{\cal U}_V^{\bf n}tq}_L\right)^2 + \left(g^{{\cal U}_V^{\bf n}tq}_R\right)^2\right] \left(\frac{m^2_t}{\Lambda_{\cal U}^2}\right)^{\du-1} \frac{m_t}{4} \left[ \frac{  a -4 + \left( a+2 \right) d_{\cal U}}{\left(\du -2\right) \,\left( \du -1\right)\,\du \,\left(\du+1\right)} \right] \label{decay-width_du_gt_2}
\eea
\end{widetext}
However if the scale invariance is broken at a scale $\mu \neq 0$, one can evaluate the decay width in the region  $1 < \du <2$.  As mentioned earlier inclusion of such mass gap also protects the theory from realizing continuous  spectrum of unparticles through the decay process. Introduction of such a scale however does not modify the cross sections  for $\mu\ll \mttb $. We  depict the exclusion contours  with various choices of the breaking scales at 1 GeV, 10 GeV and at $m_Z$ = 91.18 GeV corresponding to the color octet unparticles on $\left\vert g_L^{ut}\right\vert  - \left\vert g_R^{ut}\right\vert $ plane. These contours show the  allowed range of these couplings  constrained from the observed total top decay width $\Gamma_t=\Gamma_{SM} +\Gamma_{\cal U}$ at D\O .  The nature and behaviour of the contribution from the color singlet flavor violating unparticles are similar to that of the octet. Therefore we do not provide the corresponding contours separately.   We find from figure \ref{fig:dky_contour_oct} that the parameter region which  contributes  to $t \bar t$ events considerably along with large positive \afbt , shrinks and hence is much more constrained. We also observe that the increase in the scale invariance breaking scale relaxes the bound on the couplings.
\section{Analysis and Summary}
\label{analysis}
\subsection{\mttb distribution of \afbt }
\begin{figure*}[!ht]
 \centering
 \subfloat[FC Vector Singlet: $\lambda_{RR} \ne 0$ and $\lambda_{LR}=\lambda_{RL}=\lambda_{LL} =0 $ ]{\label{fig:FC_vec_sing_mttbar}\includegraphics[width=0.5\textwidth]{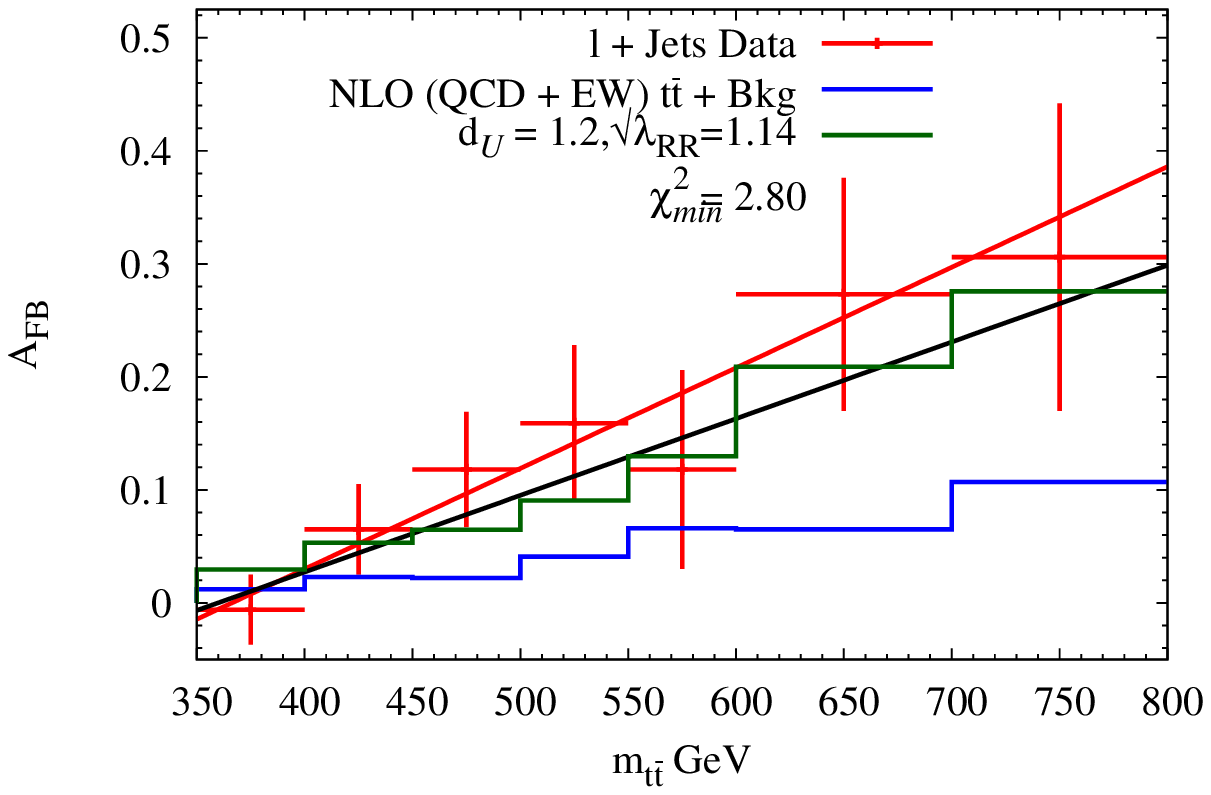}}
\subfloat[ FC Vector Octet: $\lambda_{RR} \ne 0$ and $\lambda_{LR}=\lambda_{RL}=\lambda_{LL} =0 $]{\label{fig:FC_vec_oct_mttbar}\includegraphics[width=0.5\textwidth]{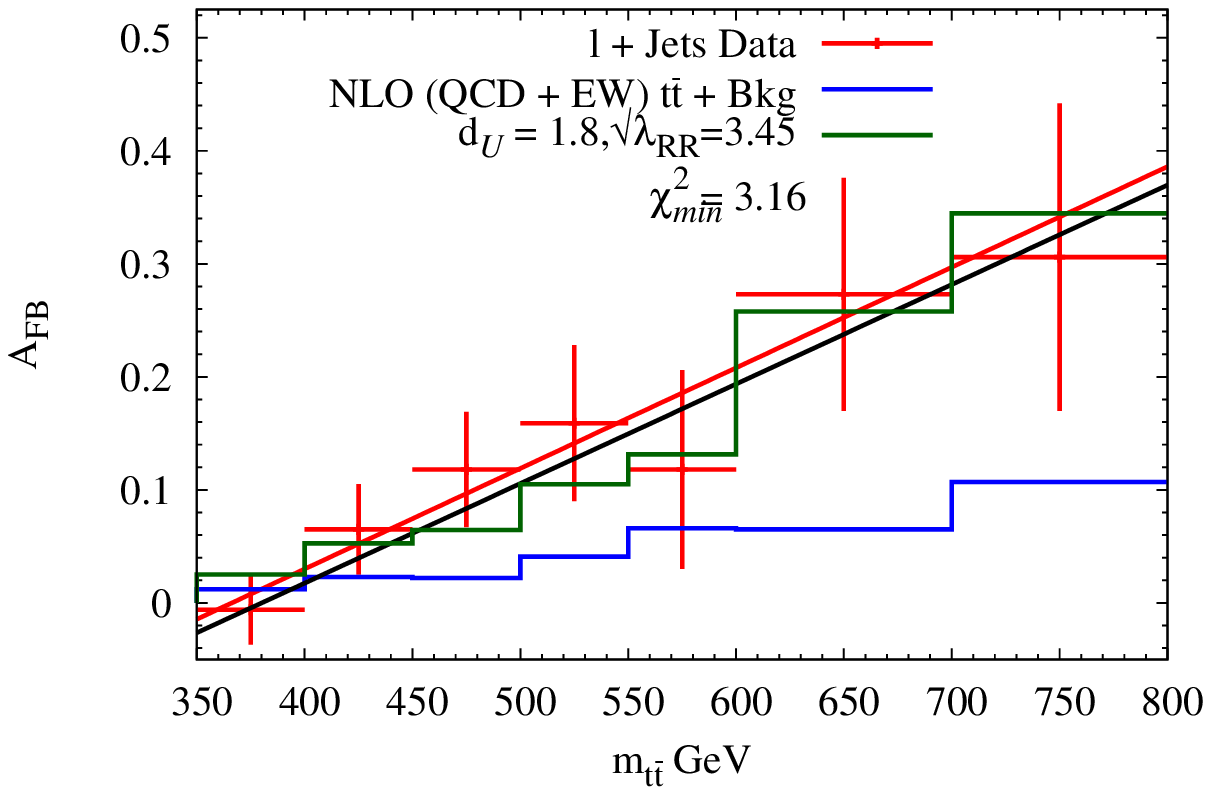}}\\
  \subfloat[ FC Tensor Octet: $\lambda_{LL} = \lambda_{RR} = -\lambda_{RL}  = -\lambda_{LR} =\lambda_{AA}$]{\label{fig:FC_ten_oct_mttbar}\includegraphics[width=0.5\textwidth]{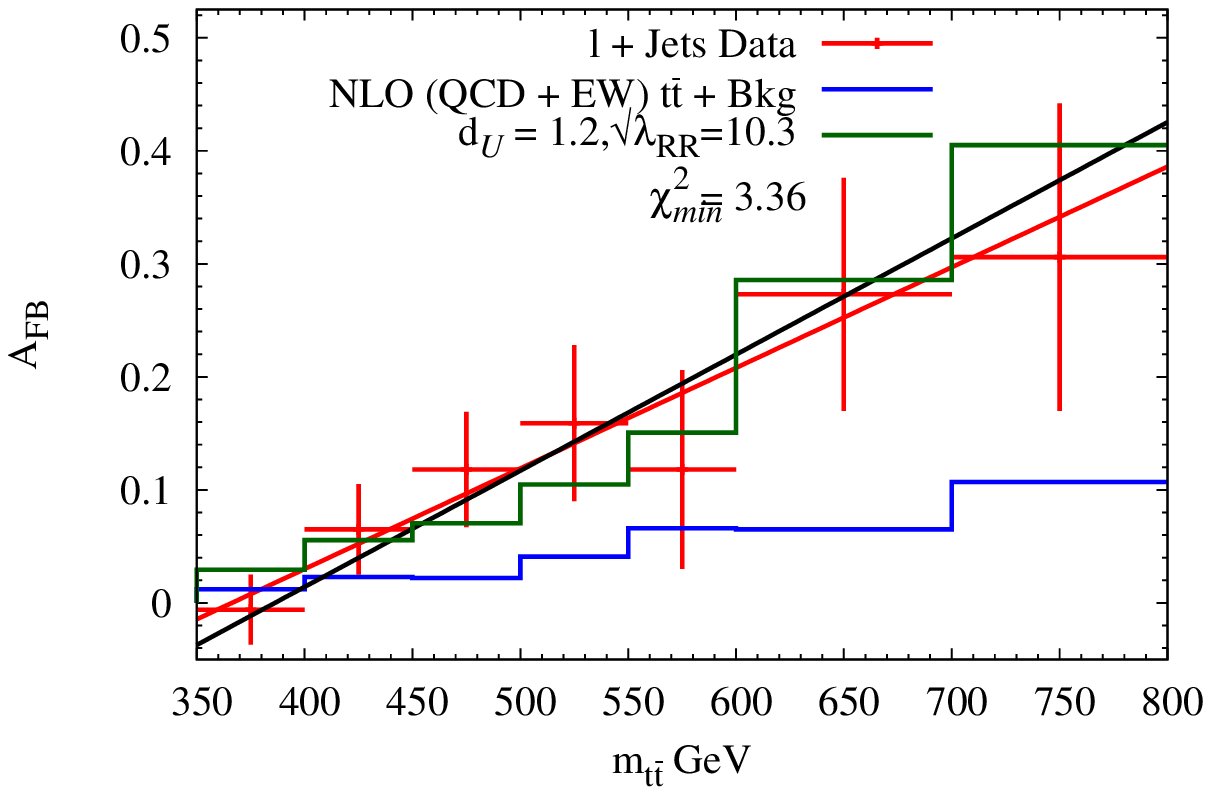}}
\subfloat[FV Vector Octet: $\lambda_{RR} \ne 0$ and $\lambda_{LR}=\lambda_{RL}=\lambda_{LL} =0 $]{\label{fig:FV_vec_oct_mttbar}\includegraphics[width=0.5\textwidth]{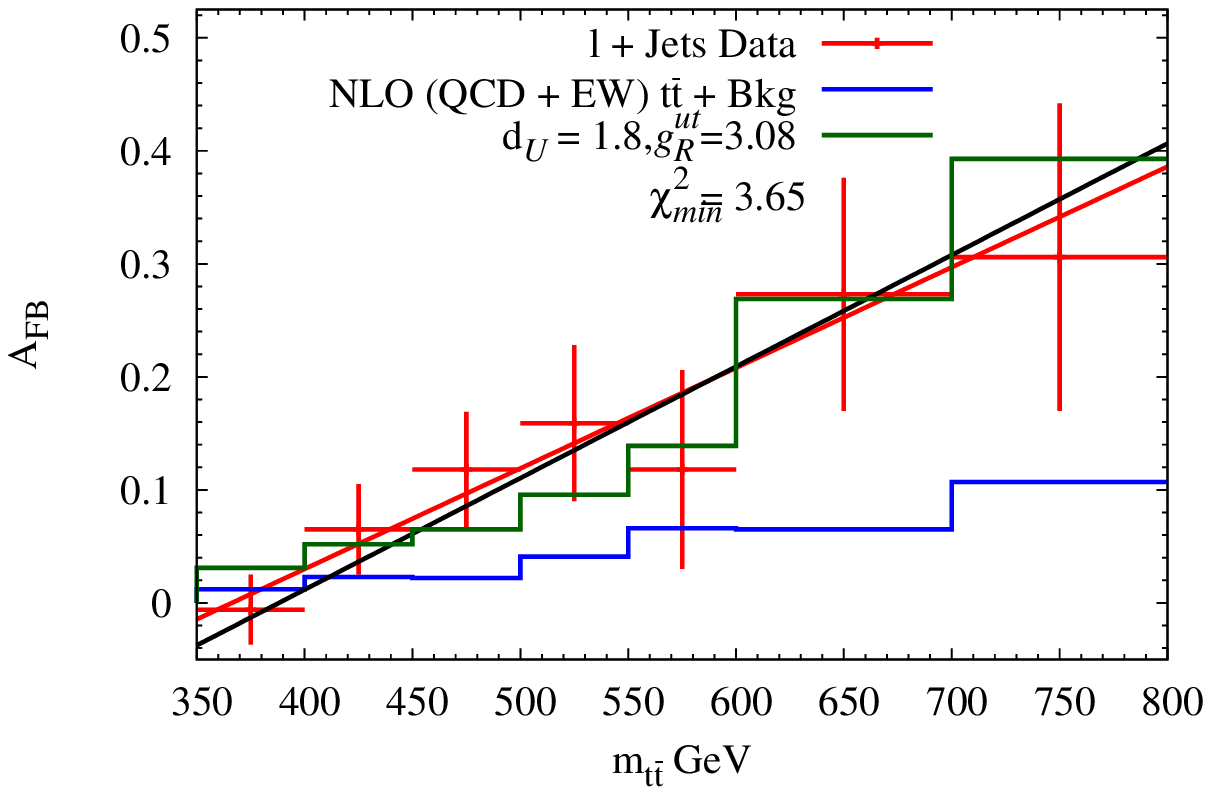}}
  \caption{\small \em{ $\mtt$ distribution of \afbt at $\chi^2_{\rm min.}$ for the four favorable point-sets in parameter space at fixed $\Lambda_{\cal U}=1$ TeV. The histogram corresponding to the best-fit model parameters are shown by green steps, the experimental data points are shown with its errors in red, while the SM NLO (QCD + EW) with backgrounds are shown in the blue shaded histogram. The red straight line in all graphs is the best-fit line with the experimental data from ref.~\cite{afbt-cdf-mar2012} while the black straight line depicts the best-fit line with the model (unparticles) plus SM NLO (QCD + EW) with backgrounds data.}}
  \label{fig:mttbar}
\end{figure*}
\par We  scan our parameters and perform  $\chi^2$ analysis for both  FC and FV cases and predict the set of best parameters  which can possibly  explain the \afbt anomaly. To perform this analysis we take into account the \afbt distribution   over $m_{t\bar t}$ bins  from  the full Run II TeVatron Dataset \cite{afbt-cdf-mar2012}. We define the  $\chi^2$ as
\bea
\chi^2=\sum_{i=1}^{n}\frac{({\cal O}^{th}_i- {\cal O}^{exp}_i)^2}{(\delta {\cal O}^{exp}_i)^2}
\label{chisq}
\eea
where $i$ is $m_{t\bar t}$ bin index, ${\cal O}^{th}_i$, ${\cal O}^{exp}_i$ and $\delta {\cal O}^{exp}_i$ are the SM+model estimate, experimental measurement and its error in the corresponding $i^{\rm th}$ bin respectively. Following the reference \cite{afbt-cdf-mar2012}, we quote and use the experimental data based on the complete dataset of Run II at an  integrated luminosity 8.7 fb$^{-1}$ ( given in the table \ref{afbmttdata})  to compute $\chi^2_{min.}$. The first column reads the bin size while the  second and third column of the table \ref{afbmttdata} gives experimentally observed value of \afbt in each of these bins  with its error and the expected number from NLO (QCD+EW) with backgrounds respectively.
We also add the observed cross-section as an eighth observable in equation (\ref{chisq})  $\sigma (p\bar p\to t\bar t) =
7.5 \pm 0.31 ({\rm stat}) \pm 0.34 ({\rm syst}) \pm 0.15 ({\rm Z \, theory})$ pb \cite{cdf-top-cross}.  
The two dimensional parameter space (($\sqrt{\lambda_{LL}}$,\, $\sqrt{\lambda_{RR}}$) for FC cases and ($g^{ut}_L,\, g^{ut}_R$) for the FV case) for fixed \du  and  $\Lambda_{\cal U}$ is scanned leading to the  minimum value of the $\chi^2 \equiv\chi^2_{\rm min.}$. 
 \par At $\chi^2_{\rm min.}$ the corresponding parameter points are likely to be  the best possible model parameters in the unparticle physics at $\Lambda_{\cal U}=1$ TeV, which are consistent with all the observations. Fourth$-$seventh columns in the table \ref{afbmttdata} exhibit the \mttb \, spectrum of \afbt corresponding to the four distinct cases of best fitted model parameters.  In  the figure \ref{fig:mttbar}, we plot  histograms   showing   the $\mttb$ spectrum of \afbt for all the four  cases. We have shown and compared the slope of our best-fit line with that from the experimental data in this figure and table \ref{afbmttdata}.
\subsection{Effect of Mass gap}
\label{mass_gap}
In our study we have not introduced any infra-red cut-off in the theory of unparticles modulo top decay width. The
existence of massless fields in a theory gives rise to severe modification in the low energy
phenomenology which is successfully explained by SM alone. To be precise, if there is no cut-off in the theory of unparticles there would be massive production of color singlet/ octet unparticles in colliders. 
\par A standard way to treat the breaking of conformal symmetry is to introduce a mass gap in the spectral density and thereby removing modes with energy less than the infrared cutoff scale $\mu$ in the spectral density. Thus the propagator for vector unparticle given in equation~\eqref{gprop-vec} will be modified to
\begin{widetext}
\begin{eqnarray}
  \Delta^{\mu\nu}(p)=\frac{-iA_{d_{\cal U}}}
  {2\sin(d_{\cal U}\pi)}[-(p^2-\mu^2)]^{d_{\cal U}-2}
  \left(- g^{\mu\nu}+ a \,\frac{p^\mu p^\nu }{p^2}\right),
\label{prop-vec-scale}
\end{eqnarray}
with
\bea
[(-(p^2-\mu^2)]^{d_{\cal U}-2} = 
\begin{cases}
& \left\vert p^2-\mu^2\right\vert^{\du-2} \qquad\quad {\rm for }\,\, p^2<\mu^2   \\
& \left\vert p^2-\mu^2\right\vert^{\du-2} e^{i\du\pi} \,\,\, {\rm for }\,\, p^2>\mu^2.
\end{cases}
\label{def-P1sq}
\eea
\end{widetext}
It reduces to~\eqref{gprop-vec} in the limit $\mu \to 0$. From equation \eqref{def-P1sq} it follows  that one
 can ignore the existence of the mass gap as long as all the momentum invariants involved with the unparticle propagators 
 are much larger than the conformal symmetry breaking scale. For $\mu =$ 1 GeV, 10 GeV and $m_Z$ the effects are negligibly small. Since the minimum of these momentum invariants in case of top pair production is the threshold \mttb, the  suppression of   the partonic cross-section  can be at most  $\left(1-\mu^2/\hat s_{\rm min.}\right)^{\du -2} = \left(1-\mu^2/\mttb^2 \right)^{\du -2}$ and  $\left(1-\mu^2/\hat t_{\rm min.}\right)^{\du -2} = \left(1-\mu^2/\mttb^2\right)^{\du -2}$ for   $s$ and $t$ channel processes respectively. We have checked the stability and consistency of  our results corresponding to the above mentioned choices of the breaking scales for  all $s$ and $t$ channel processes. 
\par Similarly computation of the same sign top pair production mediated by the flavor violating unparticles do not have any impact due to introduction of the breaking scale as long as the scale is sufficiently smaller than the same sign top pair threshold $\simeq$ 350 GeV.

\par The presence of FV couplings leads to the decay process like $t \to u + {\cal U}$ which  diverges for $\du>2$~\cite{Choudhury:2007cq, Choudhury:2007js} unless one introduces a mass gap. Therefore, the adoption of  scale invariance breaking is inevitable in the region $1<\du <2$ for the top pair production and same sign top pair production processes induced  by FV couplings. However, once scale invariance is broken at a scale $\mu$, the states with momenta $p^2<\mu^2$ are removed from phase space resulting in a finite and positive decay width even for $\du<2$~\cite{Barger:2008jt}.

\par Introduction of color octet unparticles are likely to  produce  them in plenty at hadron colliders through $q\bar q\to {\cal U}^{V/T}+ {\cal U}^{V/T}$ and $gg\to {\cal U}^{V/T}+ {\cal U}^{V/T}$. These processes mediate through $s$  and $t$ channels. Gluon flux being low at TeVatron, production of such colored unparticles from gluon-gluon initial state  may not have large bearing in our analysis. Production of colored scalar unparticles   has been studied by Cacciapaglia {\it et al} ~\cite{Cacciapaglia:2007jq}. There they showed that the pair production cross-section of color scalar unparticles is suppressed by a factor of (2-\du ) with respect to the particle pair production. One would expect the production of the vector unparticles should follow the same suit with respect to the vector particles.
But then these channels are also  constrained by observed di-jet cross-section at TeVatron. The upper limit on  the allowed cross-section  through these processes translates to an upper bound on   the couplings of the respective light quark with colored vector/ tensor unparticles. 
\par Throughout our study  including the $\chi^2$ analysis of \mttb \, distribution of \afbt, the cross-sections depends on  the product of couplings $g^{{\cal U}_{(V/T)}^{\bf n}\bar q q}_i\, g^{{\cal
    U}_{(V/T)}^{\bf n}\bar t t}_j=\lambda^{(V/T)}_{ij}$ involving light quarks and top  quark with flavor conserving color vector/ tensor unparticles. In light of our assumption $g^{{\cal U}_{(V/T)}^{\bf n}\bar
  q q}_{(L/R)} << g^{{\cal U}_{(V/T)}^{\bf n}\bar t t}_{(L/R)}$ ( mentioned  in subsection \ref{subsec:FC}),    our analysis is consistent  with respect to the observation from di-jet cross-section.  
\par Although introduction of flavor violating interactions involving  colored unparticles, up quark and top quark in a $\hat t$  channel process do not encounter these shortcomings but it can still copiously produce  unparticles in the colliders {\it via} top decay. The contribution of the decay channel to the partial top decay width constraints this coupling for $\du >2$.
\par However, all these problems  can be eased out if  we can model the color unparticle theory  by requiring it to have an infrared cut-off. Fairly large $\mu $ is likely to suppress the copious production of color unparticles through all channels discussed here. 
Even this allows to constrain the couplings from the decay process $t\to u+ {\cal U}^V$ in the region $1<\du <2$ corresponding to specific choice of $\mu $ from the measurement of the top decay width.
 \subsection{Observations and Conclusions}
\label{conclusion}
 We have studied the interactions of vector and tensor color singlet/ octet unparticles in the top sector at TeVatron, through FC and FV couplings. In this way we have made an attempt to address the existing anomaly in the \afbt at TeVatron keeping the cross-section and spin-correlation of $t\bar t$ consistent with the data.  We have also studied the contribution of these unparticles to the top decay width and the same sign top pair production induced by FV couplings. 

\begin{figure*}[!ht]
 \centering
\subfloat[\du values allowed by completely conformal theory ]{\label{fig:unitarity-conformal}\includegraphics[width=0.5\textwidth]{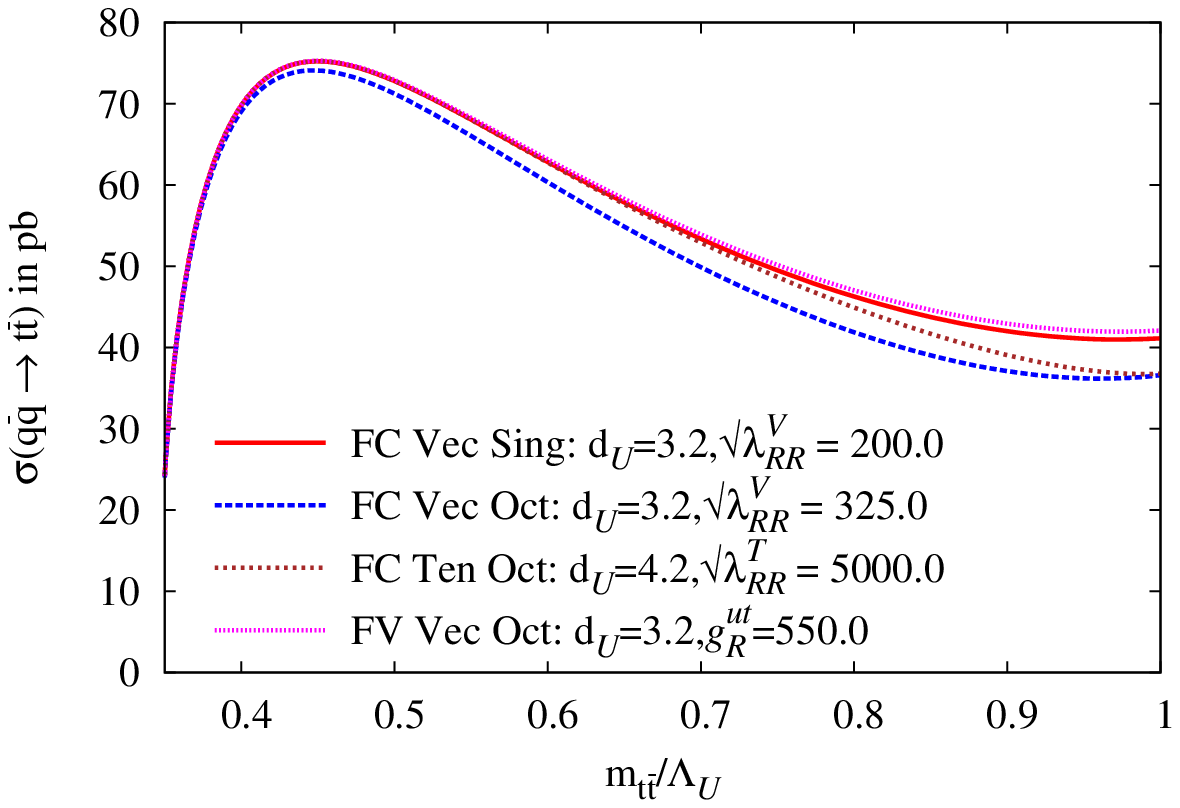}}
 \subfloat[ Four favorable point sets based on $\chi^2_{\rm min}$ ]{\label{fig:unitarity-favcases}\includegraphics[width=0.5\textwidth]{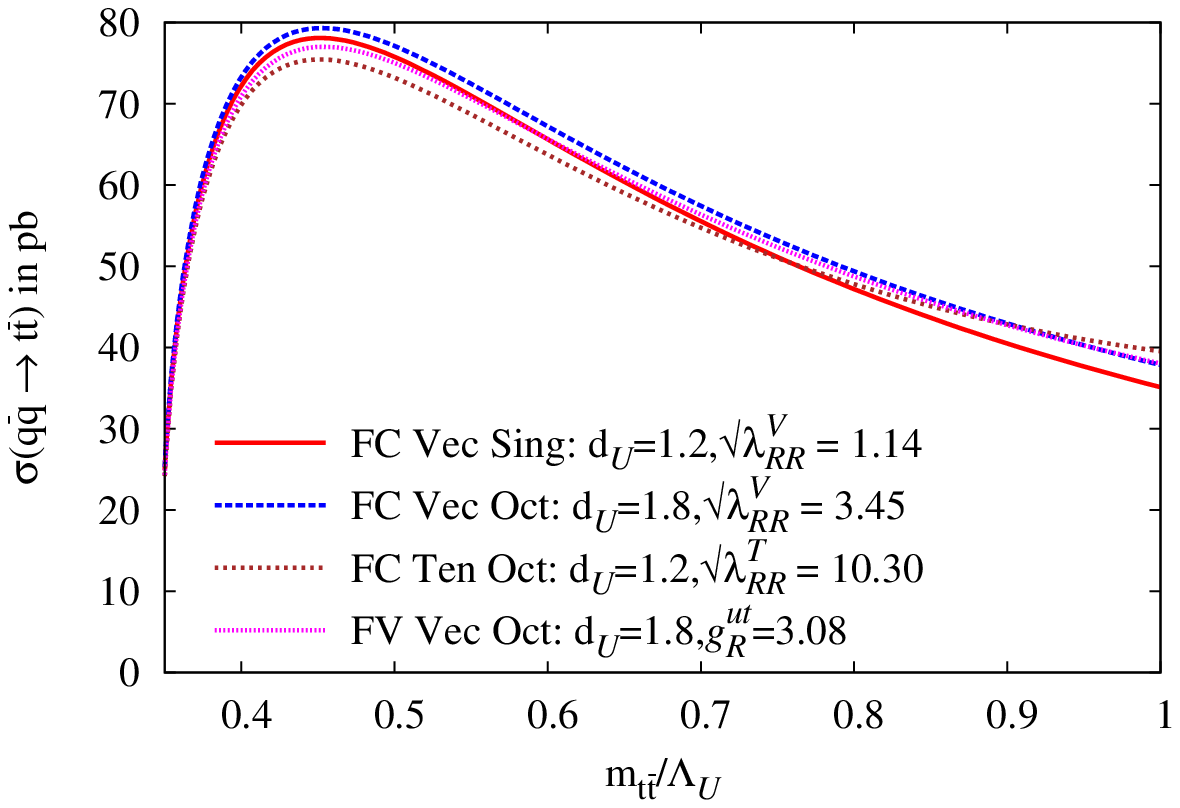}}
  \caption{\small \em{ Variation of the cross-section for the process $q \bar q \to t \bar t$ with $\sqrt{\hat s}\equiv$ \mttb in units of $\Lambda_{\cal U}$. Panel (a) gives the behaviour
 corresponding to the maximum allowed interaction strengths such that
 unitarity is not violated until 1~TeV. Panel (b) shows the $\sqrt{\hat s}$ variation for the four focus points identified in figure~\ref{fig:mttbar} illustrating that the parameter values that explain the \afbt anomaly preserve unitarity.}  }
  \label{fig:unitarity}
\end{figure*}

We summarize our observations here:
\ben
\item In the
range of $\du$ allowed by completely conformal theory, i.e., $\du>3$,
appreciable positive \afbt is obtained, albeit for very high values of
unparticle couplings which may not be in the perturbative regime. 
\par
 In figure~\ref{fig:unitarity}, we show the behaviour of
 cross section for the process $q \bar q \to t \bar t$ as a function of the parton center of mass energy $\sqrt{\hat s} \equiv m_{t\bar t}$ varying from the threshold value to 1 TeV. 
Investigating the perturbative nature of the couplings for the fully conformal theory in
 figure~\ref{fig:unitarity-conformal}, we plot the behaviour
 corresponding to the maximum allowed interaction strengths such that
 unitarity is not violated until 1~TeV. These upper limits on the interaction strengths may be read off
 directly from the figure e.g.  the interaction strength  $\sqrt{\lrr}/(\lau)^{\du -1} < 200$~(TeV)$^{1-\du}$  in case of FC vector singlet.
 For a given \du, this limit can translate into an upper bound on the
 coupling  for fixed \lau or a lower bound on \lau for a fixed $\sqrt{\lrr}$.
 Although these values contribute to the cross section \crtt within the
 experimental limits but are unable to explain the observed large value
 of \afbt. Further, as mentioned earlier, at such high values of \du, the
 SM contact interactions cannot be ignored. Hence these values of \du
 are not phenomenologically interesting. 
\item The \afbt anomaly in top pair production can be explained by the
process mediated by FC vector color singlet unparticles when only the
RH(or LH) couplings are present (i.e. the case(a) or (b) in our
article). When $1<\du<2$, one gets appreciable positive \afbt keeping
the other observables, namely, \crtt and \spincorr within the CDF
observed values. \afbt is higher for values closer to $\du = 1$. 

\item In case of FC vector color octet, the cases (a) and (b) give
  favorable values of \afbt for a large range of \du values in the
  region $1<\du<2$ and for the couplings that do not modify the cross
  section and spin correlation appreciably from the SM values. Higher
  the value of \du, smaller is the value of \afbt for a given
  coupling. If the FC octet unparticles have axial vector couplings (case(d))
  then positive \afbt is obtained only for $1.5<\du<2$. To explain the
  \afbt anomaly for \du -values in the region $1<\du<1.5$ for case (d), one has to
  assume that couplings of unparticle with light quarks are of
  opposite sign to those with top quark, just like in the case of
  axigluon models. With this non-universal choice of couplings,
  appreciable positive \afbt may be obtained even for $3<\du<3.5$.
\item With unparticle operators assumed to be vector having FC couplings,
whether being singlet or octet under $SU(3)_C$, the spin FB asymmetry \cfbt = \afbt.
\item Additional factors of $\bt$ and $\Lambda_{\cal U}^{-2}$ in the cross section
suppress the tensor unparticle effects compared to vector
unparticles. Assuming the tensor unparticle to be FC singlet, there is no
parameter space that can explain the \afbt anomaly at the same time
keeping the other observables within the experimental limits. If,
however, tensor unparticles are color octets, appreciable \afbt is
obtained for cases (a) and (c) with $1<\du<1.5$ (closer is \du to 1,
higher is \afbt ). Moreover, with the presence of tensor unparticles,
$\cfbt \ne \afbt $.
\item In the presence of FV couplings of vector unparticle involving first and
third generation quarks only, color octet unparticle
gives the  appreciable positive \afbt for $1<\du<2$. Color singlet gives
sufficiently positive \afbt only for values of \du very close to 1.
With both same and opposite helicity amplitudes contributing to \afbt
in presence of FV couplings, \afbt $\ne$ \cfbt.

\item We observe that although a large parameter region in FV sector  is consistent with CDF and D\O\, data, it gets  constrained from the same sign top/anti-top production.
\par  Considering the recent measurement of the top decay width, the parameter region shrinks to a large extent. Unparticles are likely to escape the  detection accounting for the missing energy and transverse momentum or they will show up as pair of light quark jets/leptons. Then one would like to compare these processes with the experimentally constrained FCNC partial decay width $t\to Z+ jet$, which is $\le$ 3.7\% from CDF~\cite{Aaltonen:2008ac} and $\le$ 3.2\% from D\O\ ~\cite{Abazov:2011qf} respectively. However, the partial decay width of $t\to {\cal U}\, u$  is expected to be larger than $t\to Z+jet$ as it's phase space allows the continuum spectrum for unparticles unlike $Z$ boson.

\item Our results and analysis are consistent with the inception of  non-zero infrared cut-off as long as $\mu^2 \ll$  momentum invariants involving unparticles in all $\hat s$, $\hat t$ and $\hat u$ channels at the parton level.
\item We have identified some focus points of the model which  can  
explain the \afbt anomaly  (given in table~\ref{afbmttdata}) based on the 
$\chi^2$ analysis performed with two independent parameters of the theory 
\du and the coupling $\sqrt{\lambda_{ij}}/g^{ut}_{i}$ for a given  $
\Lambda_{\cal U}$. Following the scanning of the unconstrained parameter 
region we computed the  $\chi^2_{\rm min.}$ {\it w.r.t} the deviation 
from the experimentally observed \mttb distribution of \afbt spread over 
seven bins at the TeVatron for fixed  $\Lambda_{\cal U}=1$ TeV. We find 
that all  four points corresponding  to the $\chi^2_{\rm min.}$ neither 
conflict with the other measured  observables of  $p\bar p\to t\bar t$ 
nor do they transgress  the  allowed upper limit of same sign top pair 
production cross-section and the observed total top decay width.
\par We investigated further and established that all these focus points 
preserved the unitarity which is illustrated in figure
\ref{fig:unitarity-favcases}.
\item The new physics effect can either show up at the top pair
production or in the top decay channels. However, in an experiment the 
top and anti-top are reconstructed from all observed decay products using 
the SM template  which contains the SM  $tWb$ vertices. Therefore unparticle induced  
decay channels of top/antitop  to the visible spectrum and missing 
energy will add up to the cross-section reported in the experiment. We 
have, in this article, considered the unpartcle contribution only to the 
top pair production with SM decay of top/antitop.  The total cross section
including the additional top/antitop decay {\it via} unparticles is likely
 to reach the present upper bound of allowed one sigma band at much 
lower values of the couplings in comparison to that  shown in figures \ref
{fig:FC_vec_sing_sigma}, \ref{fig:FC_vec_oct_sigma}, \ref
{fig:FC_ten_sing_sigma},\ref{fig:FC_ten_oct_sigma}, \ref
{fig:FV_vec_sing_sigma-1-3}, \ref{fig:FV_vec_sing_sigma-3.2} and \ref
{fig:FV_vec_oct_sigma-1-3}. 
\item The inclusion of the down sector in the Lagrangian \eqref{lag-V2}
induces the $bq{\cal U}$ vertices in various modes $B$ decay and in
the $b\bar b$ production at the hadron colliders. Since  our analysis
involves the flavor violation among the first and third generation quarks,
the flavor-violating couplings from rare decay modes $B^0 \to K^0 \bar K^0$
and $B^{\pm,0} \to \phi \pi^{\pm,0}$ (which have only $b\to d$ penguin
contributions in the SM) constrains the left chiral $bd{\cal U}$
couplings.~\cite{Mohanta:2007zq} The study of $CP$ phase of $B$ meson
mixing also constraints the left chiral vector unparticle current~\cite{
Parry:2008sr}. SU(2) gauge invariance would then impose the constraints
on the left chiral couplings in the up quarks sector as well. Thus the
case with the pure right handed couplings remains unconstrained.
Therefore our analysis for the flavor violating focus point which
arises from the  combination involving  $g_L^{ut}=0$ and $g_R^{ut}\neq 0$
remains unconstrained from the $B$ physics.
\een
It is worthwhile to probe the contribution of unparticles at LHC  and
correlate findings from the top sector at TeVatron. We expect that the
present and forthcoming measurements with higher luminosity data for
top pair cross section, charge asymmetry and spin correlation will
severely constrain the parameter space of flavor conserving and violating
unparticle interactions. In addition the $\mtt$ distribution of the cross
section~\cite{:2012txa,Chatrchyan:2012cx} and charge asymmetry will
contrain the new physics possiblitities. With the improved $b$-tagging
efficiency, the unparticle contribution to the $b \bar b$ production is
likely to influence the model analysis. The contribution of unparticles
to the light quark dijet production is likely to put an upper bound on
the flavor conserving light quark-unparticle interaction strength
\cite{EXO-10-001-pas,QCD-10-011,ATLAS-CONF-2012-134}.
\par
Recently the measurement of same sign top pairs at 
CMS has ruled out the favored $Z^\prime$ model of
$\afbt$~\cite{Chatrchyan:2011dk}. Therefore one expects that the same 
sign top pair production through unparticles will narrow down the allowed 
flavor violating parameter space.
 In the same spirit, the impact of the constraints on the FCNC decays of top quarks observed at ATLAS~\cite{Aad:2012ij} and 
CMS~\cite{:2012sd} needs to be studied in this model. 
\par Presently this analysis and an estimate of unparticle contribution to all these processes for LHC is in progress. 
\begin{widetext}
\begin{center}
\begin{table}[h]
\begin{tabular}{ccccccc}\hline\hline
$m_{t\bar t}$ & \afbt ($\pm$ stat.) & {\small NLO (QCD+EW)} & {\small FC Vector Singlet} & {\small  FC Vector Octet} & 
{\small FC Tensor Octet}    &{\small FV Vector Octet} \\
&&$t\bar t$+Bkg &case (a)&case (a)&case (d)&case (a)\\\hline
$<$ 400 GeV/c$^2$ & -0.006 $\pm$ 0.031 & 0.012  & 0.03 & 0.03 & 0.03 &  0.03  \\
400 $-$ 450 GeV/c$^2$ &0.065 $\pm$ 0.040 &0.023 & 0.05 & 0.05 & 0.06 &  0.05 \\
450 $-$ 500 GeV/c$^2$ &0.118 $\pm$ 0.051 &0.022 & 0.06 & 0.06 & 0.07 &  0.07  \\
500 $-$ 550 GeV/c$^2$ &0.159 $\pm$ 0.069 &0.041 & 0.09 & 0.10 & 0.10 &  0.10  \\
550  $-$600 GeV/c$^2$ &0.118 $\pm$ 0.088 &0.066 & 0.13 & 0.13 & 0.15 &  0.14  \\
600 $-$ 700 GeV/c$^2$ &0.273 $\pm$ 0.103 &0.065 & 0.21 & 0.26 & 0.29 &  0.27  \\
$\geq$  700 GeV/c$^2$& 0.306 $\pm$ 0.136& 0.107 & 0.28 & 0.34 & 0.41 &  0.39  \\
\hline\hline
$\chi^2_{\rm min.}$ & -& -&2.80 &3.16&3.36&3.65\\ 
\hline\hline 
Slope of &&&&&&\\
  Best-Fit Line & $\left(8.9\pm2.3\right)\times 10^{-4}$ & $\left(2.2\pm2.3\right)\times 10^{-4}$ & $6.7\times 10^{-4}$ & $8.8\times 10^{-4}$ & $1\times 10^{-3}$ & $9.8\times 10^{-4}$\\ 

\hline\hline
\end{tabular}

\caption{\small \em{The first three columns give the bin limits of the \mttb, the observed \afbt  with error and the NLO (QCD+EW) generated \afbt respectively \cite{afbt-cdf-mar2012}. The next four consecutive columns provide the differential \afbt  corresponding to the model parameters (given in figure \ref{fig:mttbar}) leading to $\chi^2_{\rm min.}$ at fixed $\Lambda_{\cal U}=1$ TeV. The penultimate line gives the  $\chi^2_{\rm min.}$ for  respective cases. The last line in the table gives the slope of the best fit line with the simulated data.}}
\label{afbmttdata}
\end{table}

\end{center}
\end{widetext}
\section*{Acknowledgments}
The authors would like to thank D. Choudhury, Mukesh Kumar and A. Goyal for fruitful discussions. We
acknowledge the partial support from DST, India under grant SR/S2/HEP-12/2006. SD and RI  will like to acknowledge 
the UGC research award   and CSIR  JRF respectively for the partial financial support.  MD would like to thank the IUCAA, Pune for hospitality while part of this
work was completed. We also thank RECAPP, HRI for local hospitality where this work was initiated.
We would like to thank D. Ghosh and M.~Perez-Victoria for  valuable comments.
\appendix
\def\theequation{\thesection.\arabic{equation}}
\setcounter{equation}{0}
\section{Computation of Helicity Amplitudes}
\label{app:helamp}
The  generalized Lagrangian for SM and the $J=1,2$ neutral current via color singlets and octets are given  as
\bea
&&{\cal L}_{\rm QCD}=-g_s\left\{T^a\right\}_{ji}  A^a_\mu \sum_f \bar q_{fj} \gamma^\mu q_{fi}, \label{lag-QCD}
\eea
where $g_s$ is the QCD coupling, $f$ is the flavor index and $i,j$ are the color indices.
\bea
&&{\cal L}_{\rm NC}= - \sum_f \bar q_f \gamma^\mu \Big[eQ_f A_\mu \nn\\
&&\qquad\qquad + \frac{e}{s_W c_W}\left(T^3_f P_L - s^2_W Q_f \right) Z_\mu \Big] q_f \label{lag-NC},
\eea
where $e$ = electromagnetic coupling, $Q_f$ = charge of quark $q_f$ in units of $e$, $s_W = \sin\theta_W, c_W = \cos\theta_W$, $\theta_W$ = Weinberg angle.
\bea
&&{\cal L}_{J=1}^{(s)}=\frac{g_s}{\la_{\cal U}^{\du-1}}\sum_{\substack{\alpha=L,R\\\bf n=1,8}} g^{{\cal U}^{\bf n}_V\bar q q}_\alpha\left\{T^a_{\bf n}\right\}_{ji}  {\cal O}^{{\bf n}, a}_\mu \nn\\
&&\qquad\qquad\qquad\times \sum_f \bar q_{fj} \gamma^\mu P_\alpha q_{fi} + h.c. \label{lag-V1}
\eea
\bea
&&{\cal L}_{J=1}^{(t)}=\!\frac{g_s}{\la_{\cal U}^{\du-1}}\sum_{\substack{\alpha=L,R\\\bf n=1,8}} {\cal O}^{{\bf n}, a}_\mu \left\{T^a_{\bf n}\right\}_{ji}
                  \left[g^{{\cal U}_V^{\bf n} \bar t u}_\alpha \bar t_j  \gamma^\mu P_\alpha q_{fi} \right. \nn\\
&&\qquad\qquad\left . + g^{{\cal U}_V^{\bf n} \bar b u}_\alpha \bar b_j  \gamma^\mu P_\alpha q_{fi}+h.c.\right] \label{lag-V2}
\eea
\bea
&&{\cal L}_{J=2}^{(s)}= \frac{-g_s}{4\,\la_{\cal U}^{\du}}\sum_{\substack{\alpha=L,R\\\bf n=1,8}} g^{{\cal U}^{\bf n}_T\bar q q}_\alpha\left\{T^a_{\bf n}\right\}_{ji} {\cal O}^{{\bf n}, a}_{\mu\nu}\nn\\
&&\qquad\times \sum_f \bar q_{fj}
i (\gamma_\mu \partial_\nu + \gamma_\nu \partial_\mu ) P_\alpha q_{fi} + h.c. \label{lag-T1} 
\eea
\bea
&&{\cal L}_{J=2}^{(t)}= \frac{-g_s}{4\,\la_{\cal U}^{\du}}\sum_{\substack{\alpha=L,R\\\bf n=1,8}} {\cal O}^{{\bf n}, a}_{\mu\nu} \left\{T^a_{\bf n}\right\}_{ji} \nn\\&&\qquad\qquad
\times \,i\,\left[ g^{{\cal U}_T^{\bf n} \bar t u}_\alpha \bar t_j  
 (\gamma_\mu \partial_\nu + \gamma_\nu \partial_\mu ) P_\alpha q_{fi} \right.\nonumber\\
&&\qquad\quad\left. + g^{{\cal U}_T^{\bf n} \bar b u}_\alpha \bar b_j  (\gamma_\mu \partial_\nu + \gamma_\nu \partial_\mu ) P_\alpha q_{fi} + h.c\right]. \label{lag-T2}
\end{eqnarray}
 The matrix element is computed in this appendix fixing the color flow of the particles involved in the process $ q\bar q \rightarrow t \bar t $ as following
\begin{eqnarray}
\left\{q(p_q)\right\}_i\,\, \left\{\bar q(p_{\bar q})\right\}_j \rightarrow \left\{t(p_t)\right\}_k\,\, \left\{\bar t(p_{\bar t})\right\}_l
\end{eqnarray}
\par The matrix elements corresponding to the given Lagrangian in eq. \eqref{lag-QCD}-\eqref{lag-T2}
\begin{widetext}
\begin{eqnarray}
\sum_i i{\cal M}_i&=& i{\cal M}_1({\rm QCD}) + i{\cal M}_2({\rm NC}) + i{\cal M}_3(\hat s;J=1;{\bf 1}) + i{\cal M}_4(\hat s;J=1;{\bf 8}) + i{\cal M}_5(\hat t;J=1;{\bf 1}) + i{\cal M}_6(\hat t;J=1;{\bf 8})\nn\\
&&+ i{\cal M}_7(\hat s;J=2;{\bf 1}) + i{\cal M}_8(\hat s;J=2;{\bf 8}) + i{\cal M}_9(\hat t;J=2;{\bf 1}) + i{\cal M}_{10}(\hat t;J=2;{\bf 8}) 
\end{eqnarray}
\begin{eqnarray}
&&\sum_i i{\cal M}_i = C_{\bf QCD} \Big[ \bar v(p_{\bar q},\lambda_{\bar q})_i i\gamma^\mu u(p_q,\lambda_q)_{l}\Big] \frac{-ig_{\mu\nu}}{\hat s} \Big[ \bar u(p_t,\lambda_t)_k i\gamma^\nu v(p_{\bar t},\lambda_{\bar t})_j\Big]\nn\\
&& + C_{\bf NC} \Big[ \bar v(p_{\bar q},\lambda_{\bar q})_i i\gamma^\mu u(p_q,\lambda_q)_{l}\Big] \frac{-ig_{\mu\nu}}{\hat s} \Big[ \bar u(p_t,\lambda_t)_k i\gamma^\nu v(p_{\bar t},\lambda_{\bar t})_j\Big]\nn\\
&& + C_{\bf NC} \Big[ \bar v(p_{\bar q},\lambda_{\bar q})_i i\gamma^\mu \left(T^3_q P_L - s^2_W Q_q \right) u(p_q,\lambda_q)_{l}\Big] \frac{-ig_{\mu\nu}}{\hat s - m_Z^2 + i m_Z \Gamma_Z} \Big[ \bar u(p_t,\lambda_t)_k i\gamma^\nu \left(T^3_t P_L - s^2_W Q_t \right) v(p_{\bar t},\lambda_{\bar t})_j\Big]\nn\\
&+&\sum_{{\bf n}\equiv{\bf 1},{\bf 8}} C_{\bf n}^{s} \Big[ \bar v(p_{\bar q},\lambda_{\bar q})_i\frac{i\gamma^\mu}{\Lambda^{\du - 1}} \left(g^{{\cal U}^{\bf n}_V\bar qq}_LP_L+g^{{\cal U}^{\bf n}_V\bar qq}_R P_R \right)  u(p_q,\lambda_q)_{l}\Big] \frac{iA_{d_{\cal U}}(-\hat s)^{d_{\cal U}-2}}{2\sin(d_{\cal U}\pi)} \Big[g_{\mu\nu} - a \frac{(p_q + p_{\bar q})^\mu (p_q + p_{\bar q})^\nu}{(p_q + p_{\bar q})^2} \Big] \nn\\
&&\qquad\qquad\qquad \times \Big[ \bar u(p_t,\lambda_t)_k\frac{i\gamma^\nu}{\Lambda^{\du - 1}}\left(g^{{\cal U}^{\bf n}_V\bar tt}_L  P_L + g^{{\cal U}^{\bf n}_V\bar tt}_R  P_R\right) v(p_{\bar t},\lambda_{\bar t})_j\Big]\nn
\eea

\bea
&-&\sum_{{\bf n}\equiv {\bf 1},{\bf 8}} C_{\bf n}^{t} \Big[ \bar v(p_{\bar q},\lambda_{\bar q})_i\frac{i\gamma^\mu}{\Lambda^{\du - 1}} \left(g^{{\cal U}^{\bf n}_V\bar t \bar q}_LP_L+g^{{\cal U}^{\bf n}_V\bar t \bar q}_R P_R \right) v(p_{\bar t},\lambda_{\bar t})_{j}\Big] \frac{iA_{d_{\cal U}}(-\hat t)^{d_{\cal U}-2}}{2\sin(d_{\cal U}\pi)} \Big[g_{\mu\nu} - a \frac{(p_q - p_t)^\mu (p_q - p_t)^\nu}{(p_q - p_t)^2} \Big] \nn\\
&&\qquad\qquad\qquad\times \Big[ \bar u(p_t,\lambda_t)_k\frac{i\gamma^\nu}{\Lambda^{\du - 1}} \left(g^{{\cal U}^{\bf n}_V tq}_LP_L+g^{{\cal U}^{\bf n}_V tq}_R P_R \right) q(p_{ q},\lambda_{ q})_l\Big]\nn\\
&+&\sum_{{\bf n}\equiv{\bf 1},{\bf 8}} C_{\bf n}^{s} \Big[ \bar v(p_{\bar q},\lambda_{\bar q})_i\frac{-i\left[\gamma^\mu(p_q - p_{\bar q})^\nu + \gamma^\nu(p_q - p_{\bar q})^\mu\right]}{4\Lambda^{\du}} \left(g^{{\cal U}^{\bf n}_T\bar qq}_LP_L+g^{{\cal U}^{\bf n}_T\bar qq}_R P_R \right)  u(p_q,\lambda_q)_{l}\Big] \frac{iA_{d_{\cal U}}(-\hat s)^{d_{\cal U}-2}}{2\sin(d_{\cal U}\pi)} {\cal T}_{\mu\nu,\alpha\beta} \nn\\
&&\qquad\qquad\qquad \times \Big[ \bar u(p_t,\lambda_t)_k\frac{i\left[\gamma^\alpha (p_t - p_{\bar t})^\beta + \gamma^\beta(p_t - p_{\bar t})^\alpha\right]}{4\Lambda^{\du}}\left(g^{{\cal U}^{\bf n}_T\bar tt}_L  P_L + g^{{\cal U}^{\bf n}_T\bar tt}_R  P_R\right) v(p_{\bar t},\lambda_{\bar t})_j\Big]\nn\\
&-&\sum_{{\bf n}\equiv {\bf 1},{\bf 8}} C_{\bf n}^{t} \Big[ \bar v(p_{\bar q},\lambda_{\bar q})_i\frac{-i\left[\gamma^\mu(p_t + p_q)^\nu + \gamma^\nu(p_t + p_q)^\mu\right]}{4\Lambda^{\du}} \left(g^{{\cal U}^{\bf n}_T\bar t \bar q}_LP_L+g^{{\cal U}^{\bf n}_T\bar t \bar q}_R P_R \right) v(p_{\bar t},\lambda_{\bar t})_{j}\Big] \frac{iA_{d_{\cal U}}(-\hat t)^{d_{\cal U}-2}}{2\sin(d_{\cal U}\pi)} {\cal T}_{\mu\nu,\alpha\beta} \nn
\end{eqnarray}
\begin{eqnarray}
&&\qquad\qquad\qquad\times \Big[ \bar u(p_t,\lambda_t)_k\frac{i\left[\gamma^\alpha(p_{\bar q} + p_{\bar t})^\beta + \gamma^\beta(p_{\bar q} + p_{\bar t})^\alpha\right]}{4\Lambda^{\du}} \left(g^{{\cal U}^{\bf n}_T tq}_LP_L+g^{{\cal U}^{\bf n}_T tq}_R P_R \right) q(p_{ q},\lambda_{ q})_l\Big]\nn\\ 
\end{eqnarray}
\label{matel}
\end{widetext}
Here $C^{\bf s}_{\bf n}$ is the color propagator for the color octet vectors/tensors in $s$ channel and $C_{\bf n}^{t}$ is color propagator  for the color octet ($n\equiv 8$) and singlet ($n\equiv 1$)  vectors/tensors in $t$ channel. To compute the squared and interference terms with these color propagators, we provide the color factors in the following table.
\begin{center}
\begin{table*}[h,t]
\begin{tabular}{|c|c|c|c|c|}\hline\hline
Color Factor&FC Octet  &FV Octet &   FC  singlet&FV  singlet \\
$q(p_q)_i \bar q(p_{\bar q})_j \rightarrow $
&$\left\{j,i\right\}\,\,\,\left\{k,l\right\}$& $\left\{i,k\right\}\,\,\,\left\{l,j\right\}$&$\left\{j,i\right\}\,\,\,\left\{k,l\right\}$&$\left\{i,k\right\}\,\,\,\left\{l,j\right\}$\\
$ t(p_t)_k\bar t(p_{\bar t})_l$&$s$ chan.&$t$ chan.&$s$ chan.&$t$ chan.\\
&$C^s_{\bf 8}$=$C_{ \bf QCD}$ &$C^{t}_{\bf 8}$ &$C^{ s}_{\bf 1}$=$C_{ \bf NC}$ & $C^{ t}_{\bf 1}$\\
\hline
Int. with QCD &2&-2/3 &0&4\\
\hline
Int. with NC &0&0 &9&3\\
\hline
Squared term & 2 &2 & 9&9    \\
\hline\hline
\end{tabular}
\caption{Color Factors for the interference and the squared terms of $s$ and $t$  channels for color singlet and octet vectors and tensors.}
\end{table*}
\label{col-fac}
\end{center}
We compute the helicity amplitudes in the center of mass frame $q\bar q$. The momentum assignments are 
\bea
p_q= \frac{\sqrt{\hat s}}{2} (1,0,0,1),&&p_{\bar q} = \frac{\sqrt{\hat s}}{2} (1,0,0,-1), 	\notag\\
p_t= \frac{\sqrt{\hat s}}{2} (1,\beta_t s_\theta,0,\beta_t c_\theta), &&
p_{\bar t} = \frac{\sqrt{\hat s}}{2} (1,-\beta_t s_\theta,0,-\beta_t c_\theta)\nn\\
\eea
where $s_\theta\equiv\sin\theta;\,\, c_\theta\equiv\cos\theta ,\beta_t = \sqrt{1 - \frac{4m_t^2}{\hat s}}$ and $\theta$ is the angle between $q$ and $t$ momenta.

\subsection {\bf  Helicity amplitudes for $q \bar q \to t \bar t$ in Standard Model}

The helicity amplitudes for $q \bar q \to t \bar t$ in Standard Model are given by 
\begin{eqnarray}
&&{\cal M}_{1,2}(+ - \pm \pm)
  = \nn\\
  &&\quad
  \mp g_R
  \left[g_L + g_R\right]
  \frac{1}{2} \sqrt{1 - \beta_t^2}
  s_\theta \label{app-sm1}\eea
\bea
&&{\cal M}_{1,2}(+ - \pm \mp)
 = \nn\\
  && \quad
  \pm g_R
  \left[g_L \left(1 \mp \beta_t\right) + g_R \left(1 \pm \beta_t\right)\right]
  \frac{1}{2}
  (1 \pm c_\theta)\nn\\ &&\label{app-sm2}\eea
\bea
&&{\cal M}_{1,2}(- + \pm \pm)
 = \nn\\
  &&\quad
  \mp g_L
  \left[g_L + g_R\right]
  \frac{1}{2} \sqrt{1 - \beta_t^2}
  s_\theta\label{app-sm3}\eea
\bea
&&{\cal M}_{1,2}(- + \pm \mp)
  = \nn\\
  && \quad
  \mp g_L
  \left[g_L \left(1 \mp \beta_t\right) + g_R \left(1 \pm \beta_t\right)\right]
  \frac{1}{2}
  (1 \mp c_\theta).\nn\\ &&\label{app-sm4}
\end{eqnarray}

In above if the helicity amplitudes are gluon mediating then they are in units of $g_s^2$, if photon mediating then they are in units of $(eQ_q)(eQ_t)$ and if $Z$-boson mediating then they are in units of $(e/s_W c_W)^2 \,\, s/\left(s-m_Z^2\right)$.  Here for gluon and photon mediating process $g_L = g_R = 1$ whereas for $Z$-boson mediating process $g_L = T^3_q - s_W^2 Q_q, g_R = - s_W^2 Q_q$.

\subsection {\bf  Helicity amplitudes for  $q \bar q \to t \bar t$ via flavor
conserving vector unparticles}

The helicity amplitudes mediated by the flavor conserving vector unparticle ${\cal M}_{3,4}(\hat s;J=1;{\bf n})$ in units of $g_s^2(-1)^{d_{\cal U}-3} \frac{A_{d_{\cal U}}}{2 \sin (d_{\cal U} \pi)} \left[\frac{\hat s}{\Lambda_{\cal U}^2}\right]^{d_{\cal U}-1}$ are given by 

\begin{eqnarray}
&&{\cal M}_{3,4}(+ - \pm \pm)
  = \nn\\
  &&\quad
  \mp g^{{\cal U}_V^{\bf n}\bar q q}_R
  \left[g^{{\cal U}_V^{\bf n}\bar tt}_L + g^{{\cal U}_V^{\bf n}\bar tt}_R\right]
  \frac{1}{2} \sqrt{1 - \beta_t^2}
  s_\theta \label{app-vec-fc1}\eea
\bea
&&{\cal M}_{3,4}(+ - \pm \mp)
 = \nn\\
  && \quad
  \pm g^{{\cal U}_V^{\bf n}\bar q q}_R
  \left[g^{{\cal U}_V^{\bf n}\bar tt}_L \left(1 \mp \beta_t\right) + g^{{\cal U}_V^{\bf n}\bar tt}_R \left(1 \pm \beta_t\right)\right]
  \frac{1}{2}
  (1 \pm c_\theta)\nn\\ &&\label{app-vec-fc2}\eea
\bea
&&{\cal M}_{3,4} (- + \pm \pm)
  =
  {\cal M}_{3,4} (+ - \pm \pm)|_{L \leftrightarrow R} 
\label{app-vec-fc3}
 \eea
\bea
&&{\cal M}_{3,4} (- + \pm \mp)
  =
  {\cal M}_{3,4} (+ - \mp \pm)|_{L \leftrightarrow R}.
\label{app-vec-fc4}
\end{eqnarray}

\subsection {\bf  Helicity amplitudes for  $q \bar q \to t \bar t$ via flavor violating vector unparticles}
The following helicity amplitudes ${\cal M}_{5,6}(\hat t;J=1;{\bf n})$ are given in units of $ g_s^2 \frac{A_{d_{\cal U}}}{2  \sin (d_{\cal U} \pi)} 
  \left[\frac{\hat t}{\Lambda_{\cal U}^2}\right]^{d_{\cal U}-1}
  \frac{\hat s}{\hat t}$.
\begin{eqnarray}
&&{\cal M}_{5,6} (++\pm\pm) 
 = \nn\\
  && \quad
  \pm g^{{\cal U}_V^{\bf n}\bar tq}_L g^{{\cal U}_V^{\bf n}\bar tq^\prime}_R  (1\pm\beta_t)
	\left[1 \pm a \frac{m_t^2}{4\hat t}(1 \pm c_\theta)\right]
	\label{app-vec-fv1}\eea
\bea
&&{\cal M}_{5,6} (++\pm\mp) 
 = \nn\\
  &&\quad 
  g^{{\cal U}_V^{\bf n}\bar tq}_L g^{{\cal U}_V^{\bf n}\bar tq^\prime}_R a \frac{m_t^2}{4\hat t}\sqrt{1 - \beta_t^2} s_\theta
  \label{app-vec-fv2}\eea
\bea
&&{\cal M}_{5,6} (+-\pm\pm) 
 =\nn\\
  &&\quad
  - g^{{\cal U}_V^{\bf n}\bar tq}_R g^{{\cal U}_V^{\bf n}\bar tq^\prime}_R  \sqrt{1-\beta_t^2}\frac12 s_\theta \left[1 + a \frac{m_t^2}{4\hat t}\right] \nn\\ &&
  \label{app-vec-fv3}\eea
\bea
&&{\cal M}_{5,6} (+-\pm\mp) 
 =\nn\\
  &&\quad
  - g^{{\cal U}_V^{\bf n}\bar tq}_R g^{{\cal U}_V^{\bf n}\bar tq^\prime}_R  (1\pm\beta_t)\frac12(1\pm c_\theta) \left[1 + a \frac{m_t^2}{4\hat t}\right]\nn\\ &&  \label{app-vec-fv4}\eea
\bea
&&{\cal M}_{5,6} (-+\pm\pm) 
 = {\cal M}_{5,6} (+-\mp\mp)|_{L \leftrightarrow R}
	\label{app-vec-fv5}\eea
\bea
&&{\cal M}_{5,6} (-+\pm\mp) 
 = {\cal M}_{5,6} (+-\mp\pm)|_{L \leftrightarrow R}
	\label{app-vec-fv6}\eea
\bea
&&{\cal M}_{5,6} (--\pm\pm) 
 = {\cal M}_{5,6} (++\mp\mp)|_{L \leftrightarrow R} 
	\label{app-vec-fv7}\eea
\bea
&&{\cal M}_{5,6} (--\pm\mp) 
 = {\cal M}_{5,6} (++\mp\pm)|_{L \leftrightarrow R}.
	\label{app-vec-fv8}
\end{eqnarray}
\subsection {\bf  Helicity amplitudes for  $q \bar q \to t \bar t$ via flavor conserving tensor unparticles}
The following helicity amplitudes ${\cal M}_{7,8}(\hat s;J=2;{\bf n})$ are given in units of $g_s^2(-1)^{d_{\cal U}-2} \frac{A_{d_{\cal U}}}{2 \sin (d_{\cal U} \pi)} 
  \left[\frac{\hat s}{4\Lambda_{\cal U}^2}\right]^{d_{\cal U}}\, 4 \du (\du - 1)$.
\begin{eqnarray}
&&{\cal M}_{7,8} (+ - \pm \pm)
  =\nn\\
  &&\quad
  \mp 
  g^{{\cal U}_T^{\bf n}\bar qq}_R
  [g^{{\cal U}_T^{\bf n}\bar tt}_L + g^{{\cal U}_T^{\bf n}\bar tt}_R]
  \beta_t \sqrt{1 - \beta_t^2}
  s_\theta c_\theta 
\label{app-ten-fc1}\eea
\bea
{\cal M}_{7,8} (+ - \pm \mp)
  &=&
  g^{{\cal U}_T^{\bf n}\bar qq}_R
  \left[g^{{\cal U}_T^{\bf n}\bar tt}_L \left(1 \mp \beta_t\right) + g^{{\cal U}_T^{\bf n}\bar tt}_R \left(1 \pm \beta_t\right)\right] \nn\\
  && \quad \times
  \frac{1}{2}\beta_t
  (1 \pm c_\theta) (1 \mp 2 c_\theta) \nn \\
 \label{app-ten-fc2}\eea
\bea
&&{\cal M}_{7,8} (- + \pm \pm)
  =
  {\cal M}_{7,8} (+ - \pm \pm)|_{L \leftrightarrow R} 
\label{app-ten-fc3} \eea
\bea
&&{\cal M}_{7,8} (- + \pm \mp)
  =
  {\cal M}_{7,8} (+ - \mp \pm)|_{L \leftrightarrow R}.
\label{app-ten-fc4}
\end{eqnarray}

\subsection {\bf  Helicity amplitudes for  $q  q \to t  t$ via flavor violating vector unparticles}\label{app:same_sign_hel_amp}
The following helicity amplitudes of $t$-channel diagram are given in units of $g_s^2 \frac{A_{d_{\cal U}}}{2\sin (d_{\cal U} \pi)} 
  \left[\frac{\hat t}{\Lambda_{\cal U}^2}\right]^{d_{\cal U}-1}
  \frac{\hat s}{\hat t}$. The symbol $a$ is defined in \eqref{gprop-vec}
\begin{eqnarray}
{\cal M}^t_{++\pm\pm}
	 &= & - 
  g^{qt}_R g^{qt}_R  (1 \pm \beta_t) \left[1 + a \frac{m_t^2}{p^2} \frac12(1\pm c_\theta) \right]\nn \\
 \eea
\bea
{\cal M}^t_{++\pm\mp}
	 &= & 
  g^{qt}_R g^{qt}_R \sqrt{1 - \beta_t^2} \frac12 s_\theta\ a \frac{m_t^2}{p^2}  \eea
\bea
{\cal M}^t_{-+\pm\pm}
	 &= & \mp 
  g^{qt}_L g^{qt}_R  \sqrt{1 - \beta_t^2} \frac12 s_\theta \left[1 + a \frac{m_t^2}{p^2} \right]  \eea
\bea
{\cal M}^t_{-+\pm\mp}
	 &= & \pm 
  g^{qt}_L g^{qt}_R  (1 \mp \beta_t) \frac12(1\mp c_\theta) \left[1 + a \frac{m_t^2}{p^2} \right] \nn\\
 \eea
\bea
{\cal M}^t_{+-\pm\pm}
	 &= & {\cal M}^t_{-+\pm\pm}|_{L\to R}  \eea
\bea
{\cal M}^t_{+-\pm\mp}
	 &= & {\cal M}^t_{-+\mp\pm}|_{L\to R}  \eea
\bea
{\cal M}^t_{--\pm\pm}
	 &= & {\cal M}^t_{++\mp\mp}|_{L\to R}  \eea
\bea
{\cal M}^t_{--\pm\mp}
	 &= & - {\cal M}^t_{++\mp\pm}|_{L\to R}
\end{eqnarray}

The helicity amplitudes of $u$-channel diagram are given in units of $g_s^2 \frac{A_{d_{\cal U}}}{2\sin (d_{\cal U} \pi)} 
  \left[\frac{\hat t}{\Lambda_{\cal U}^2}\right]^{d_{\cal U}-1}
  \frac{\hat s}{\hat u}$.
\begin{eqnarray}
{\cal M}^u_{++\pm\pm}
	 &= & - {\cal M}^t{++\pm\pm}|_{\theta \to \pi + \theta}
 \eea
\bea
{\cal M}^u_{++\pm\mp}
	 &= & - {\cal M}^t{++\pm\mp}|_{\theta \to \pi + \theta}  \eea
\bea
{\cal M}^u_{-+\pm\pm}
	 &= & - {\cal M}^t{-+\pm\pm}|_{\theta \to \pi + \theta}  \eea
\bea
{\cal M}^u_{-+\pm\mp}
	 &= & {\cal M}^t{-+\pm\mp}|_{\theta \to \pi + \theta}
 \eea
\bea
{\cal M}^u_{+-\pm\pm}
	 &= & {\cal M}^u_{-+\pm\pm}|_{L\to R}  \eea
\bea
{\cal M}^u_{+-\pm\mp}
	 &= & {\cal M}^u_{-+\mp\pm}|_{L\to R}  \eea
\bea
{\cal M}^u_{--\pm\pm}
	 &= & {\cal M}^u_{++\mp\mp}|_{L\to R}  \eea
\bea
{\cal M}^u_{--\pm\mp}
	 &= & - {\cal M}^u_{++\mp\pm}|_{L\to R}
\end{eqnarray}

\section {\bf  Partial Decay Width for  $t\to q{\cal U}$ via flavor violating vector unparticles}
\label{app:decaywidth}
The matrix element of $t\to q{\cal U}$ is given by ${\cal M} = g_s\,\Lambda_{\cal U}^{1-\du}\,
		\bar u(p_q) \gamma^\mu \left(g^{{\cal U}_V^{\bf n}tq}_L P_L + g^{{\cal U}_V^{\bf n}tq}_R P_R\right) u(p_t)
		\epsilon_\mu(p_{\cal U})$.
Averaging over spin and color the matrix element squared reads as  
\bea
\overline{|{\cal M}|^2}
             &=&\frac{N_C}{6}\frac{2g_s^2}{\Lambda_{\cal U}^{2(\du-1)}} \left[\left(g^{{\cal U}_V^{\bf n}tq}_L\right)^2 + \left(g^{{\cal U}_V^{\bf n}tq}_R\right)^2\right] m_t p_q^0 \nonumber\\
&&\,\times
		\left[\frac{(a+2)m_t - 4 p_q^0}{(m_t - 2 p_q^0)}\right]
\eea
The  differential decay width for the top decaying to an unparticle and quark is given as

\begin{widetext}
\bea
d\Gamma &=& \frac{N_C}{6}\frac{\overline{|{\cal M}|^2}}{2m_t} \frac{A_{\du}}{16\pi^3}
	 \frac{p_q^0 dp_q^0 d\Omega}{(m^2_t - 2m_t p_q^0)^{2-\du}}  \nn\\
	 &=& \frac{N_C}{6}\frac{g_s^2}{\Lambda_{\cal U}^{2(\du-1)}} \frac{A_{\du}}{16\pi^3}\left[\left(g^{{\cal U}_V^{\bf n}tq}_L\right)^2 + \left(g^{{\cal U}_V^{\bf n}tq}_R\right)^2\right] \left[\frac{(a+2)m_t^2 - 4m_t p_q^0}{(m_t^2 - 2m_t p_q^0)}^{3-\du}\right]  (p_q^0)^2\,\, \theta(m_t -2p_q^0) \,\, dp_q^0 d\Omega\nn\\
&=& \frac{N_C}{6} \frac{A_{\du}}{4\pi^2} g_s^2 \left[\left(g^{{\cal U}_V^{\bf n}tq}_L\right)^2 + \left(g^{{\cal U}_V^{\bf n}tq}_R\right)^2\right] \left(\frac{m^2_t}{\Lambda_{\cal U}^2}\right)^{\du-1} m_t
	    \left[(a+2) - 4 x\right]
	  (1 - 2x)^{\du-3} x^2 \,\,\theta(1-2x)\,\,dx \, .\label{diff-decay-width}
\eea
In \eqref{diff-decay-width} we have taken $x = p_q^0/m_t$. Integrating \eqref{diff-decay-width} w.r.t. $x$ in the limit $[0,1/2]$ we  get the total decay width of top quark for $\du >2$. The color factor $N_C=3,4$ for singlet and octet unparticle respectively. On introduction of the mass gap $\mu\neq 0$ equation \eqref{diff-decay-width} becomes 
\bea
d\Gamma &=&\frac{N_C}{6} \frac{A_{\du}}{4\pi^2} g_s^2 \left[\left(g^{{\cal U}_V^{\bf n}tq}_L\right)^2 + \left(g^{{\cal U}_V^{\bf n}tq}_R\right)^2\right] \left(\frac{m^2_t}{\Lambda_{\cal U}^2}\right)^{\du-1} m_t
	    \left[(a+2) - 4 x\right]
	  (1 - 2x-x_0)^{\du-2} \, \frac{x^2}{(1-2x)} \,\,\theta(1-2x-x_0)\,\,dx\, , \nn\\
\label{diff-decay-width_massgap}
\eea
where $x_0=\mu^2/m_t^2$. Evaluating the integral in the limit of $[0, (1-x_0)/2]$, we get the partial top decay width in the region $\du >1$.  
\end{widetext}


\end{document}